\documentclass{pasa}

\title[High-$z$ Universe with {\it SPICA}]{Probing the High-Redshift
  Universe with SPICA: \\ Toward the Epoch of Reionization and
  Beyond}

\author[E.~Egami et al.]{E.~Egami$^1$,
S.~Gallerani$^2$,
R.~Schneider$^3$,
A.~Pallottini$^{2,4,5,6}$,
L.~Vallini$^{7}$,
E.~Sobacchi$^{2}$,
A.~Ferrara$^2$,
S.~Bianchi$^8$,
M.~Bocchio$^8$,
S.~Marassi$^{9}$,
L.~Armus$^{10}$,~
L.~Spinoglio$^{11}$,
A.~W.~Blain$^{12}$,
M.~Bradford$^{13}$,
D.~L.~Clements$^{14}$,
H.~Dannerbauer$^{15,16}$,
J.~A.~Fern\'andez-Ontiveros$^{11,15,16}$,
E.~Gonz\'alez-Alfonso$^{17}$,
M.~J.~Griffin$^{18}$,
C.~Gruppioni$^{19}$,
H.~Kaneda$^{20}$, 
K.~Kohno$^{21}$, 
S.~C.~Madden$^{22}$,
H.~Matsuhara$^{23}$,
P.~Najarro$^{24}$,
T.~Nakagawa$^{23}$,
S.~Oliver$^{25}$,
K.~Omukai$^{26}$,
T.~Onaka$^{27}$,
C.~Pearson$^{28}$,
I.~Perez-Fournon$^{15,16}$,
P.~G.~P\'erez-Gonz\'alez$^{29}$,
D.~Schaerer$^{30}$,
D.~Scott$^{31}$,
S.~Serjeant$^{32}$,
J.~D.~Smith$^{33}$,
F.~F.~S.~van der Tak$^{34,35}$,
T.~Wada$^{24}$,~
and~H.~Yajima$^{36}$ 
\affil{$^1$Steward Observatory, University of Arizona, 933 N.~Cherry Ave., Tucson, AZ 85721, USA}
\affil{$^2$Scuola Normale Superiore, Piazza dei Cavalieri 7, I-56126, Pisa, Italy}
\affil{$^3$Dipartimento di Fisica ``G. Marconi'', Sapienza Universit\'a di Roma, P.le A. Moro 2, 00185 Roma, Italy}
\affil{$^4$Kavli Institute for Cosmology, University of Cambridge, Madingley Road, Cambridge CB3 0HA, UK}
\affil{$^5$Cavendish Laboratory, University of Cambridge, 19 J. J. Thomson Ave., Cambridge CB3 0HE, UK}
\affil{$^6$Centro Fermi, Museo Storico della Fisica e Centro Studi e Ricerche ``Enrico Fermi'', Piazza del Viminale 1, Roma, 00184, Italy}
\affil{$^7$Leiden Observatory, Leiden University, P.O. Box 9513, NL-2300 RA Leiden, The Netherlands}
\affil{$^8$INAF, Osservatorio Astrofisico di Arcetri, Largo E. Fermi 5, 50125 Firenze, Italy}
\affil{$^9$INAF, Osservatorio Astronomico di Roma, Via di Frascati 33, I-00040 Monte Porzio Catone, Italy}
\affil{$^{10}$IPAC, California Institute of Technology, Pasadena, CA 91125, USA}
\affil{$^{11}$INAF, Istituto di Astrofisica e Planetologia Spaziali, Via Fosso del Cavaliere 100, I-00133 Roma, Italy}
\affil{$^{12}$Department of Physics and Astronomy, University of Leicester, University Road, Leicester LE1 7RH, UK}
\affil{$^{13}$Jet Propulsion Laboratory, California Institute of Technology, Pasadena, CA 91109, USA}
\affil{$^{14}$Blackett Lab, Imperial College, London, Prince Consort Road, London SW7 2AZ, UK}
\affil{$^{15}$Instituto de Astrof\'isica de Canarias (IAC), E--38205 La Laguna, Tenerife, Spain}
\affil{$^{16}$Universidad de La Laguna, Dpto. Astrof\'isica, E-38206 La Laguna, Tenerife, Spain}
\affil{$^{17}$Universidad de Alcal\'a, Departamento de F\'isica y Matem\'aticas, Campus Universitario, E-28871 Alcal\'a de Henares, Madrid, Spain}
\affil{$^{18}$School of Physics \& Astronomy, Cardiff University, The Parade, Cardiff CF24 3AA, UK}
\affil{$^{19}$INAF, Osservatorio Astronomico di Bologna, via Ranzani 1, I-40127 Bologna, Italy}
\affil{$^{20}$Graduate School of Science, Nagoya University, Furo-cho, Chikusa-ku, Nagoya 464-8602, Japan}
\affil{$^{21}$Institute of Astronomy, University of Tokyo, 2-21-1 Osawa, Mitaka, Tokyo 181-0015, Japan}
\affil{$^{22}$Laboratoire AIM, CEA/IRFU/Service d'Astrophysique, Universit\'e Paris Diderot, Bat. 709, F-91191 Gif-sur-Yvette, France}
\affil{$^{23}$Institute of Space Astronautical Science, Japan Aerospace Exploration Agency, Sagamihara, Kanagawa 252-5210, Japan}
\affil{$^{24}$Centro de Astrobiolog\'ia (CSIC/INTA), ctra. de Ajalvir km. 4, 28850 Torrej\'on de Ardoz, Madrid, Spain}
\affil{$^{25}$Astronomy Centre, Department of Physics and Astronomy, University of Sussex, Brighton BN1 9QH, UK}
\affil{$^{26}$Astronomical Institute, Tohoku University, Aoba, Sendai 980-8578, Japan}
\affil{$^{27}$Department of Astronomy, Graduate School of Science, The University of Tokyo, 7-3-1 Hongo, Bunkyo-ku, Tokyo 113-0033, Japan}
\affil{$^{28}$RAL Space, CCLRC Rutherford Appleton Laboratory, Chilton, Didcot, Oxfordshire OX11 0QX, UK}
\affil{$^{29}$Departamento de Astrof\'isica, Facultad de CC. F\'isicas, Universidad Complutense de Madrid, E-28040 Madrid, Spain}
\affil{$^{30}$Observatoire de Gen\'eve, Universit\'e de Gen\'eve, 51 Ch. des Maillettes, 1290, Versoix, Switzerland}
\affil{$^{31}$Department of Physics and Astronomy, University of British Columbia, 6224 Agricultural Road, Vancouver BC V6T 1Z1, Canada}
\affil{$^{32}$School of Physical Sciences, The Open University, MK7 6AA, Milton Keynes, UK}
\affil{$^{33}$Ritter Astrophysical Research Center, University of Toledo, 2825 West Bancroft Street, M. S. 113, Toledo, OH 43606, USA}
\affil{$^{34}$SRON Netherlands Institute for Space Research, Landleven 12, 9747 AD Groningen, The Netherlands}
\affil{$^{35}$Kapteyn Astronomical Institute, University of Groningen, The Netherlands}
\affil{$^{36}$Center for Computational Sciences, University of Tsukuba, Ten-nodai, 1-1-1, Tsukuba, Ibaraki 305-8577, Japan}
}

\jid{PASA}
\doi{10.1017/pas.\the\year.xxx}
\jyear{\the\year}

\bibpunct{(}{)}{;}{a}{}{,}
\usepackage{graphicx}
\usepackage{aas_macros}
\usepackage{hyperref} 
\usepackage{dblfloatfix}
\hypersetup{colorlinks,citecolor=blue,linkcolor=blue,urlcolor=blue}
\usepackage{color}

\usepackage{macros_egami}

\begin{document}

\begin{frontmatter}

\maketitle

\end{frontmatter}

\begin{frontmatter}


\begin{abstract}

  With the recent discovery of a dozen dusty star-forming galaxies and
around 30 quasars at $z$\,$>$\,5 that are hyper-luminous in the
infrared ($\mu$\,$L_{\rm IR}$\,$>$\,$10^{13}$ \lsun, where $\mu$ is a
lensing magnification factor), the possibility has opened up
for \spica, the proposed ESA M5 mid-/far-infrared mission, to extend
its spectroscopic studies toward the epoch of reionization and
beyond.  In this paper, we examine the feasibility and scientific
potential of such observations with \spica's far-infrared spectrometer
SAFARI, which will probe a spectral range (35--230 \micron) that will
be unexplored by ALMA and \jwst.  Our simulations show that SAFARI is
capable of delivering good-quality spectra for hyper-luminous infrared
galaxies (HyLIRGs) at $z$\,$=$\,5--10, allowing us to sample spectral
features in the rest-frame mid-infrared and to investigate a host of
key scientific issues, such as the relative importance of star
formation versus AGN, the hardness of the radiation field, the level
of chemical enrichment, and the properties of the molecular gas.  From
a broader perspective, SAFARI offers the potential to open up a new
frontier in the study of the early Universe, providing access to
uniquely powerful spectral features for probing first-generation
objects, such as the key cooling lines of low-metallicity or
metal-free forming galaxies (fine-structure and \hh\ lines) and
emission features of solid compounds freshly synthesized by
Population~III supernovae.  Ultimately, SAFARI's ability to explore
the high-redshift Universe will be determined by the availability of
sufficiently bright targets (whether intrinsically luminous or
gravitationally lensed).  With its launch expected around
2030, \spica\ is ideally positioned to take full advantage of upcoming
wide-field surveys such as LSST, SKA, \euclid, and \wfirst, which are
likely to provide extraordinary targets for SAFARI.

\end{abstract}


\end{frontmatter}

\section*{Preface}

\noindent
The following set of articles describe in detail the science goals of
the future Space Infrared telescope for Cosmology and Astrophysics
(\spica).  The \spica\ satellite will employ a 2.5-m telescope,
actively cooled to below 8\,K, and a suite of mid- to far-infrared
spectrometers and photometric cameras, equipped with state-of-the-art
detectors.  In particular, the \spica\ Far Infrared Instrument
(SAFARI) will be a grating spectrograph with low ($R$\,$=$\,300) and
medium ($R$\,$=$\,3000--11000) resolution observing modes
instantaneously covering the 35--230\,$\mu$m wavelength range.  The
\spica\ Mid-Infrared Instrument (SMI) will have three operating modes:
a large field of view (12\arcmin$\times$10\arcmin) low-resolution
17--36\,$\mu$m spectroscopic ($R$\,$=$\,50--120) and photometric
camera at 34\,$\mu$m, a medium resolution ($R$\,$=$\,2000) grating
spectrometer covering wavelengths of 18--36\,\micron\ and a
high-resolution echelle module ($R$\,$=$\,28000) for the
12--18\,\micron\ domain.  A large field of view 
  (160\arcsec$\times$160\arcsec)\footnote{Some other
  \spica\ papers refer to this POL field of view as
  80\arcsec$\times$80\arcsec, but it is
  160\arcsec$\times$160\arcsec\ according to the latest design.},
three channel (110, 220, and 350\,\micron) polarimetric camera (POL)
will also be part of the instrument complement.  These articles will
focus on some of the major scientific questions that the
\spica\ mission aims to address; more details about the mission and
instruments can be found in \citet{Roelfsema2018}.

\section{Introduction}

Through a series of multi-wavelength observations from the UV to radio over the last few decades, it has been shown that the ``observed'' UV star-formation rate density (SFRD, without any dust-extinction correction) is an order of magnitude smaller than that in the infrared at 0\,$<$\,$z$\,$<$\,2 \citep[e.g.,][]{Madau2014}.  This indicates that in the redshift range where robust measurements of the far-infrared luminosity density exist, most of cosmic star formation took place in dusty/dust-obscured environments, which absorb UV light from young stars and reradiate in the infrared.  Although this is not necessarily a surprise if we consider that stars form in dusty molecular clouds locally, it suggests the likelihood that optical/near-infrared observations may miss a significant fraction of galaxies at high redshift due to dust extinction.

A case in point is HDF~850.1, the brightest submillimeter source discovered in the very first deep 850-\micron\ map of the sky, taken over the Hubble Deep Field North (HDF-N) with SCUBA on JCMT \citep{Hughes1998}.  Despite its brightness (7\,mJy at 850\,\micron), it took 14 years to localize this source and determine its redshift, which turned out to be $z$\,$=$\,5.18 based on the CO and [\ion{C}{2}] line detections \citep{Walter2012}.  This is because its counterpart is not seen in the deep \hst\ optical and near-infrared images.  At $z$\,$>$\,5, even near-infrared observations are sampling the rest-frame UV light, and are therefore susceptible to dust extinction.  Such optical and near-infrared dropout sources have also been discovered with deep \spitzer/IRAC survey data, indicating the presence of a substantial population of massive dusty star-forming galaxies at $z$\,$>$\,3 \citep[e.g.,][]{Wang2016}.

Note that the star-formation rate (SFR) of HDF~850.1 is quite large, 850 \msun\ yr$^{-1}$, as derived from the total infrared luminosity (\lir) of 8.7$\times$$10^{12}$\,\lsun (conventionally defined as the integrated luminosity over 8--1000\,\micron; see \citealt{Sanders1996}), which qualifies this source as an ultra-luminous infrared galaxy (ULIRG: \lir\,$=$\,$10^{12}$--$10^{13}$\,\lsun).  This clearly illustrates that even such an intrinsically luminous galaxy could be completely missed by optical/near-infrared observations due to dust extinction.  Note, however, that not all $z$\,$>$\,5 infrared-luminous galaxies\footnote{Also often referred to as ``submillimeter galaxies'' \citep[SMGs;][]{Blain2002} or ``dusty star-forming galaxies'' \citep[DSFGs;][]{Casey2014}.  Here, we adopt the term ``infrared-luminous galaxies'' for much of this paper, which refers to the property in the galaxy's rest-frame and includes AGN galaxies like quasars in the definition.  See \citet{Sanders1996} for an earlier review.}  are so optically faint.  For example, AzTEC-3 at $z$\,$=$\,5.30, the first submillimeter galaxy (SMG) that has been identified at $z$\,$>$\,5 \citep{Riechers2010,Capak2011}, has a counterpart with $i$\,$\sim$\,26 mag, whose optical spectrum shows a \lya\ emission line as well as a rest-frame UV continuum with metal absorption lines.  This suggests that in some high-redshift infrared-luminous galaxies, UV-bright star-forming regions coexist with those that are heavily dust-obscured.

Recent ALMA observations have further reinforced the view that the infrared-luminous galaxy population plays an important role in the cosmic history.  For example, ALMA 1.3-mm imaging of the Hubble Ultra Deep Field (HUDF) has indicated that about 85\,\% of the total star formation at $z$\,$\simeq$\,2 is enshrouded in dust, about 65\,\% of which is occurring in high-mass galaxies with an average obscured to unobscured star formation ratio of 200 \citep{Dunlop2017}.  A subsequent analysis of these HUDF ALMA sources as well as those detected in a wider GOODS-S area (26~arcmin$^{2}$) has shown a surprisingly large X-ray AGN fraction \citep{Ueda2018}, suggesting a possible connection between the dusty phase of massive galaxy evolution and growth of super-massive black holes (SMBHs).  On the high-redshift front,  ALMA has started to discover $z>8$ galaxies through the detection of the \oiii\ 88-\micron\ line, such as MACS0416-Y1 at $z$\,$=$\,8.31 \citep{Tamura2018}, A2744-YD at $z$\,$=$\,8.38 \citep{Laporte2017}, and MACS1149-JD1 at $z$\,$=$\,9.11 \citep{Hashimoto2018}, and surprisingly the first two galaxies were also detected in dust continuum with corresponding infrared luminosities of 1--2$\times$$10^{11}$\,\lsun.  ALMA dust-continuum detections also exist for a few $z$\,$=$\,7--8 galaxies, such as B14-65666 at $z$\,$=$\,7.15 \citep{Hashimoto2018b} and A1689-zD1 at $z$\,$=$\,7.5 \citep{Knudsen2017}, with corresponding infrared luminosities of 2--6$\times$$10^{11}$\,\lsun.  These recent discoveries confirm the existence of dusty, infrared-luminous ($>$\,$10^{11}$\,\lsun) galaxies well inside the epoch of reionization, only about half a billion years after the Big Bang.  These recent exciting developments clearly indicate the importance of probing the high-redshift Universe in the infrared, which will allow us to obtain the full picture of the earliest phases of galaxy evolution by mitigating the effects of dust extinction/obscuration.

\spica\ is a proposed ESA M5 mission, whose main scientific goal is to explore the dusty/dust-obscured Universe, both near and far, by conducting sensitive imaging and spectroscopic observations in the mid-/far-infrared \citep{Roelfsema2018}.  \spica\ is expected to revolutionize a wide spectrum of research areas in astronomy/astrophysics, and it will be especially powerful for probing the dusty/dust-obscured Universe at high redshift through spectroscopy.  On the extragalactic side, a key goal of the \spica\ mission is to conduct large spectroscopic surveys of galaxies at $z$\,$=$\,1--4 and characterize their physical properties through the analysis of spectral features in the mid-/far-infrared.  For example, a 2000-hr SAFARI survey will obtain low-resolution (LR; $R$\,$=$\,300) spectra for over 1000 galaxies up to $z$\,$\simeq$\,4 \citep{Spinoglio2017} while a 600-hr SMI survey will identify about 50,000 galaxies in a 10\,deg$^{2}$ area through $R$\,$=$\,50--120 spectroscopy of PAH emission features \citep{Kaneda2017}.  Such data sets will enormously advance our understanding of galaxy/AGN evolution, and will shed light on key science topics such as chemical evolution/metal enrichment \citep{Fernandez2017} and molecular outflows/inflows \citep{Gonzalez2017}.  Note that the great power of \spica\ mainly resides in such spectroscopic observations, especially in the far-infrared ($>$\,100\,\micron), where \herschel/SPIRE has already achieved confusion-limited broad-band imaging sensitivities with a 3.5-m telescope.  Another area of \spica's strength is its ability to conduct deep and wide imaging surveys with the SMI's slit-viewer camera at 34\,\micron, where the confusion limit will be significantly lower \citep{Gruppioni2017}.

The goal of this paper is to examine \spica's potential for extending infrared spectroscopic studies toward the epoch of reionization and beyond.  More specifically, we will assess SAFARI's ability to obtain high-quality galaxy spectra (similar to those obtained by \spitzer/IRS at lower redshift) in a redshift range of $z$\,$=$\,5--10.  A redshift of 5 defines a natural boundary for SAFARI because at $z$\,$>$\,5 the 6.2\,\micron\ PAH feature is redshifted into the SAFARI band, making SAFARI data sets self-sufficient for a variety of mid-infrared spectral analyses.  In the current design, SAFARI will deliver LR ($R$\,$=$\,300) spectra covering 35--230\,\micron\ with a line-flux sensitivity of around 5$\times$$10^{-20}$\,W\,m$^{-2}$ (5\,$\sigma$, 1\,hour).  Based on this sensitivity estimate and recent discoveries of infrared-luminous galaxies/quasars at $z$\,$>$\,5, we will examine the detectability of various types of galaxies by simulating their SAFARI spectra, and will discuss the scientific potential of such observations (Section~\ref{sec:gal_qso}).  In addition, we will extend our discussion to a few exploratory science programs that are significantly more challenging but have the potential to open up a new frontier in the study of the early Universe (Section~\ref{sec:exploratory}).  In the final section (Section~\ref{sec:targets}), we will review a variety of existing and future wide-field data, which can be used to select SAFARI targets effectively.

Throughout the paper, we assume a $\Lambda$CDM cosmology with $H_{0}$\,$=$\,70\,km\,s$^{-1}$\,Mpc$^{-1}$, $\Omega_{\rm m}$\,$=$\,0.3, and $\Omega_{\Lambda}$\,$=$\,0.7.

\section{Probing the \lowercase{$z$}\,$>$\,5 Universe}

\label{sec:gal_qso}

As has been demonstrated by the large body of work with \iso\ and
\spitzer, as reviewed by \citet{Genzel2000} and \citet{Soifer2008},
respectively, the rest-frame mid-infrared spectral range is extremely
rich in diagnostic information, with a variety of atomic
fine-structure lines, molecular hydrogen (\hh) lines, PAH features,
and silicate emission/absorption features (e.g., see
\citealt{Genzel1998} for \iso\ and \citealt{Armus2007} for
\spitzer\ results, as well as the companion papers by
\citealt{Spinoglio2017} and \citealt{Tak2018}).  Some galaxies are so
embedded in dust that rest-frame mid-infrared spectroscopy is crucial
for identifying the dominant luminosity source (whether star formation
or AGN).  Without such spectral information, it is impossible to fully
capture the landscape of the dust-obscured Universe at high redshift.
Although ALMA and \jwst\ will undoubtedly make great progress in the
near future, they will leave the 30--300\,\micron\ spectral range
unexplored, i.e., the rest-frame mid-infrared at $z$\,$=$\,5--10,
requiring an infrared space mission like \spica\ to fill this
information-rich gap.

\subsection{Dusty Star-Forming Galaxies}

One recent crucial development, which has opened up \spica's potential
to probe the $z$\,$>$\,5 Universe, was a series of discoveries finding
that a significant fraction of the brightest submillimeter/millimeter
sources in a random blank sky field corresponds to
gravitationally-lensed infrared-luminous galaxies at high redshift
(except for nearby galaxies and bright AGN).  The discovery of the
Cosmic Eyelash galaxy at $z$\,$=$\,2.3 \citep{Swinbank2010}, which was
the first of such {\it super-bright}
($S_{870}$\,$>$\,100\,mJy)\footnote{$S_{870}$ denotes the flux density
  at 870 \micron.  Similar notations will be used to indicate flux
  densities at specific wavelengths.} lensed infrared-luminous
galaxies to be found, allowed a variety of multi-wavelength
observations even with those observing facilities that normally do not
have the sensitivity to probe beyond the low-redshift Universe.

Although this first discovery was serendipitous, wide-field surveys
with \herschel, South Pole Telescope (SPT), Atacama
  Cosmology Telescope (ACT), and \planck\ quickly followed with more
discoveries of similarly bright infrared-luminous galaxies
\citep[e.g.,][]{Negrello2010, Combes2012, Vieira2013, Weiss2013,
  Marsden2014, Canameras2015, Harrington2016}, a small number of which
have turned out to be at $z$\,$>$\,5.  Due to lensing, these
$z$\,$>$\,5 galaxies are all substantially brighter than HDF~850.1 and
AzTEC-3, so their redshifts were easily measured by blind CO searches.
At the time of writing, the discoveries of ten such lensed
infrared-luminous galaxies have been reported at $z$\,$>$\,5,
(Table~\ref{tab:highz}; SPT0311-58~W and SPT0311-58~E are
  counted as one), with the highest-redshift galaxy at $z$\,$=$\,6.90
\citep{Strandet2017,Marrone2018}.  Even when corrected for lensing
magnification ($\mu$), many of these objects are
hyper-luminous infrared galaxies (HyLIRGs;
$L_{\rm IR}$\,$>$\,$10^{13}$\,\lsun) but without any sign of a strong
AGN, leading to their classification as dusty star-forming galaxies
(DSFGs; \citealt{Casey2014}).  Note that non-lensed $z>5$
  galaxies that are significantly more luminous than HDF~850.1 and
  AzTEC-3 are also being discovered (Table~\ref{tab:highz}).  With a
variety of wide-field surveys being conducted/planned (see
Section~\ref{sec:targets}), the list of such HyLIRGs at
$z$\,$=$\,5--10, whether gravitationally lensed or intrinsically
luminous, will grow rapidly over the coming years, providing excellent
targets for SAFARI.

\begin{table*}[!t]

\caption{Currently known infrared-luminous galaxies ($\mu$\,$L_{\rm
    IR}$\,$\gtrsim$\,$10^{13}$\,\lsun) at $z$\,$>$\,5 (non-quasars)}
\label{tab:highz}

\begin{tabular}{@{}lcccccll@{}}
\hline\hline
Object        & $z$  & $S_{500}$ & $S_{870}$ & $\mu$\,$L_{\rm IR}^a$    & $\mu^b$ & Survey & Ref \\
              &      & (mJy)     & (mJy)     & ($10^{13}$\,\lsun) &         &        &     \\
\hline
\multicolumn{8}{l}{Gravitationally-lensed galaxies:} \\
SPT0311$-$58 W  & 6.90 &  50    &  35$^c$  & 7.3               &  2.2  & SPT               &  1, 2 \\
SPT0311$-$58 E  &      &   5    &  4$^c$   & 0.6               &  1.3  &                   &       \\
HFLS3           & 6.34 &  47    &  33$^d$  & 4.2               &  2.2  & \herschel/HerMES  &  3, 4 \\
HATLAS~J0900    & 6.03 &  44    &  36$^e$  & 3.5               &  9.3  & \herschel/HATLAS  &  5  \\
SPT2351$-$57    & 5.81 &  74    &  35      & 11$^h$ & $\sim$10$^h$ & SPT        &  6, 7 \\ 
SPT0243$-$49    & 5.70 &  59    &  84      & 4.5               &  9.8  & SPT               &  7, 8, 9, 10, 11 \\ 
SPT0346$-$52    & 5.66 & 204    &  131     & 16                &  5.6  & SPT               &  7, 8, 9, 10, 11, 12, 13 \\ 
SPT2353$-$50    & 5.58 &  56    &  41      & 7.8$^h$ & $\sim$10$^h$ & SPT        &  6, 7 \\ 
SPT2319$-$55    & 5.29 &  49    &  38      & 2.5$^i$           & 20.8  & SPT               &  6, 7, 10 \\ 
HLSJ0918        & 5.24 & 212    &  125$^d$ & 16                &  9    & \herschel/HLS     &  14, 15 \\
HELMS\_RED\_4   & 5.16 & 116    &  65$^f$  & \dots             & \dots & \herschel/HerMES  &  16 \\
\multicolumn{8}{l}{\noindent Non-lensed galaxies:} \\
CRLE            & 5.67 &  31    &  17$^e$  & 3.2               & 1     & ALMA/COSMOS       &  17 \\
ADFS-27         & 5.65 &  24    &  25      & 2.4               & 1     & \herschel/HerMES  &  18 \\
AzTEC-3         & 5.30 &  $<32$ &  9$^g$   & 1.6               & 1     & AzTEC/COSMOS      &  19, 20 \\
HDF 850.1       & 5.18 &  $<14$ &  7       & 0.65              & 1     & SCUBA/HDF-N       &  21 \\
\hline\hline
\end{tabular}

  \tabnote{
$^a$Infrared luminosity \lir(8--1000\micron) without a lensing correction.
$^b$Magnification factor.  
$^c$At 869\,\micron\ with ALMA.}
  \tabnote{
$^d$At 880\,\micron\ with SMA.  
$^e$At 850\,\micron\ with SCUBA-2.
$^f$At 920\,\micron\ with CSO/MUSIC.  
$^g$At 890\,\micron\ with SMA. \\
$^h$J.\ Spilker 2018, private communication.
$^i$\lir(42--500 \micron).
}
  \tabnote{{\bf References:} (1) \citet{Strandet2017}; (2) \citet{Marrone2018}; (3) \citet{Riechers2013}; (4) \citet{Cooray2014};} 
  \tabnote{(5) \citet{Zavala2018}; (6) \citet{Strandet2016}; (7) \citet{Spilker2016}; (8) \citet{Vieira2013}; (9) \citet{Weiss2013};} 
  \tabnote{(10) \citet{Gullberg2015}; (11) \citet{Aravena2016}; (12) \citet{Ma2015}; (13) \citet{Ma2016}; (14) \citet{Combes2012};} 
  \tabnote{(15) \citet{Rawle2014}; (16) \citet{Asboth2016}; (17) \citet{Pavesi2018}; (18) \citet{Riechers2017}; (19) \citet{Younger2007};} 
  \tabnote{(20) \citet{Smolcic2015}; (21) \citet{Walter2012}.}

\end{table*}

\begin{figure*}[!h]

\vspace*{0.5cm}

\hspace*{-0.2cm}\includegraphics[width=7.0in]{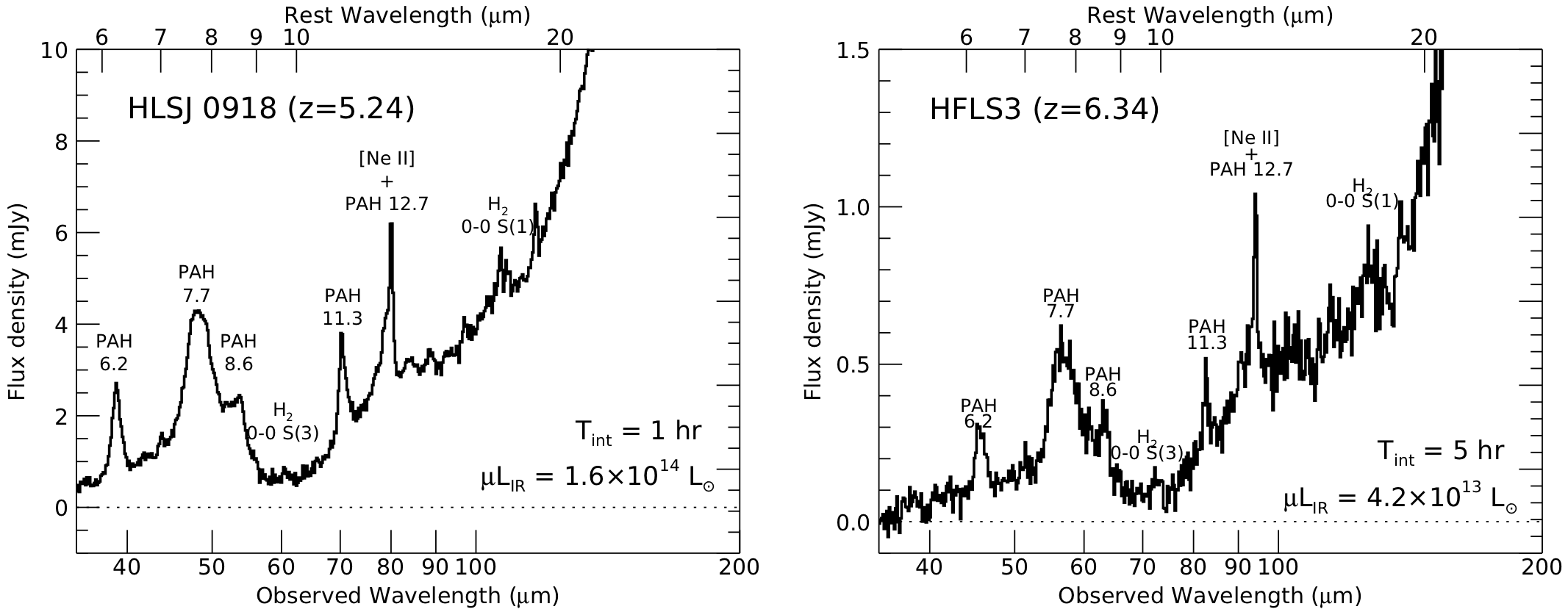}

\caption{Simulated SAFARI spectra of HLSJ0918 
  \citep[$z$\,$=$\,5.24, $\mu$\,$=$\,9:][]{Combes2012,Rawle2014} and
  HFLS3 \citep[$z$\,$=$\,6.34, $\mu$\,$=$\,2:][]{Riechers2013,
    Cooray2014} are shown in the left and right panels, respectively.
  The average local galaxy SED templates \citep{Rieke2009} of $L_{\rm
    IR}$\,$=$\,$10^{11.75}$ and $10^{12.50}$\,\lsun\ were used,
  respectively, which produce good fits to the observed rest-frame
  far-infrared SEDs of these galaxies.  The template SEDs were first
  scaled to the infrared luminosities without a lensing correction
  ($\mu$\,$L_{\rm IR}$ in each panel) and then fit with PAHFIT
  \citep{Smith2007} with a pixel sampling of $R$\,$=$\,600.  These
  PAHFIT-produced model spectra were then redshifted and noise-added
  for corresponding integration times ($T_{\rm int}$ in each panel).
  Finally, the resultant spectra were resampled with $R$\,$=$\,300
  pixels.  However, the effective resolution of these simulated
  spectra is less than $R$\,$=$\,300 due to the low resolution
  ($R$\,$\approx$\,60--130) of the \spitzer/IRS data used by
  \citet{Rieke2009} to build templates.
  Note that the actual mid-infrared spectra of these $z$\,$>$\,5
  galaxies may significantly differ from those of local LIRGs/ULIRGs
  (see Section~\ref{EoR} for more discussion).}

\label{fig:HyLIRGs}

\end{figure*}

To illustrate the power of SAFARI, we show in Figure~\ref{fig:HyLIRGs}
the simulated SAFARI spectra of two gravitationally-lensed
infrared-luminous galaxies from Table~1, HLSJ0918 at $z$\,$=$\,5.24
\citep{Combes2012, Rawle2014} and HFLS3 at $z$\,$=$\,6.34
\citep{Riechers2013,Cooray2014}.  These galaxies were discovered as
\herschel\ sources showing red colors in the three SPIRE bands
($S_{250}$\,$<$\,$S_{350}$\,$<\,$$S_{500}$), a technique that has
proved to be effective for finding $z$$>$4 DSFGs
\citep[e.g.,][]{Dowell2014}.  The figure clearly shows that SAFARI is
capable of detecting main spectral features in the rest-frame
mid-infrared at these redshifts if the infrared luminosities of target
galaxies are $>$$10^{13}$\,\lsun.  If the PAH features and
fine-structure lines in these galaxies are as strong as those seen at
lower redshift, SAFARI will be able to detect them clearly, and the
measured PAH strengths can be used to estimate SFRs.  Compared to
other SFR indicators, PAH features have the advantage of being less
vulnerable to dust extinction (e.g., compared to \ha) and being more
luminous (e.g., compared to \neii\ 12.8\,\micron).

PAH equivalent widths are also a powerful diagnostic for assessing the
AGN contribution to the rest-frame mid-infrared continuum emission
\citep[e.g.,][]{Pope2008,Riechers2014a}.  Considering that many of the
lensed infrared-luminous galaxies listed in Table~1 are HyLIRGs even
intrinsically (i.e., when corrected for the lensing magnification), it
is important to examine if they harbor luminous AGN and therefore
exhibit smaller PAH equivalent widths.  For the detection of AGN,
especially those heavily obscured by dust, the rest-frame mid-infrared
range is optimal as the AGN contribution becomes most conspicuous
there.  Other mid-infrared spectral features that can be used to
detect the presence of AGN are high excitation lines such as
\nev\ 14.3/24.3\,\micron\ and \oiv\ 25.9\,\micron, which can be used
to estimate the AGN contribution and black hole accretion rates
\citep[e.g.,][]{Spinoglio2017}.

Other prominent mid-infrared spectral features include atomic
fine-structure lines such as \neii/\neiii\ 12.8/15.6\,\micron, and
molecular hydrogen (\hh) lines such as 0--0 S(1)/0--0 S(3)
17.0/9.66\,\micron\ (some of these lines are not seen in
  Figure~\ref{fig:HyLIRGs} because of their faintness and the low
  resolution of the template spectra).  The \neii/\neiii\ lines, for
example, will serve as excellent indicators of SFRs and the hardness
of ionizing radiation \citep[e.g.,][]{Thornley2000, Ho2007} while \hh\
lines allow us to measure the temperature and mass of warm
($T$\,$\gtrsim$\,100\,K) molecular hydrogen gas directly
\citep[e.g.,][]{Rigopoulou2002, Higdon2006}.

Note that high-redshift HyLIRGs are likely more luminous in the
rest-frame mid-infrared than the local ones, which will help SAFARI
detections of submillimeter/millimeter-selected DSFGs like those
listed in Table~\ref{tab:highz}.  This is because at
$z$\,$\gtrsim$\,1, many star-forming HyLIRGs/ULIRGs are spatially
extended over kpc scales, exhibiting flatter and colder infrared SEDs
that are more similar to those of local luminous infrared galaxies
(LIRGs: \lir\,$=$$10^{11}$--$10^{12}$\,\lsun; see
\citealt{Rujopakarn2013} and references therein).  Indeed, the
\herschel-observed far-infrared SEDs of HLSJ0918 and HFLS3 take shapes
consistent with those of galaxies with much lower infrared
luminosities (see the caption of Figure~\ref{fig:HyLIRGs}), supporting
the validity of such an assumption.

\subsection{UV-Bright Star-Forming Galaxies}

\label{sec:LBG}

At $z$\,$\gtrsim$\,5, the majority of galaxies have been selected
through robust optical (broad-band/narrow-band) color selections and
identified either as Lyman break galaxies (LBGs) or Lyman-alpha
emitters (LAEs).  LBGs and LAEs are inherently UV-bright star-forming
galaxies because they are selected through the detections of the Lyman
break at 912\,\AA\ and/or \lya\ break/emission at 1216\,\AA.  Unlike
DSFGs discussed above, which can be extremely faint in the rest-frame
UV (e.g., HDF~850.1), LBGs/LAEs are less dust-obscured as populations,
especially at $z$\,$\gtrsim$\,5 where many of LBGs/LAEs are seen to
exhibit extremely blue UV continuum slopes
\citep[e.g.,][]{Bouwens2012,Dunlop2012,Finkelstein2012,Jiang2013}.

At $z$\,$\sim$\,3, \spitzer/IRS spectra exist for a small number of
bright gravitationally-lensed LBGs, such as MS1512-cB58 at
$z$\,$=$\,2.73 \citep{Siana2008} and the Cosmic Eye at $z$\,$=$\,3.07
\citep{Siana2009}, giving a glimpse of what the mid-infrared spectra
of UV-selected star-forming galaxies look like.  The mid-infrared
spectra of these particular LBGs are similar to those of typical
infrared-luminous galaxies like those in Figure~\ref{fig:HyLIRGs},
showing strong PAH features and resembling those of infrared-selected
lensed galaxies at comparable redshift \citep{Rigby2008}.  This is
probably not surprising, considering that these LBGs are LIRGs in
terms of their infrared luminosities and therefore are probably among
the more infrared-luminous members of the LBG population.  In fact, a
significant fraction of $z$\,$\simeq$\,3 LBGs are thought to be
infrared-luminous despite their rest-frame UV selection
  \citep[e.g.,][]{Coppin2015, Koprowski2016}.  A recent \herschel\
stacking analysis of about 22,000 $z$\,$\simeq$\,3 LBGs indicates that
these galaxies are LIRGs on average \citep{Alvarez2016}.  It has also
been shown that some of the $z$\,$\simeq$\,3 LBGs are even ULIRGs
\citep[e.g.,][]{Oteo2013, Magdis2017}.  Even at
  $z$\,$\simeq$\,7, bright LBGs are thought to be LIRGs
  on average \citep{Bowler2018}.

Because of the simple color selection criteria, LBGs are known to
constitute a heterogeneous sample of galaxies with a wide spectrum of
physical properties, from dusty infrared-luminous galaxies to luminous
LAEs with little dust extinction.  One exciting prospect for SAFARI is
that it will be able to detect the latter population (which likely
dominates in number), making it possible to study both populations in
a uniform way, using the same set of mid-infrared diagnostics.

In this context, particularly interesting are low-mass,
low-metallicity, unreddened galaxies with strong emission lines at
$z$\,$\sim$\,2, which may be better analogs of $z$\,$\gtrsim$\,5
galaxies \citep[e.g.,][]{Erb2010,Stark2014}.  These galaxies may be
similar to low-metallicity blue compact dwarfs (BCDs) in the local
Universe \citep[e.g.,][]{Watson2011}, and if so, their mid-infrared
spectra are likely distinctly different from those of typical
infrared-luminous galaxies shown in Figure~\ref{fig:HyLIRGs}.  We will
discuss the mid-infrared spectra of these local BCDs in
Section~\ref{EoR}.

\subsection{Quasars/AGN}

\label{sec:qso}

\begin{figure}[!htb]
\begin{center}
\hspace*{-0.6cm}\includegraphics[width=3.8in]{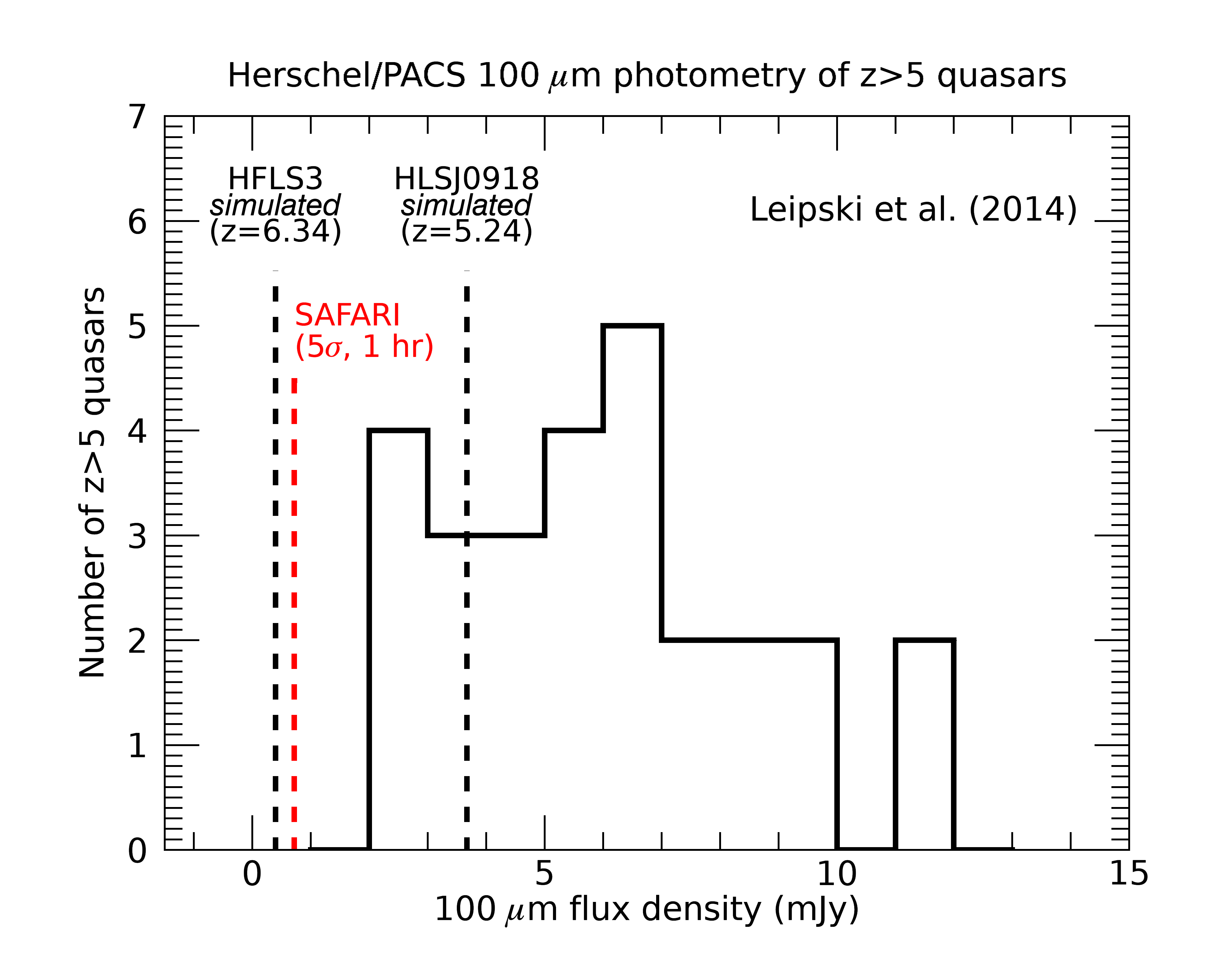}

\caption{\herschel/PACS 100-\micron\ photometry of 27 $z$\,$=$\,5--6.4
  Type-1 quasars reported by \citet{Leipski2014}.  In comparison, the
  simulated 100-\micron\ flux densities of HLSJ0918 and HFLS3 (see
  Figure~\ref{fig:HyLIRGs}) are also shown, as well as the sensitivity
  of SAFARI LR-mode (0.7\,mJy at 100 \micron, 5\,$\sigma$ in 1 hour).
  These $z$\,$>$\,5 quasars are bright enough to be observable with
  SAFARI in under an hour (each), providing details about the dust
  composition and distribution of dust around their nuclei.}

\label{fig:qso}
\end{center}
\end{figure}

Compared to star-forming galaxies, quasars have much flatter infrared
SEDs because of the power-law continuum produced by the central AGN.
As a result, they are significantly brighter in the rest-frame
mid-infrared, and are easier to observe with SAFARI.
Figure~\ref{fig:qso} shows the 100-\micron\ flux-density distribution
of 27 $z$\,$>$\,5 Type-1 quasars (up to $z$\,$=$\,6.4) based on the
\herschel/PACS photometry reported by \citet{Leipski2014}.  Note that
the PACS 100-\micron\ band directly measures the source brightness in
the wavelength range that SAFARI will cover.  The measured
100-\micron\ flux densities range from 2 to 12\,mJy, indicating that
SAFARI, with a 5\,$\sigma$ continuum sensitivity of 0.7\,mJy in
1\,hour, will be able to obtain high-quality spectra for these quasars
quickly.

Mid-infrared spectra of low-redshift Type-1 AGN are often
characterized by a power-law continuum, silicate emission/absorption
features, and PAH emission features
\citep[e.g.,][]{Siebenmorgen2005,Hao2005,Shi2006, Shi2007, Shi2009,
  Shi2014}.  The power-law continuum seen in the rest-frame
mid-infrared is thought to be produced by the dusty torus around the
central AGN \citep[e.g.,][]{Leipski2014}, allowing us to study the
properties and geometry of the circumnuclear material.  For example,
the strengths of silicate emission/absorption features (at 9.7 and
18\,\micron) are thought to correlate (at least in the first order)
with the orientation of the dusty torus (i.e., edge-on $\rightarrow$
absorption; face-on $\rightarrow$ emission), and can be used to infer
the structure of the torus in the framework of unification models
\citep[e.g.,][]{Shi2006}.  At $z$\,$\gtrsim$\,6, some quasars are found to
be deficient in hot dust, suggesting that their dusty tori are not
fully developed or are even absent \citep{Jiang2010,Leipski2014}.
SAFARI spectroscopy of $z$\,$>$\,5 quasars therefore offers the
possibility to investigate, through observations and modeling, the
physical conditions and formation/evolution processes of AGN dusty
tori.

Since the $z$\,$>$\,5 quasars plotted in Figure~\ref{fig:qso} are so
bright, the quality of SAFARI spectra will be high enough to examine
the composition of dust grains.  For example, using the \spitzer/IRS
data for 93 AGN at $z$\,$\lesssim$\,0.5 that exhibit the 9.7 and
18\,\micron\ silicate emission features, \citet{Xie2017} have
determined that 60 of these AGN spectra can be well reproduced by
``astronomical silicates'', while 31 sources favor amorphous olivine
(Mg$_{1.2}$Fe$_{0.8}$SiO$_4$) and two sources favor amorphous pyroxene
(Mg$_{0.3}$Fe$_{0.7}$SiO$_3$).  They also concluded that all sources
require micron-sized dust grains, which are significantly larger than
the submicron-sized dust grains found in the Galactic ISM.  By
measuring the central wavelength, width, and relative intensity of the
two silicate features, SAFARI will allow us to infer the chemical
composition and grain properties of the circumnuclear dust around AGN
at $z$\,$>$\,5.  (see the companion paper by \citealt{Fernandez2017} for a
further discussion of quasar mid-infrared spectra).

The PAH emission features, on the other hand, reveal star-forming
activities in the quasar host galaxies
\citep[e.g.,][]{Shi2007,Shi2009}.  Among the sample of
\citet{Leipski2014}, there are seven $z$\,$>$\,5 Type-1 quasars that
have been detected at 500\,\micron.  Although a significant fraction
of $z$\,$>$\,5 quasar far-infrared luminosities are thought to be
produced by AGN (estimated to be 30--70\,\% by \citealt{Schneider2015b}
and \citealt{Lyu2016}), the infrared luminosities powered by star
formation could still be larger than
$10^{13}$\,\lsun\ \citep{Leipski2014}.  Such infrared luminosities are
comparable to that of HLFS3 ($z$\,$=$\,6.34) shown in
Figure~\ref{fig:HyLIRGs}, suggesting that SAFARI will likely detect
PAH emission features in many of these seven $z$\,$>$\,5 quasars
superposed on the power-law AGN continuum.  Quasars with vigorously
star-forming hosts may also allow us to examine the interplay between
AGN and star formation at these early epochs.

\subsection{Galaxies in the Epoch of Reionization}

\label{EoR}

\begin{figure*}[!htb]

\begin{minipage}{0.5\linewidth}

  \hspace*{-0.65cm}\includegraphics[width=3.8in]{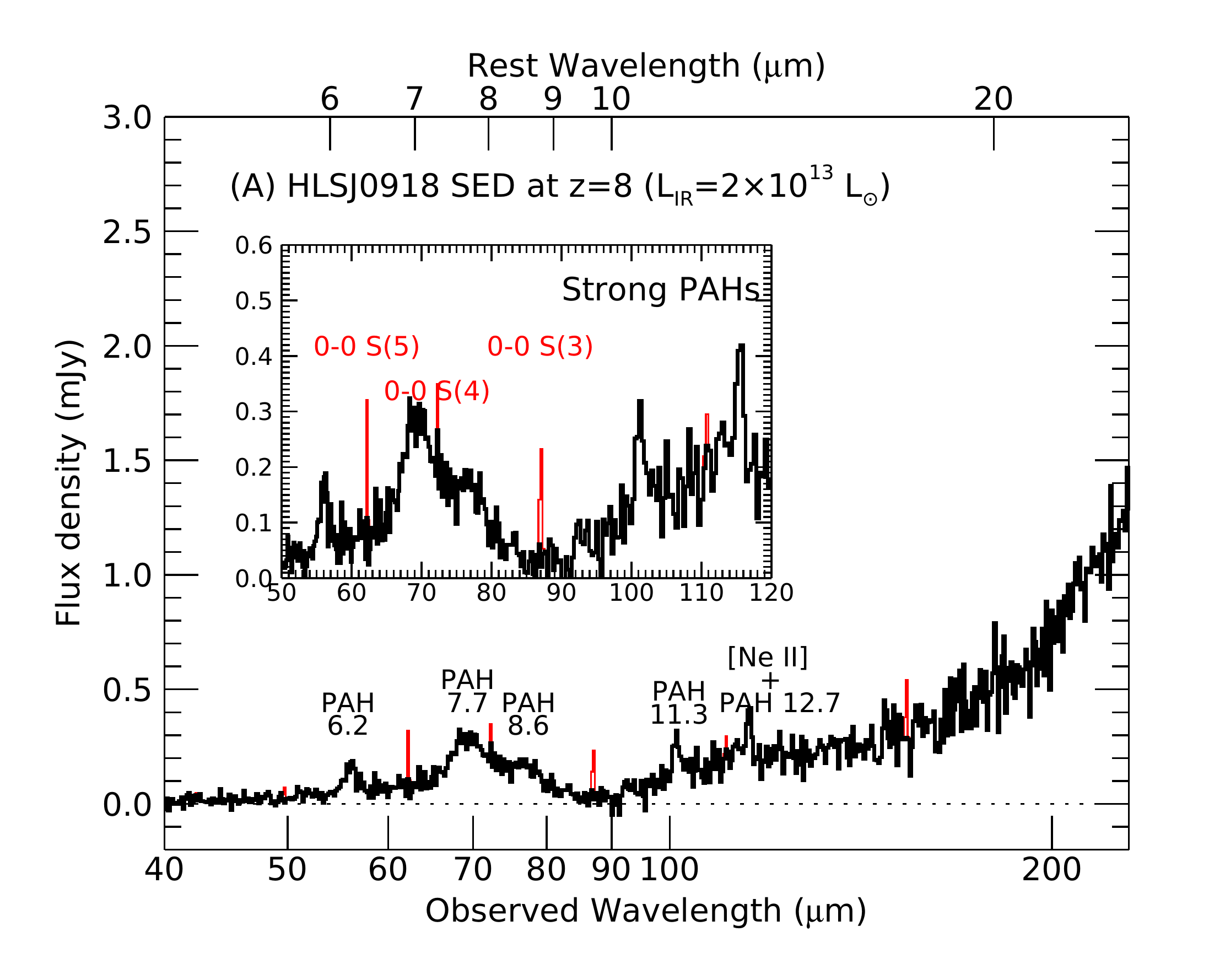}

\end{minipage}\hfill
\begin{minipage}{0.5\linewidth}

  \hspace*{-0.25cm}\includegraphics[width=3.8in]{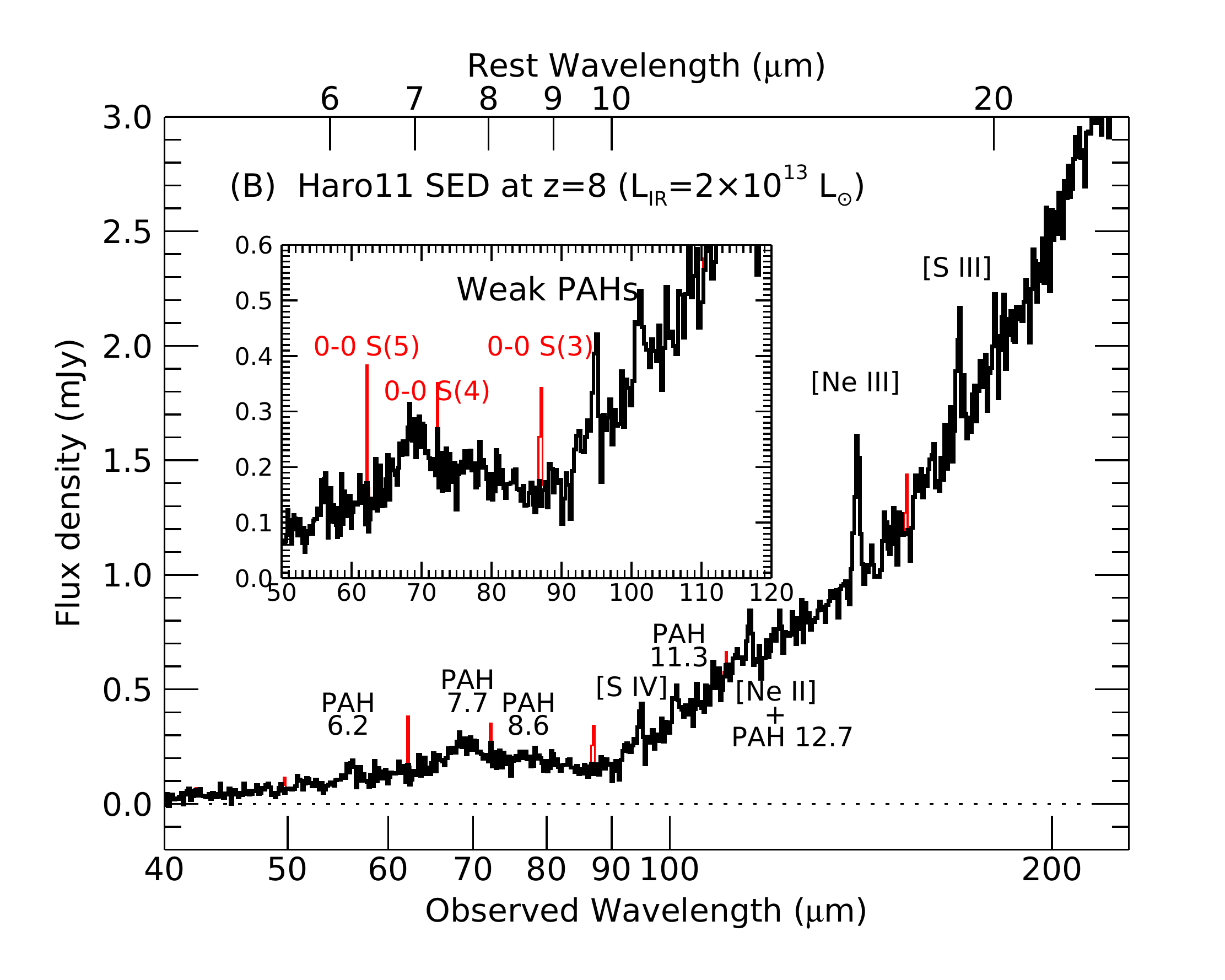}

\end{minipage}

\begin{minipage}{0.55\linewidth}

  \hspace*{-0.6cm}\includegraphics[width=3.8in]{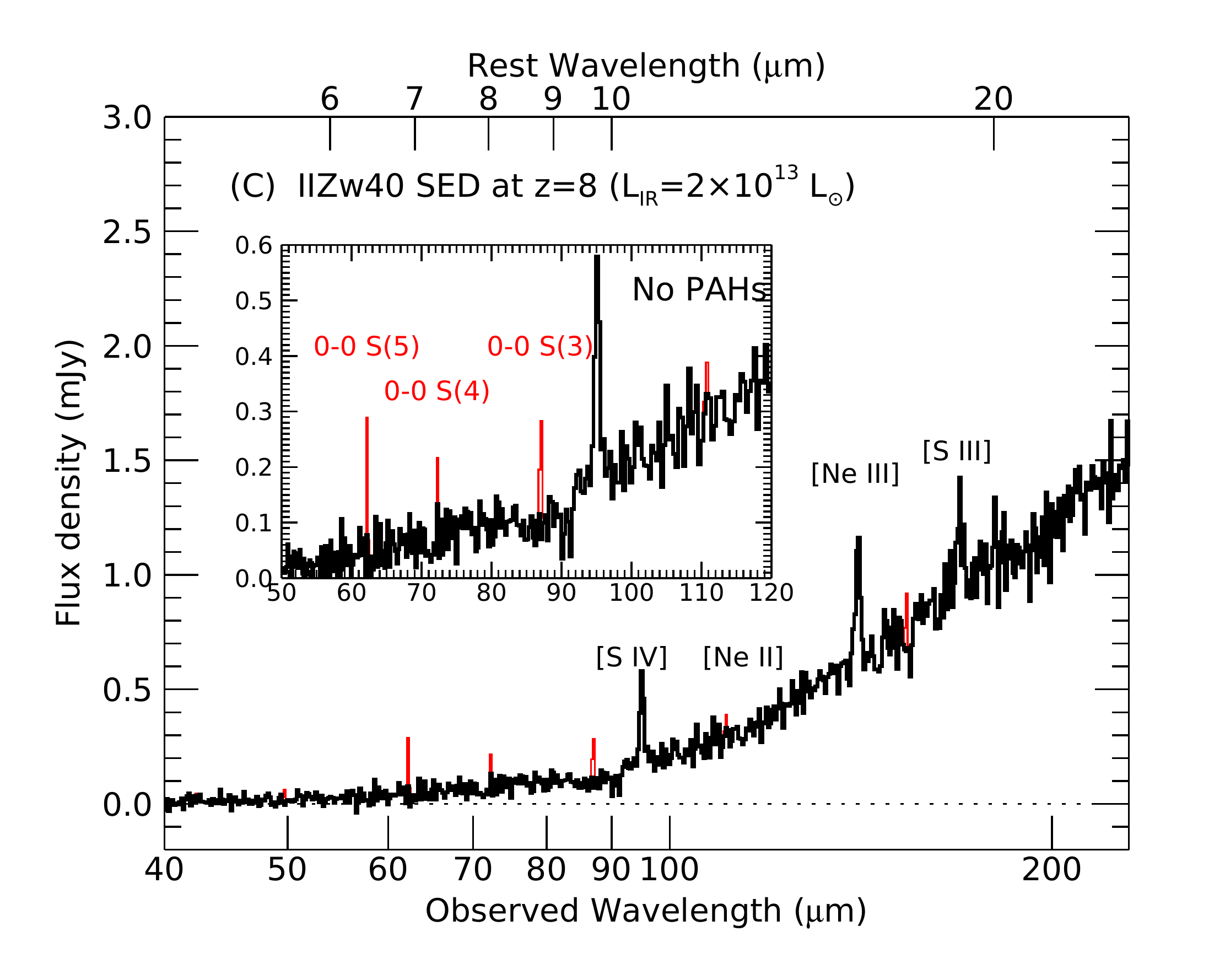}

\end{minipage}\hfill
\begin{minipage}{0.45\linewidth}

  \vspace*{0.5cm}

  \caption{SAFARI 10-hr LR ($R$\,$=$\,300) spectra for $z$\,$=$\,8
    galaxies simulated for the following three-types of galaxies: (A)
    HLSJ0918, a HyLIRG at $z$\,$=$\,5.24 (see Figure~\ref{fig:HyLIRGs}
    and Table~\ref{tab:highz}); (B) Haro~11, a low-metallicity
    infrared-luminous local BCD; and (C) II~Zw~40, another
    low-metallicity local BCD that is not infrared-luminous.  For
    HLSJ0918, the \lir\,$=$\,$10^{11.75}$\,\lsun\ LIRG SED from
    \citet{Rieke2009} was used as in Figure~\ref{fig:HyLIRGs}, while
    for the two BCDs, the fully-processed \spitzer/IRS low-resolution
    spectra were obtained from the Combined Atlas of Sources with
    \spitzer\ IRS Spectra (CASSIS; \citealt{Lebouteiller2011}).  The
    infrared luminosities of these SEDs have been scaled to
    2$\times$$10^{13}$\,\lsun, comparable to the intrinsic luminosity
    of HFLS3.  See the caption of Figure~\ref{fig:HyLIRGs} for how
    these SAFARI spectra were simulated.  The red lines show simulated
    $H_{2}$ emission lines (assumed to be unresolved) produced by
    2$\times$$10^{10}$\,\msun\ of $T$\,$=$\,200\,K gas and
    2$\times$$10^{8}$\,\msun\ of $T$\,$=$\,1000\,K gas under the LTE
    assumption (an ortho-to-para ratio of 3:1 is also assumed).  These
    \hh\ lines are hardly visible in the original galaxy spectra.
    \label{fig_z8}}

\end{minipage}

\end{figure*}

As shown in Figure~\ref{fig:HyLIRGs}, SAFARI will be able to deliver
good-quality rest-frame mid-infrared spectra for HyLIRGs at least up
to $z$\,$\sim$\,6.  The next question, therefore, is how much farther
we can push SAFARI in redshift.  The answer to this question depends
on whether or not there exist HyLIRGs at $z$\,$>$\,6 that are
sufficiently massive and luminous to be detectable with SAFARI.
Note that such high-redshift HyLIRGs are not explicitly
  included in some models of infrared-luminous galaxy evolution.  For
  example, the model by \citet{Bethermin2017}, one of the most
  advanced and up-to-date, applies a sharp SFR limit of
  $<1000$\,\msun\ yr$^{-1}$, excluding HyLIRGs like those listed in
  Table~\ref{tab:highz}.

In this respect, the discoveries of HFLS3 at $z$\,$=$\,6.34
\citep{Riechers2013} and SPT0311$-$58 at $z$\,$=$\,6.9
\citep{Strandet2017,Marrone2018} are encouraging.  The halo masses
(\mhalo) of these high-redshift DSFGs have been estimated to be
$\gtrsim$\,$10^{12}$\,\msun\ \citep{Marrone2018}, and therefore their
existence provides a proof that such massive infrared-luminous
galaxies do exist at $z$\,$\sim$\,6--7, possibly marking the rare
density peaks that would become present-day galaxy clusters and have a
space density of only $10^{-3}$--$10^{-4}$ times that of typical
$z$\,$\sim$\,6 LBGs \citep{Riechers2013}.

Though rare, the existence of massive and luminous DSFGs is expected
in overdense regions at $z$\,$\sim$\,6--7.  For example, the
simulation by \citet{Yajima2015} has shown that overdense regions
evolve at a substantially accelerated pace at high redshift, being
able to produce DSFGs at $z$\,$\sim$\,6 inside a halo with a mass of
\mhalo\,$\sim$\,$10^{12}$\,\msun.  This simulation, however, failed to
reproduce the observed infrared luminosity of HFLS3, falling short by
a factor of about 10.  One possible explanation is that HFLS3 is
experiencing a powerful starburst that boosts the infrared luminosity.
The same simulation also predicts the existence of
\lir\,$\sim$\,6$\times$$10^{11}$\,\lsun\ galaxies at $z$\,$\sim$\,10,
and if their infrared luminosities are similarly enhanced by a strong
starburst (i.e., by a factor of about 10), ULIRG-type galaxies may
exist in some exceptional overdense regions even at $z$\,$\sim$\,10.

Figure~\ref{fig_z8} shows simulated 10-hour spectra of $z$\,$=$\,8 galaxies
using the spectra/SEDs of the following three objects.
\begin{enumerate}

  \setlength{\itemsep}{5pt}

  \item {\bf HLSJ0918}: The $z$\,$=$\,5.24 gravitationally-lensed
    infrared-luminous galaxy shown in Figure~\ref{fig:HyLIRGs}
    \citep{Combes2012, Rawle2014}.

  \item {\bf Haro\,11}: Local ($D$\,$\approx$\,90\,Mpc)
    infrared-luminous (\lir\,$\approx$\,2$\times$$10^{11}$\,\lsun)
    low-metallicity ($Z$\,$\approx$\,1/3\,\zsun) BCD
    \citep[e.g.,][]{Cormier2012, Lyu2016}.

  \item {\bf II\,Zw\,40}: Another local ($D$\,$\approx$\,10\,Mpc)
    low-metallicity ($Z$\,$\approx$\,1/5\,\zsun) BCD with a
    significantly lower infrared luminosity of
    \lir\,$\approx$\,3$\times$$10^{9}$\,\lsun\ (see
    \citealt{Consiglio2016} and \citealt{Kepley2016} for recent ALMA
    studies and references).  II~Zw~40 is one of the two \ion{H}{2}
    galaxies (along with I~Zw~18) studied by \citet{Sargent1970},
    which have defined BCDs as a distinct class of galaxies.

\end{enumerate}
The \spitzer/IRS spectra of the two BCDs were analyzed by
\citet{Hunt2005} and \citet{Wu2006}, while their broad-band SEDs
(covering from near-infrared to submillimeter) were presented by
\citet{Remy-Ruyer2015}.  As already mentioned in
Section~\ref{sec:LBG}, these local BCDs are often thought to be good
analogs of high-redshift low-metallicity galaxies (although the
metallicities of actual $z$\,$=$\,8 galaxies are likely even lower).
The spectra of these BCDs were scaled up by assuming an infrared
luminosity of 2$\times$$10^{13}$\,\lsun, comparable to the
lensing-corrected luminosity of HFLS3.

As Figure~\ref{fig_z8} shows, 10-hr integration SAFARI spectra will
allow us to characterize the physical properties of such HyLIRGs at
$z$\,$\simeq$\,8 in terms of the following characteristics: (i) PAH
feature strengths; (ii) fine-structure line strengths; and (iii)
underlying continuum shapes.  For example, SAFARI will be able to test
whether or not many of $z$\,$>$\,5 galaxies are scaled-up versions of
local low-metallicity BCDs.  The mid-infrared spectra of
low-metallicity BCD are distinctly different from those of normal
infrared-luminous galaxies because of weak PAH features, strong
high-excitation lines (e.g., \neiii~15.5\,\micron\ and
\siv~10.5\,\micron), and a sharply rising red continuum, as first
reported by \citet{Madden2006} based on \iso\ observations.  Weak (or
even absent) PAH features are a common characteristic of
low-metallicity galaxies while strong high-excitation lines are likely
due to a harder UV radiation field \citep{Hunt2005,Wu2006}.  The
latter also explains the presence of the strong \oiii\ 88-$\mu$m line
(even more luminous than the \cii\ 158-$\mu$m line) in low-metallicity
dwarf galaxies, as recently observed by \herschel/PACS spectroscopy
\citep{Cormier2015}.  A powerful way to discriminate between
low-metallicity BCDs, ``normal'' (solar-metallicity) starburst
galaxies, and AGN has also been derived from specific mid-infrared
line ratios as presented by \citet{Fernandez2016} (see their
Figure~11) and \citet{Spinoglio2017}.

The recent detections of high-ionization UV lines in high-redshift
galaxies \citep[e.g.,][]{Stark2015a,Stark2015b,Stark2007,Mainali2007}
suggest that their mid-infrared spectra may also exhibit
high-ionization lines like those seen in these BCD spectra, and if so,
that may support the idea that local BCDs are good analogs of
high-redshift star-forming galaxies.  A sharply increasing red
continuum indicates a significantly warmer dust temperature (46.5\,K
in the case of Haro~11 by \citealt{Lyu2016}), and SAFARI will be
effective for detecting such a warm-dust SED since its wavelength
coverage extends to $>$\,20\,\micron\ in the rest-frame even at
$z$\,$=$\,8.  The existence of such a warm-dust host galaxy has been
suggested for $z$\,$>$\,5 quasars based on their SED analysis
\citep{Lyu2016}.

\subsection{Molecular Hydrogen (\hh) Emission}

The rest-frame mid-infrared spectral range is uniquely important,
since it contains \hh\ lines originating from the lowest energy levels
(i.e., so-called \hh\ pure-rotational lines\footnote{The \hh\ line
  emission produced by transitions between two rotational energy
  states in the ground electronic ($n=0$)/vibrational ($v=0$) level,
  such as 0--0 S(0) ($v=0\rightarrow0$; $J=2\rightarrow0$) at
  28\,\micron\ and 0--0 S(1) ($v=0\rightarrow0$; $J=3\rightarrow1$) at
  17\,\micron.  Ro-vibrational lines are those that involve
  transitions between different vibrational levels, such as 1-0 S(1)
  ($v=1\rightarrow0$; $J=3\rightarrow1$) at 2.12~\micron.}), which
allow us to measure the temperature and mass of the bulk of warm
($T$\,$\gtrsim$\,100\,K) molecular hydrogen gas in galaxies directly.

In local/low-redshift LIRGs and ULIRGs, the luminosities of the
\hh\ 0--0 S(1) line (which is normally one of the brightest
pure-rotational lines) are typically around 0.005\,\% of the total
infrared luminosities (e.g., as estimated by \citealt{Egami2006b}
using the data from \citealt{Rigopoulou2002} and
\citealt{Higdon2006}).  With a 10-hr integration, SAFARI's 5\,$\sigma$
line detection limit will be $\gtrsim$\,$10^{9}$\,\lsun\ at
$z$\,$>$\,5, so this means that for a successful detection of the
\hh\ 0--0 S(1) line at $z$\,$>$\,5, we would need a galaxy with a
total infrared luminosity of $>$\,2$\times$$10^{13}$\,\lsun, i.e.,
HyLIRGs like those listed in Table~\ref{tab:highz}.

By combining {\tt CLOUDY} calculations \citep{Ferland2013} with a
zoom-in, high-resolution ($\simeq$\,30\,pc) numerical simulation, it
is now possible to examine the physical conditions and internal
structures of the inter-stellar medium (ISM) in high-redshift galaxies
including molecular hydrogen gases \citep[e.g.,][]{Vallini2012,
  Vallini2013, Vallini2015, Pallottini2017, Pallottini2017b}.  So far,
these simulations have explored the properties of {\it average}
$z$\,$\sim$\,6 LBGs, and their \hh\ line luminosities are predicted to
be well below SAFARI's detection limit (see Appendix~\ref{h2}).  For a
successful SAFARI \hh\ detection at such high redshift, we would
therefore need a more massive galaxy undergoing a more violent
\hh\ heating process.

From the observations of the nearby and lower-redshift Universe, it is
known that there exist galaxies that exhibit exceptionally strong
\hh\ emission.  Examples include the local LIRG NGC~6240
\citep[e.g.,][]{Lutz2003, Egami2006, Armus2006}, the brightest cluster
galaxy (BCG) in the center of the X-ray-luminous cluster Zwicky~3146
(Z3146; $z$\,$=$\,0.29; \citealt{Egami2006b}), and the radio galaxy
PKS1138$-$26 at $z$\,$=$\,2.16 (the Spiderweb galaxy; \citealt{Ogle2012}).
The $L$(\hh\ 0--0~S(1))/$L_{\rm IR}$ ratios of the first two galaxies
are 0.03\,\% and 0.25\,\%, respectively, significantly larger than the
typical value of 0.005\,\% quoted above.  No \hh\ 0--0~S(1) measurement
is available for the Spiderweb galaxy because of its high redshift,
but the $L$(\hh\ 0--0~S(3))/$L_{\rm IR}$ ratio is comparably high
(0.4\,\%).  Such luminous \hh\ emission lines are thought to be
generated by mechanisms involving strong shocks, such as galaxy
mergers (e.g., NGC~6240) and radio jets (e.g., the Spiderweb).

Note that some of the reported warm \hh\ gas masses are exceptionally
large, $\sim$\,$10^{10}$\,\msun\ for the Z3146 BCG and
$\sim$\,2$\times$$10^{10}$\,\msun\ for the radio galaxy 3C~433 at
$z$\,$=$\,0.1 \citep{Ogle2010}.  However, their CO observations
indicate that warm/cold \hh\ mass ratios are very different between
these two galaxies: $\sim$\,0.1 for the Zwicky 3146 BCG, which is a
typical value for infrared-luminous galaxies, while $>$\,3 for 3C~433,
likely indicating an abnormally strong \hh\ heating process.

\begin{figure}[!htb]

\begin{center}

\hspace*{-0.6cm}\includegraphics[width=3.8in]{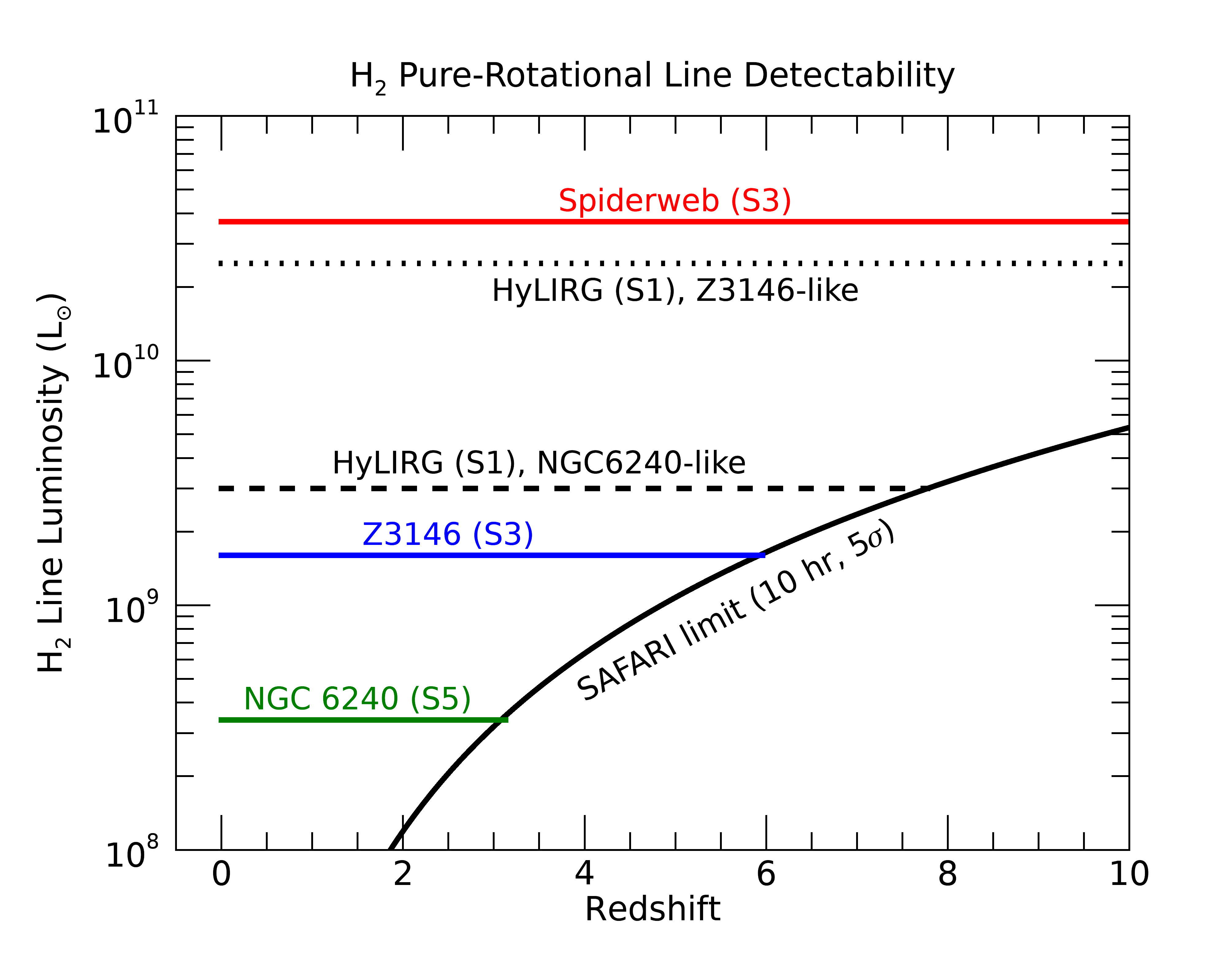}

 \caption{Detectability of \hh\ pure-rotational lines for three known
   extreme \hh\ emitters: (1) the Spiderweb radio galaxy at
   $z$\,$=$\,2.16 with $L$(0--0
   S(3))\,$=$\,3.7$\times$$10^{10}$\,\lsun\ \citep{Ogle2012}; (2)
   Z3146 BCG at $z$\,$=$\,0.29 with $L$(0--0
   S(3))\,$=$\,1.6$\times$$10^{9}$\,\lsun\ \citep{Egami2006}; (3)
   NGC~6240 at $z$\,$=$\,0.0245 with $L$(0--0
   S(3))\,$=$\,3.4$\times$$10^{8}$\,\lsun\ \citep{Armus2006}.  The
   brightest line was used for each case.  With SAFARI's line
   sensitivity, which is also plotted (10\,hours, 5\,$\sigma$), the
   Spiderweb galaxy would be visible beyond $z$\,$=$\,10, while the Z3146
   BGC would be visible up to $z$\,$\sim$\,6; NGC~6240, on the other hand,
   would drop out at $z$\,$\sim$\,3.  Also shown are the visibilities of a
   HyLIRG ($L_{\rm IR}$\,$=$\,$10^{13}$\,\lsun) through the 0--0 S(1) line
   assuming $L_{\rm 0-0~S(1)}/L_{\rm IR}$\,$=$\,0.25\,\% (Z3146-like) and
   0.03\,\% (NGC~6240-like).}

 \label{fig_H2}

\end{center}

\end{figure}

Figure~\ref{fig_H2} shows the detectability of the brightest
pure-rotational lines of three luminous \hh\ emitters
  (NGC~6240, the Z3146 BCG, and the Spiderweb galaxy) toward high
  redshift.  Although the \hh\ 0--0 S(5) line of NGC~6240 would drop
out of SAFARI detection at $z$\,$\sim$\,3, the \hh\ 0--0 S(3) line of
the Z3146 BCG would remain visible up to $z$\,$\sim$\,6, and the \hh\
0--0 S(3) line of the Spiderweb galaxy would stay well above the
SAFARI detection limit even at $z$\,$=$\,10.  The figure also shows
that if we assume an NGC~6240-like $L$(\hh\ 0--0~S(1))/$L_{\rm IR}$
ratio (i.e., 0.03\,\%), a HyLIRG with
$L_{\rm IR}$\,$=$\,$10^{13}$\,\lsun\ will produce an \hh\ 0--0 S(1)
line detectable up to $z \sim 8$, and with a Z3146 BCG-like ratio
(i.e., 0.25\,\%), it will be detectable beyond $z$\,$=$\,10, just like
the Spiderweb galaxy.  The existence of these extreme \hh\ emitters
suggests that \hh\ lines will likely serve as important probes for
galaxies at high redshift, providing crucial observational constraints
on theoretical models like the one presented for a $z$\,$\simeq$\,6
LBG in Appendix~\ref{h2}.

It should be emphasized that \hh\ emission is sensitive only to warm
($T$\,$\gtrsim$\,100\,K) \hh\ gas, meaning that it allows sampling of
only a limited fraction ($\sim$\,10--20\,\%) of the total molecular
gas mass in a typical galaxy.  However, it is also possible to
estimate the total \hh\ gas mass from the measured warm \hh\ gas mass
by making some assumptions.  For example, \citet{Togi2016} recently
proposed such a method, estimating the total \hh\ gas mass from the
observations of multiple \hh\ emission lines alone, assuming a
continuous power-law distribution of rotational temperatures down to a
certain cutoff value.  For a sample of local galaxies with reliable
CO-based molecular masses, this method has been shown to produce the
total molecular gas mass within a factor of 2 of those derived from CO
when a cutoff temperature of around 50\,K is adopted.  Though indirect
and dependent on some assumptions, methods like this have the
potential to provide useful estimates for total \hh\ gas masses,
especially for low-metallicity galaxies, for which CO-/dust-based
methods are known to underpredict molecular gas mass by a factor of
approximately 100 locally, possibly due to the presence of CO-dark
\hh\ gas \citep{Wolfire2010,Togi2016}.

Taking these known extreme \hh\ emitters as a guide, we also simulated
the spectra of \hh\ lines in Figure~\ref{fig_z8}, by making the
following two assumptions: (1) \hh\ level populations are fully
thermalized (i.e., in the local thermodynamic equilibrium; LTE); and
(2) the galaxy contains two warm \hh\ gas components, one with a gas
mass of 2$\times$$10^{10}$\,\msun\ and a gas temperature of
$T$\,$=$200\,K and the other with a gas mass of
2$\times$$10^{8}$\,\msun\ and a gas temperature of $T$\,$=$\,1000\,K.
An ortho-to-para ratio of 3:1 is also assumed.  Such a two-component
LTE model is known to produce good fits to the excitation diagrams of
\hh\ pure-rotational lines for lower-redshift galaxies
\citep[e.g.,][]{Higdon2006}, although this should probably be taken as
a simple and effective parameterization of more complex underlying gas
temperature and mass distributions.

The red lines shown in Figure~\ref{fig_z8} indicate the \hh\ emission
lines produced by such a model.  These simulated \hh\ lines have
luminosities of 0.5--1$\times$$10^{10}$\,\lsun, exceeding SAFARI's
10-hr 5\,$\sigma$ line-flux limit of 3$\times$$10^{9}$\,\lsun\ at
$z=8$.  If detected, such luminous \hh\ lines would indicate the
existence of a large {\em warm} \hh\ gas reservoir, as well as some
mechanism that heats it (e.g., shocks), possibly marking the sites of
galaxy formation/assembly.

\section{Exploratory Sciences}

\label{sec:exploratory}

From a broader perspective, the 35--230\,\micron\ window targeted by
SAFARI has a singular importance over the coming decades, as we try to
detect and study the first-generation objects that appeared in the
early Universe.  This spectral range, which samples the rest-frame
mid-infrared at $z$\,$>$\,5, is uniquely powerful for probing
first-generation objects because it contains: (1) key cooling lines of
low-metallicity or metal-free gas, especially \hh\ lines; and (2)
emission features of solid compounds that are thought to be abundant
in the remnants of Pop III supernovae (SNe).  Detections of such
spectral features, if successful, will open up a new frontier in the
study of the early Universe, shedding light on the physical properties
of the first galaxies and first stars.

As soon as the Big-Bang cosmology was validated by the detection of
the cosmic microwave background (CMB) radiation \citep{Penzias1965},
it was recognized that \hh\ molecules must have played an important
role as a coolant of pristine pre-galactic gas clouds
\citep{Saslaw1967,Peebles1968,Hirasawa1969,Matsuda1969,Takeda1969}.
In the metal-free environment that existed in the early Universe, the
only available coolants were hydrogen, helium, and molecular hydrogen;
Since the gas cooling curves of the former atomic species have a
cutoff around $10^{4}$\,K \citep[e.g.,][]{Thoul1995}, \hh\ molecules
must have been the dominant coolant in pristine primordial gas clouds
that are not massive enough (e.g., $<10^8$\,\msun\ at $z$\,$\sim$\,10)
to have a virial temperature ($T_{\rm vir}$) of $>$\,$10^{4}$\,K.  Put in
another way, \hh\ cooling determines the minimum mass of a pristine
gas cloud that can cool and contract at a given redshift
\citep[e.g.,][]{Tegmark1997}.  As a result, \hh\ lines are considered
to be the most powerful (and likely the only) probe of the first
cosmological objects that appeared in the early Universe.  In such
pristine gas clouds, cooling is dominated by \hh\ pure-rotational
lines, and at the expected formation redshift of such first-generation
objects, $z$\,$\sim$\,10--30, these \hh\ lines will fall in the
far-infrared.

From the discussion in the previous section, it is clear that SAFARI
can only detect exceptionally luminous systems at high redshift
($\mu\,L_{\rm line}$\,$>$\,$10^{9}$\,\lsun\ at $z$\,$>$\,5).  However,
the abundance and physical properties of such luminous (and therefore
likely massive) systems at $z$\,$>$\,5, not to mention those of the
first-generation objects, are barely known at present, preventing us
from making realistic predictions for what SAFARI may be able to
detect and study.  The goal of this section, therefore, is to {\it
  explore} (as opposed to {\it assess}) \spica's potential to open up
a new window toward the early Universe.  Recognizing that any current
model predictions suffer from considerable uncertainties, we discuss
various topics while allowing a gap of up to a factor of 100 between
SAFARI's expected sensitivity and model-predicted source luminosities.
This is because any theoretical prediction could easily be off by an
order of magnitude and gravitational lensing could bridge a gap of
another factor of 10 (or even more).  The aim here is to present
scientific ideas for further refinement rather than making a
quantitative assessment, which is not yet possible given the lack of
direct observational constraints.

\subsection{First Objects: Current Picture}

\label{overview}

Although first stars and galaxies are yet to be observed, they have
been a major focus of theoretical studies over the years (see
\citealt{Ciardi2005}, \citealt{Bromm2011}, \citealt{Yoshida2012},
\citealt{Bromm2013}, \citealt{Greif2015}, and \citealt{,Barkana2016}
for review).  In the framework of the standard $\Lambda$CDM model, we
expect the first (i.e., Pop~III) stars to form in dark matter (DM)
minihalos of around $10^{6}$\,\msun\ at redshifts
$z$\,$\simeq$\,20--30, cooling via \hh\ molecular lines
\citep{Haiman1996, Tegmark1997, Yoshida2003}.  The first stars formed
in such a metal-free environment are believed to be quite massive
($>100$\,\msun; e.g., \citealt{hirano2015}), and would emit strong
\hh-dissociating UV radiation \citep[e.g.,][]{Omukai1999} and produce
powerful supernova explosions \citep{Bromm2003a}, essentially shutting
off subsequent star formation.  For this reason, these
  minihalos are not regarded as ``first galaxies'' although they are
  the sites of the first-star formation.  The next generation of star
formation will then take place in more massive halos
($\sim$\,$10^{8}$\,\msun) collapsing at $z$\,$\sim$\,10, whose virial
temperature is high enough ($>$\,$10^{4}$\,K) to sustain cooling due
to atomic hydrogen \citep[e.g.,][]{Oh2002}.  These so-called ``atomic
cooling halos'' hosting the second generation of stars are often
considered as ``first galaxies'' \citep{Bromm2011}.

Note that according to this current standard picture, first galaxies
are not necessarily metal-free (Pop~III), which is often taken as the
observational definition of the first galaxies.  In fact,
  ``This popular definition of a first galaxy may be misleading and
  may render any attempts to find first galaxies futile from the very
  outset'' \citep{Bromm2011}.  This is because it is difficult to
prevent minihalos, i.e., the building blocks of first galaxies, from
forming massive Pop~III stars and chemically enriching their
surroundings through SNe explosions.  In other words, to produce
genuine Pop~III galaxies, it is necessary to inhibit star formation in
the progenitor mini-halos by suppressing the formation of molecular
hydrogen in them.  This would require \hh-dissociating Lyman-Werner
(LW) background radiation in the Far-UV (11.2--13.6\,eV photons) and
the source of such radiation before the formation of Pop~III galaxies.
This leads to a scenario in which the first galaxies that appeared in
the Universe (in the chronological sense) were mostly Pop~II galaxies;
Pop~III galaxies would appear subsequently in underdense regions where
the star formation in minihalos were suppressed by radiation emitted
by stars/galaxies formed earlier in overdense regions.  
  For this reason, Pop~III galaxies may be considered as the
  second-generation galaxies containing first-generation stars
\citep[e.g.,][]{Johnson2008, Trenti2009, Johnson2010, Stiavelli2010,
  Johnson2013}.

Adopting this theoretical picture as the baseline, we will discuss
below the rest-frame mid-infrared spectral signatures of massive
(\mhalo\,$\sim$\,$10^{11}$--$10^{12}$\,\msun) forming galaxies
containing low-metallicity or pristine (i.e., metal-free) gas clouds.
By ``forming'', we denote galaxies that are yet to form stars, meaning
that the source of line luminosities is the gravitational energy
released by the contraction of clouds under their self-gravity, rather
than stellar radiation.  

Production of the first dust by Pop~III SNe would be another important
scientific topic that could be uniquely addressed by rest-frame
mid-infrared spectroscopy.  However, considering that even larger
uncertainties are involved in such a discussion, we limit ourselves
here to providing a qualitative overview in
Appendix~\ref{sec:firstdust}, deferring a more quantitative analysis
to a forthcoming paper (R. Schneider et al., in preparation).

\subsection{Massive Forming Galaxies}

\label{sec:pop3_gal}

\begin{figure*}[!htb]

\centerline{\includegraphics[width=1.02\linewidth]{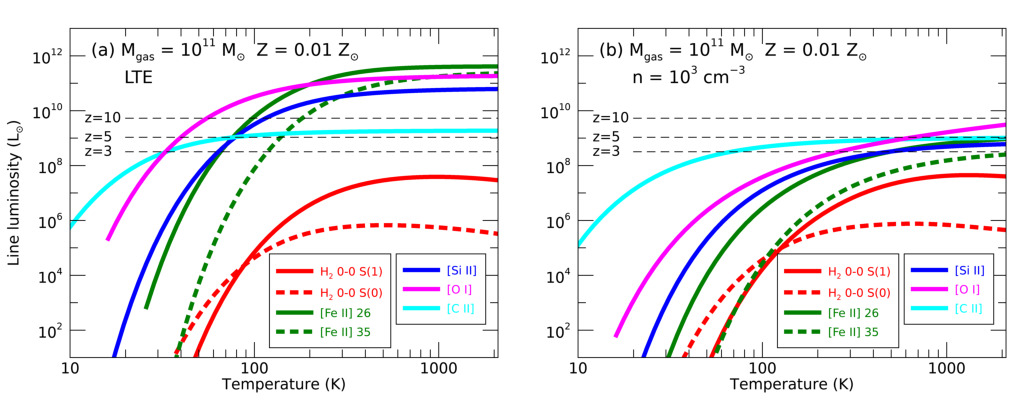}}

\caption{Luminosities of \hh\ lines (0--0 S(1) 17\,\micron\ and 0--0
  S(0) 28\,\micron) and fine-structure lines
  (\fetwo\ 25.99/35.35\,\micron, \sitwo\ 34.8\,\micron,
  \oi\ 63\,\micron, and \cii\ 158\,\micron) calculated for a
  low-metallicity ($Z$\,$=$\,0.01\,\msun) forming galaxy with a gas
  mass of $10^{11}$\,\msun.  The left panel (a) assumes a gas density
  high enough for these lines to be thermalized (i.e., in LTE) while
  the right panel (b) corresponds to the case with a gas density of
  $10^{3}$\,cm$^{-3}$.  A molecular fraction of 2$\times$$10^{-4}$ was
  assumed for both cases.  The former is similar to the calculation
  presented by \citet{Santoro2006} (see their Figure~11 for a similar
  model with a gas mass of $10^{8}$\,\msun).  The horizontal dotted
  lines indicate the nominal 5\,$\sigma$ detection limit of SAFARI
  with a 10-hour on-source integration time for $z$\,$=$\,3, 5, and
  10.  The LTE case on the left suggests that if the gas is
  sufficiently warm ($T$\,$\gtrsim$\,200\,K), fine-structure lines
  like \fetwo\ 25.99/35.35\,\micron\ and \sitwo\ 34.8\,\micron\ can be
  quite luminous (note, however, that these lines will be redshifted
  out of the SAFARI window at $z$\,$=$\,7.8/5.5, and 5.6,
  respectively).}
\label{fig:omukai}

\end{figure*}

\subsubsection{Population~II}

\label{pop2_clouds}

As already discussed above, formation of Pop~II galaxies may precede
that of Pop~III galaxies in the history of the Universe.  When the gas
metallicity exceeds a certain threshold\footnote{This critical
  metallicity ($Z_{\rm crit}$) has been estimated to be
  $\sim$\,$10^{-3.5}$\,Z$_{\odot}$ for dense gas
  ($n$\,$\gtrsim$\,$10^{3}$\,cm$^{-3}$) and
  $\sim$\,$10^{-3}$--$10^{-2}$\,Z$_{\odot}$ for lower-density
  ($n$\,$=$\,1--100\,cm$^{-3}$) gas \citep[e.g.][]{Santoro2006}.}, gas
cooling will be dominated by several key fine-structure lines, such as
\cii~158\,\micron\ and \oi~63\,\micron\ in the far-infrared
\citep{Bromm2003b} and \sitwo~34.8\,\micron\ and
\fetwo~25.99/35.35\,\micron\ in the mid-infrared \citep{Santoro2006}
if we assume that C, Si, and Fe atoms are photoionized by ambient UV
radiation.  The former far-infrared lines, when redshifted, can be
studied with submillimeter/millimeter telescopes on the ground (e.g.,
ALMA), but the detection of the latter mid-infrared lines requires a
space far-infrared telescope such as \spica\footnote{At
  $z$\,$\gtrsim$\,10, these mid-infrared lines will also be redshifted
  into the highest-frequency ALMA band (i.e., Band~10) although the
  sensitivity may be an issue.}.

For SAFARI to have any chance of detection, the target object must be
exceptionally luminous and massive.  As a maximal model, we consider
here a Pop~II forming galaxy that has a gas mass ($M_{\rm gas}$) of
$10^{11}$\,\msun.  The corresponding halo mass (\mhalo) would be
$\sim$\,$10^{12}$\,\msun, and such a massive halo has a comoving
number density of $\sim$\,$10^{-5}$\,Mpc$^{-3}$ at $z$\,$\sim$\,6.
The resultant gas mass fraction would be $\sim$10\,\%, which is still
well below the cosmic baryon fraction of $\sim$0.19
($f_{b}$\,$=$\,$\Omega_b/\Omega_c$; \citealt{Planck2016}).  Such a
classical model of a massive forming galaxy containing a uniform gas
is obviously an over-simplification, but it serves the purpose of
illustrating the parameter space SAFARI could potentially probe.

Figure~\ref{fig:omukai} plots the expected luminosities of key cooling
lines from such a massive forming galaxy assuming a metallicity of
$Z$\,$=$\,0.01\,\zsun.  Figure~\ref{fig:omukai}a shows that if the gas
is sufficiently warm ($T$\,$\gtrsim$\,200\,K) and dense ($n_{\rm
  H}$\,$\sim$\,$10^{5}$--10$^{6}$\,cm$^{-3}$) to thermalize all these
lines (i.e., in LTE), the mid-infrared fine-structure lines such as
the \fetwo~25.99/35.35\,\micron\ and \sitwo~34.8\,\micron\ lines can
be quite luminous ($\gtrsim$\,$10^{11}$\,\lsun) and detectable with
SAFARI to high redshift (Note, however, that these lines will
  be redshifted out of the SAFARI window at $z$\,$=$\,7.8/5.5, and
  5.6, respectively).  On the other hand, if the typical gas density
is more like $\sim$\,$10^{3}$\,cm$^{-3}$, these mid-infrared
fine-structure lines will become undetectable beyond $z$\,$\sim$\,5
even for such a massive forming galaxy (Figure~\ref{fig:omukai}b).

As discussed by \citet{Santoro2006}, \hh\ lines in these models are
significantly fainter when compared to the fine-structure lines.  This
is because the molecular fraction was assumed to be
2$\times$$10^{-4}$ here, resulting in a total \hh\ gas mass of
only 2$\times$$10^{7}$\,\msun.  As already shown in Figure~\ref{fig_z8},
for SAFARI to detect \hh\ lines at high redshift, the corresponding
\hh\ gas mass needs to be $\gtrsim$\,$10^{10}$\,\msun.
Note that the luminosities of \hh\ lines, as well as that of
\cii~158\,\micron, do not change much between the two cases because a
gas density of 10$^{3}$\,cm$^{-3}$ is close to the critical densities
of these transitions.

\subsubsection{Population~III}

\label{sec:giant_gal}

For the discussion of Pop~III forming galaxies, we refer to the model
calculations by \citet{Omukai2003} as a guide.  Because of the reduced
cooling efficiency due to the absence of metal lines, Pop~III systems
with \mhalo\,$\gtrsim$\,$10^{12}$\,\msun\ cannot cool appreciably
within the halo merging timescale, constantly heated by infalling
subhalos.  As a result, the most \hh-luminous Pop~III forming galaxies
are expected to be those with a halo mass of
$\simeq$\,$10^{11}$\,\msun; \hh\ line luminosities will decline
sharply for higher-mass systems.  The $z$\,$=$\,8
\mhalo\,$=$\,$10^{11}$\,\msun\ Pop~III forming galaxy
model\footnote{The fiducial model with $f_{\rm trb}$\,$=$\,0.25.}
by \citet{Omukai2003} predicts the luminosity of the
  brightest \hh\ line (0--0 S(3)) to be
    3.3$\times$$10^{7}$\,\lsun, still two orders of magnitude below
  the SAFARI's nominal detection limit
  ($\sim$3$\times$$10^{9}$\,\lsun\ at $z$\,$=$\,8, 10 hours,
  5\,$\sigma$).  Other theoretical studies predicted similar \hh\ line
  luminosities \citep[e.g.,][]{Mizusawa2005,Gong2013}, indicating that
  a successful detection of \hh\ lines will require some mechanism to
  boost the line luminosities (see Section~\ref{h2_boost}) as well as
  magnification by lensing (see
  Section~\ref{surface_density})\footnote{\citet{Mizusawa2005}
    presented a more optimistic view about the detectability of \hh\
    lines because they assumed a line-flux sensitivity of
    approximately $10^{-22}$\,W m$^{-2}$, which is about
      two orders of magnitude deeper than what we are assuming here
    for SAFARI.}.

Note that the virial temperature of a
\mhalo\,$\simeq$\,$10^{11}$\,\msun\ halo is high enough
($>$\,$10^4$~K) to sustain \ion{H}{1} atomic cooling.  As a result,
these massive forming galaxies are also expected to be strong
\lya\ emitters (The source of \lya\ emission here is the
  release of gravitational energy due to the contraction of pristine
  gas clouds and not the radiation from young stars).  In fact, the
\lya\ line is more luminous (4.9$\times$$10^7$\,\lsun) than any of the
individual \hh\ lines, although the total \hh\ line luminosity is
$\sim$\,$10^{8}$\,\lsun, exceeding that of \lya.  Even with \jwst,
however, such a \lya\ line will not be easy to detect.  The estimated
observed \lya\ line flux will be 2.4$\times$$10^{-22}$\,W m$^{-2}$ at
$z$\,$=$\,8 while the line sensitivity of \jwst/NIRSpec at the
wavelength of the redshifted \lya\ ($\sim$1\,\micron) will be
$\simeq$\,2.5$\times$$10^{-21}$\,W m$^{-2}$ (5\,$\sigma$, 1\,hour,
with the $R$\,$=$\,1000 grating), which is one order of magnitude
brighter.  Furthermore, \lya\ emission may be resonantly scattered and
absorbed by the intervening neutral IGM.

\subsubsection{Other \hh\ Excitation Mechanisms}

\label{h2_boost}

The calculations above indicate that gravitational contraction of
massive forming galaxies alone is unlikely to release enough
energy to produce \hh\ lines that are detectable with SAFARI at high
redshift.  The next question is therefore if there could be any other
\hh\ excitation mechanisms that would produce even more luminous
\hh\ lines.  Below, we discuss a few possibilities:

{\bf Pop~III Supernovae}: Explosions of Pop~III SNe may produce strong
\hh\ lines by blowing away the gas content of the parent galaxy and
collecting it into a cooling shell where \hh\ rapidly forms
\citep{Ciardi2001}.  Such SN blowouts are expected to happen in
low-mass Pop III galaxies, whose gravitational potential is shallow
\citep{Ferrara1998}.  According to the calculation by
\citet{Ciardi2001}, a Pop III galaxy with a halo mass of
$10^{8}$--$10^{9}$\,\msun\ could emit \hh\ lines with luminosities
reaching $10^{6}$--$10^{7}$\,\lsun\ at $z$\,$\simeq$\,8--10,
corresponding to $\sim$\,10\,\% of the explosion energy.  These line
luminosities are comparable to those of the
\mhalo\,$=$\,$10^{11}$\,\msun\ Pop~III forming galaxy discussed in
Section~\ref{sec:giant_gal}, but the halo mass here is 10--100$\times$
smaller, signifying the power of SN explosions to enhance
\hh\ luminosities.  Given the significant uncertainties associated
with the simple analytic model of \citet{Ciardi2001}, a further
theoretical investigation is needed to assess this model more
quantitatively.

One interesting aspect of this Pop~III SNe explosion model is that the
ro-vibrational line 1--0 S(1) (2.12\,\micron) is expected to be
significantly brighter than the pure-rotational line 0-0 S(1)
(17\,\micron), reflecting a hotter temperature of the \hh-emitting
gas.  Strong ro-vibrational lines would make shorter-wavelength
observations effective.  For example, the 1--0 S(1) line would be
redshifted to 19\,$\mu$m at $z$\,$=$\,8, which is shortward of the
SAFARI wavelength coverage but within those of \jwst/MIRI and
\spica/SMI.  However, the line-luminosity predicted by
\citet{Ciardi2001} is $\sim$\,$10^{-21}$\,W m$^{-2}$, still not bright
enough for these instruments to make an easy detection.

{\bf Merging of Massive Halos}: In the high-redshift Universe, where
halos are constantly merging to create more massive galaxies,
\hh\ formation and excitation due to strong shocks are likely
important.  Such a mechanism has been seen to be at work in some
systems in the local Universe.  One particularly interesting example
is the Stephan's Quintet, located in a compact group at 94 Mpc, which
exhibits exceptionally luminous \hh\ emission
(2$\times$$10^{8}$\,\lsun\ for 0-0 S(0) through S(5) combined)
spreading over an area of approximately 50$\times$35\,kpc$^2$
\citep{Appleton2006, Cluver2010, Appleton2013, Appleton2017}.  It is
believed that in this system, one high-velocity ``intruder'' galaxy is
colliding with the intergroup medium and generating shocks
\citep{Sulentic2001}.  According to the model presented by
\citet{Guillard2009}, \hh\ molecules form out of the shocked gas, and
\hh\ emission is powered by the dissipation of kinetic turbulent
energy of the \hh\ gas.  A similar mechanism will likely generate
luminous \hh\ lines through merging of massive halos (i.e., major
mergers) at high redshift.

\subsection{Pop III Objects at Lower Redshift}

\label{lowz_pop3}

Although detecting genuine Pop~III objects at high redshift will
likely require strong boosting of \hh\ line luminosities by some
mechanism as well as lensing amplification, SAFARI may be able to
probe the properties of such objects through the observations of
similarly metal-poor objects that may exist at lower redshift.  For
example, a number of studies have suggested that Pop III star
formation may continue toward low redshift and maybe even down to
$z$\,$\sim$\,3 \citep[e.g.,][]{scannapieco2003, jimenez2006,
  schneider2006a, tornatore2007, ricotti2008, Trenti2009, Johnson2010,
  Johnson2013, Pallottini2014, Pallottini2015}.  According to the
recent {\it Renaissance Simulations} \citep{xu2016}, only 6\,\% of the
volume and 13\,\% of the gas mass is enriched to [Z/H] $>$\,$-4$ at
$z$\,$=$\,7.6 in the comoving survey volume of 220\,Mpc$^{3}$,
indicating that there is a large amount of pristine gas available for
Pop III star formation at $z$\,$<$\,7.6 (although much of it likely
resides in low-density diffuse IGM).  Observationally, however, there
has been no secure identification of a Pop III galaxy so far.
Although there are indications of massive, low-metallicity stars in
the nearby lowest-metallicity galaxy I~Zw~18 \citep{Kehrig2015}, a
clear detection of Pop~III objects is still missing.  Note that, at
$z$\,$=$\,3, the line-luminosity detection limit of SAFARI will be
lower by an order of magnitude ($\sim$\,3$\times$$10^8$\,\lsun,
5\,$\sigma$ in 10\,hours) compared to that at $z=8$, potentially
enhancing the probability of detecting Pop~III objects.

\section{Finding Targets for SAFARI}

\label{sec:targets}

SAFARI will conduct spectroscopy in the single-object mode, targeting
one source at a time.  Since the surface density of luminous
$z$\,$>$\,5 galaxies like those listed in Table~\ref{tab:highz} is
small (e.g., the six SPT galaxies listed were discovered by a
millimeter survey covering an area of $\sim$\,2500\,deg$^{2}$), the
success of SAFARI in the high-redshift exploration will heavily depend
on how we will be able to discover exciting targets with the existing
and future wide-field survey data.  Below, we discuss strategies for
such a target selection.

\subsection{Surface Density of Massive Halos at $z$\,$>$\,5}

\label{surface_density}

\begin{figure}[!htb]

\hspace*{-0.2cm}\includegraphics[width=3.6in]{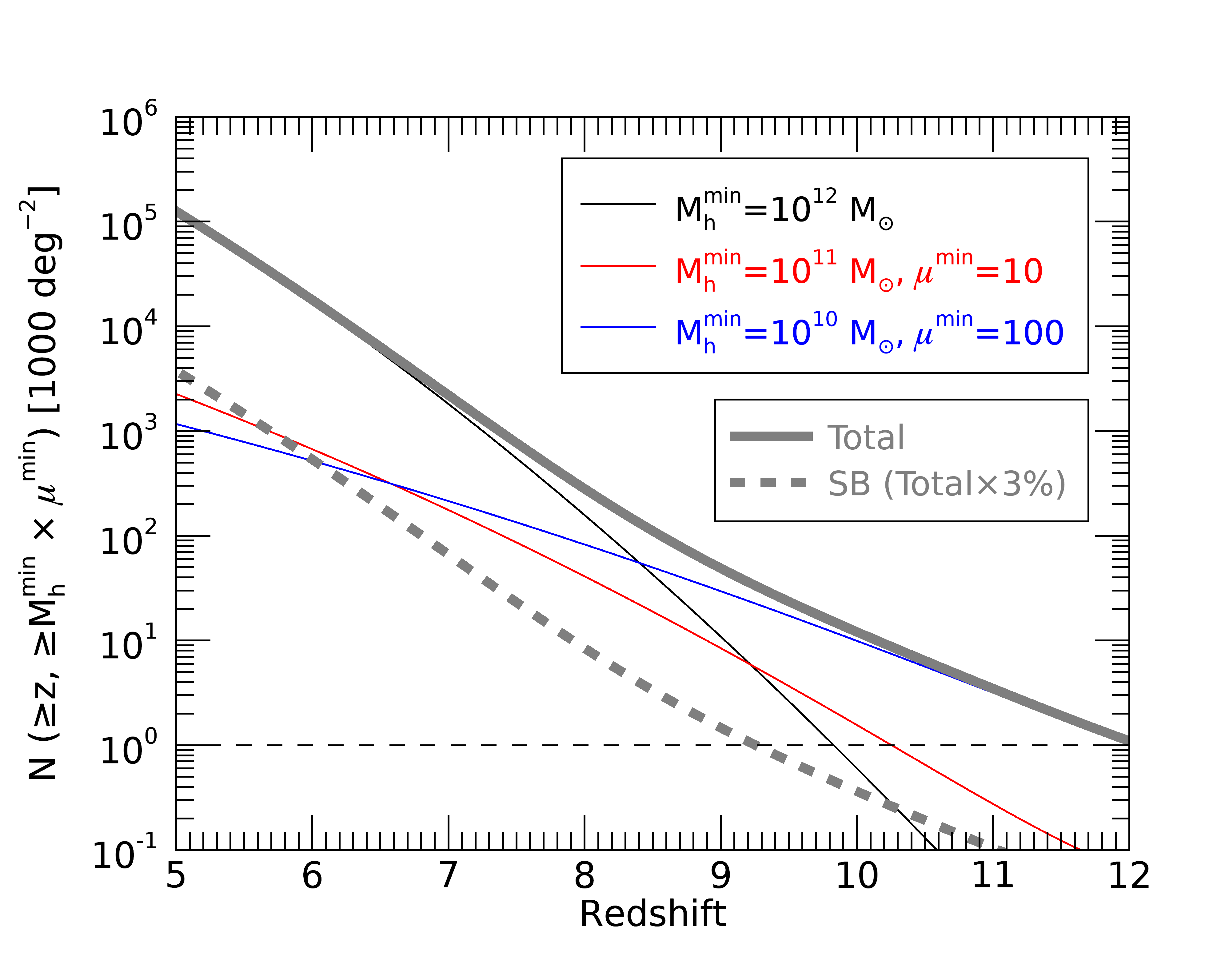}

\caption{Cumulative surface densities of DM halos per 1000 deg$^2$ as
  predicted by the the standard $\Lambda$CDM model.  The black line
  indicates the cumulative surface density of $\geq 10^{12}$
  \msun\ halos down to given redshifts while the blue and red lines
  plot the corresponding numbers for $\geq 10^{11}$ \msun\ halos
  gravitationally lensed by a factor of $\geq 10\times$ ($\mu \geq
  10$) and for $\geq 10^{10}$ \msun\ halos lensed by a factor of $\geq
  100\times$ ($\mu \geq 100$), respectively.  According to the
  calculation by \citet{Zackrisson2015}, when the source redshift is
  above $\sim$5, the corresponding lensing probability is roughly
  constant and $\sim 10^{-4}$ for $\mu \geq 10$ and $\sim 10^{-6}$ for
  $\mu \geq 100$ along an average line of sight.  These probabilities
  have been multiplied with the $\geq 10^{11}$ and $\geq 10^{10}$
  \msun\ halo surface densities.  The thick grey line plots the total
  surface density of the three halo populations combined while the
  thick grey dashed line plots 3\% of the total number, comparable to
  the starburst fraction estimated at lower redshift
  \citep[e.g.,][]{Bethermin2017}.  The halo comoving number density
  was computed with the Press-Schechter formalism \citep{Press1974}.
  The horizontal dashed line corresponds to $N=1$.
}

\label{fig:lensed}

\end{figure}

Whether we try to detect hyper-luminous galaxies like
HFLS3/SPT0311$-$58 (Section~\ref{EoR}) or a massive Pop~II forming
galaxy (Section~\ref{pop2_clouds}), we will be searching for objects
residing in massive halos with \mhalo\,$\geq$\,$10^{12}$\,\msun.
Figure~\ref{fig:lensed} shows the cumulative surface density of
$\geq$\,$10^{12}$\,\msun\ halos per 1000\,deg$^{2}$ (the black line).
Together, we also plot the surface densities of
$\geq$\,$10^{11}$\,\msun\ halos gravitationally lensed by a factor of
$\geq$\,10$\times$ (the red line) and $\geq$\,$10^{10}$\,\msun\ halos
lensed by a factor of $\geq$\,100$\times$ (the blue line).  If we make
a simple assumption that various physical properties of galaxies
hosted by these halos (e.g., SFR, \mstar) roughly scale linearly with
the halo mass at $z$\,$>$\,5 (i.e., maintaining a constant
mass-to-light ratio\footnote{Recent theoretical studies suggest that
  the SFR and \lir\ of high-redshift galaxies are roughly proportional
  to stellar mass (\mstar) \citep[e.g.,][]{Yajima2015}, and that the
  stellar mass is proportional to the halo mass (\mhalo), at least in
  the mass range of \mhalo\,$=$\,$10^{10}$--$10^{12}$\,\msun\
  \citep[e.g.,][]{Behroozi2015}, naturally leading to this assumption
  of a constant mass-to-light ratio.  Note, however, that galaxies
  undergoing a strong starburst phase could be significantly
  over-luminous for a give stellar/halo mass, such as HFLS3 as
  discussed in Section~\ref{EoR}.  At
  \mhalo\,$\gtrsim$\,$10^{12}$\,\msun, the \mstar--\mhalo\ relation
  flattens significantly with decreasing \mstar/\mhalo ratios, likely
  due to AGN feedback suppressing star formation}), these three
populations of halos would contain galaxies that have comparable
apparent brightnesses.  The cumulative surface density of the three
halo populations combined (the thick grey line) is about
2200 at $z$\,$\geq$\,7, 280 at $z$\,$\geq$\,8, 50 at $z$\,$\geq$\,9,
and 12 at $z$\,$\geq$\,10 (per 1000\,deg$^2$).  The figure also shows
that lensed populations would start to dominate in number at
$z$\,$\gtrsim$\,8.5, especially $\geq$\,$10^{10}$\,\msun\ halos
magnified by a factor of $\geq$\,100$\times$.

Note, however, that translating the surface density of halos into that
of HyLIRGs ($\mu$\,$L_{\rm}$\,$>$\,$10^{13}$\,\lsun) or massive forming
galaxies is not trivial due to a number of astrophysical processes
involved.  In this sense, Figure~\ref{fig:lensed} sets the upper limit
on the number of possible detections (i.e., there cannot be more
galaxies than there are halos).  In practice, we may assume that only
a fraction of these halos hosts luminous objects that are observable
with SAFARI.  As an illustration, we also plot a line denoting 3\,\% of
the total halo number, comparable to a starburst fraction assumed by
some models \citep[e.g.,][]{Bethermin2017}.  Although luminous objects
may be more abundant at high redshift due to increased
star/structure-formation activities, these simple calculations
indicate that it will likely be necessary to survey hundreds/thousands
of square degrees to find a handful of high-redshift objects that are
luminous enough for SAFARI to observe, which is consistent with the
outcomes of the wide-field \herschel/SPT surveys so far.

\subsection{Submillimeter/Millimeter Surveys}

\label{sec:submm_mm}

Wide-field submillimeter/millimeter surveys have proven to be
extremely effective in finding gravitationally lensed
infrared-luminous galaxies at high redshift
\citep[e.g.,][]{Negrello2010,Vieira2013,Weiss2013}.  This is because,
(1) the apparent brightnesses of infrared-luminous galaxies do not
fade much in the submillimeter/millimeter to high redshift (up to
$z$\,$\sim$\,10 at $\sim$\,1\,mm) due to a strong negative K
correction, and (2) foreground galaxies are faint in the
submillimeter/millimeter unless they contain strong AGN or they are at
really low redshift.  As a result, it is straightforward to identify
lensed infrared-luminous galaxies at high redshift by just inspecting
the brightest sources in the submillimeter/millimeter, 20--30\,\% of
which are typically lensed galaxies at $z$\,$\gtrsim$\,1
\citep{Negrello2010}.

Recent results from \herschel/SPIRE wide-field surveys show that the
surface density of bright ($S_{500}$\,$>$\,100\,mJy) lensed
infrared-luminous galaxies ranges from 0.13\,deg$^{-2}$ \citep[80
  sources over 600\,deg$^2$,][] {Negrello2017} to 0.21\,deg$^{-2}$
\citep[77 sources over 372\,deg$^2$,][] {Nayyeri2016}.  The surface
density of SPT-selected galaxies has been shown to be consistent with
that of these \herschel/SPIRE-selected galaxies \citep{Mocanu2013}.
These low ($<$\,1\,deg$^{-2}$) surface densities clearly indicate that
wide submillimeter/millimeter surveys covering hundreds/thousands of
square degrees are needed to produce a large sample of such lensed
infrared-luminous galaxies, only a limited fraction of which will be
at $z$\,$>$\,5.

The depth of \herschel/SPIRE wide surveys matches well with the
spectroscopic sensitivity of SAFARI.  HFLS3 ($z$\,$=$\,6.34) shown in
Figure~\ref{fig:HyLIRGs} has a SPIRE 500-\micron\ flux density of
47\,mJy, while the corresponding 5\,$\sigma$ confusion limit is
approximately 35\,mJy \citep{Nguyen2010}.  The figure indicates that it
will be difficult for SAFARI to obtain good-quality spectra for
sources much fainter than HFLS3.  In this sense, \spica\ will be a
well-matched spectroscopic follow-up mission for \herschel-selected
$z$\,$>$\,5 galaxies.

Note also that most of these
  submillimeter/millimeter-detected lensed galaxies are located at
moderate redshift ($z$\,$=$\,1--4).  Although these
  lower-redshift galaxies will make excellent SAFARI targets in
  general (with a typical magnification factor of 10, integration
  times will be reduced by a factor of 100), $z$\,$>$\,5 galaxies are
  much more scarce.  Compared to the \herschel\ surveys, the SPT
  survey has been more successful in finding $z$\,$>$\,5 galaxies (see
  Table~\ref{tab:highz}) presumably because the effect of the negative
  K correction extends toward higher redshift in the millimeter.
This suggests that future/on-going millimeter surveys offer great
potential for finding bright targets for SAFARI at the highest
redshifts (e.g., Advanced ACT, SPT3G, as well as new cameras on the
APEX, IRAM 30-m, and LMT telescopes).

A complementary approach would be to search for lensed
infrared-luminous galaxies in the fields of massive lensing clusters.
In fact, HLSJ0918, one of the most infrared-luminous galaxies at
$z$\,$\gtrsim$\,5 identified so far (see Table~\ref{tab:highz} and
Figure~\ref{fig:HyLIRGs}) was discovered by the \herschel\ Lensing
Survey (HLS) targeting such lensing cluster fields\footnote{The
  dominant lens of this particular lensed object turned out to be a
  foreground galaxy rather than a galaxy cluster \citep{Rawle2016}.}
\citep{Egami2010,Combes2012,Rawle2016}.

\subsection{\lya\ Survey}

Although submillimeter/millimeter surveys have been quite successful
so far for finding luminous (often lensed) infrared-luminous galaxies
at high redshift (up to $z$\,$=$\,6.9), the fraction of such an
infrared-luminous galaxy population will likely decrease at higher
redshift, where the metallicities of galaxies are significantly lower
on average.  For the detection of less dusty and therefore UV-bright
galaxies, wide-field \lya\ emitter surveys will be effective.
Although \lya\ emission can be strongly suppressed at $z$\,$>$\,6 by
increasingly neutral IGM, the line may survive if the \lya-emitting
galaxy is located in a large \ion{H}{2} bubble
\citep[e.g.,][]{Cen2000, Haiman2002}.  In fact, the tentative
detection of \lya\ emission from the $z$\,$=$\,9.1 galaxy recently
reported by \citet{Hashimoto2018} supports this idea.  In the next
subsection (Section~\ref{ska}), we will discuss how we can find such
\ion{H}{2} bubbles with the Square Kilometer Array (SKA).

Search for strong \lya\ emitters will be particularly powerful for
finding massive forming galaxies discussed in
Section~\ref{sec:pop3_gal}.  These objects will have little continuum
emission before stars are formed, so strong emission lines like
\lya\ will be the only available tracers for such objects.  As already
discussed in Section~\ref{sec:giant_gal}, luminous Pop III
\hh\ emitters are likely luminous \lya\ emitters as well.

In this sense, one particularly interesting recent example is CR7 at
$z$\,$=$\,6.6 \citep{Sobral2015}.  Its exceptionally bright but narrow
\lya\ line
(2$\times$$10^{10}$\,\lsun), coupled with the detection of a strong
\ion{He}{2} 1640\,\AA\ line and a lack of any metal lines, was
originally taken as the sign that this object might harbor a Pop~III
stellar population.  Such a Pop~III scenario, however, encountered
many difficulties because of the extreme conditions required for the
underlying Pop~III stellar population
\citep[e.g.,][]{Pallottini2015b,Yajima2017,Visbal2017} as well as of
the possible presence of strong \oiii\ 4959/5007\,\AA\ lines inferred
from the \spitzer/IRAC photometric data \citep{Bowler2017}.  Although
other interpretations were also put forth for the nature of CR7, such
as a direct collapse black hole (DCBH, \citealt{Pallottini2015b,
  Agarwal2016, Dijkstra2016, Smith2016, Agarwal2017, Pacucci2017}),
the recent ALMA observations by \citet{Matthee2017} with the detection
of the \cii~158\,\micron\ line have concluded that CR7 is a system
that is undergoing the build-up process of a central galaxy through
complex accretion of star-forming satellites.  The
  original \heii\ line detection was also not reproduced by the
  reanalysis of the same data by \citet{Shibuya2018} although a
  re-analysis done by the CR7 discovery team still detects the line
  \citep{Sobral2018}, leaving the situation unclear.

  Although CR7 is not likely to be a Pop~III galaxy,
  identifying similarly luminous LAEs may prove to be the key to
  finding luminous \hh\ emitters at high redshift.  The most luminous
  \hh\ emitters are those hosted by \hi\ atomic cooling halos, which
  would also emit comparably luminous \lya\ if the line is not
  significantly absorbed by the IGM (see Section~\ref{sec:giant_gal}).
  Since the typical line detection limit of current
  $z$\,$\sim$\,6--7 narrow-band LAE surveys is
  $\sim$\,2$\times$$10^{9}$\,\lsun\ \citep[e.g.][]{Ouchi2018}, if the
  corresponding \hh\ line luminosities are on the same order, SAFARI
  should be able to detect such \hh\ lines in $\sim$\,10 hours
  (5\,$\sigma$).  Without having the capability to conduct a sensitive
  wide-field \hh\ line survey in the mid-/far-infrared directly,
  \lya-based surveys will likely play an important role for finding
  strong \hh\ emitters (and strong mid-infrared line emitters in
  general) at high redshift.

\subsection{SKA Survey}

\label{ska}

In the coming decade, the Square Kilometer Array (SKA) will start to
provide tomographic views of the high-redshift Universe through the
$21\text{-cm}$ emission line of neutral hydrogen
\citep[e.g.,][]{Koopmans2015}.  The wide-field SKA map will be ideal
for identifying large bubbles of ionized (\ion{H}{2}) gas at high
redshift, marking the concentrations of star-forming galaxies in the
early Universe.  Multi-wavelength imaging observations targeting these
\hii\ bubbles and the area around them will likely provide interesting
targets for SAFARI.

\begin{figure}[!htb]

\hspace*{0cm}\includegraphics[width=0.52\textwidth]{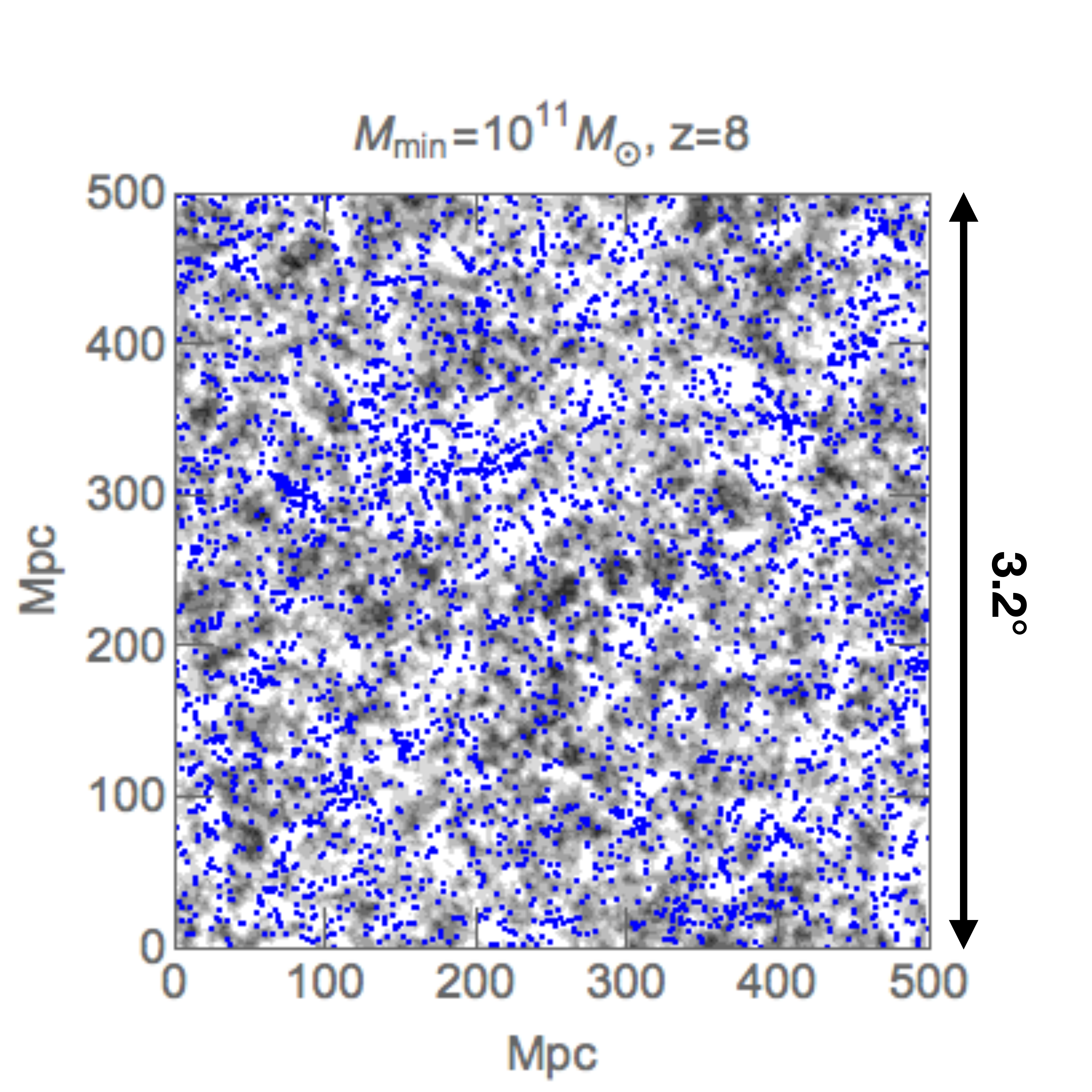}

\caption{Distribution of dark halos with
  \mhalo\,$\geq$\,$10^{11}$\,\msun\ at $z$\,$=$\,8 (blue dots)
  superimposed on the background ionization field (grey scale) taken
  from the simulated tomographic map at $z$\,$=$\,8 shown over an area
  of 500\,Mpc (comoving) on a side, which corresponds to 3.2\,deg.
  This map was smoothed over a scale of 30\,Mpc.  The darker
  (brighter) regions correspond to more neutral (ionized) regions.
  Both distributions were projected from a 100-Mpc thick slice at
  $z$\,$=$\,8.  The map of ionization field was produced based on the
  FULL model of \citet{Sobacchi2014}.}

\label{fig:maps}

\end{figure}

Figure \ref{fig:maps} shows halos with
\mhalo\,$>$\,$10^{11}$\,\msun\ (the blue dots) superimposed on the
background ionization field from the simulated tomographic map at
$z$\,$=$\,8, both projected from a $100\text{-Mpc}$ thick slice (see
\citealt{Sobacchi2014} for the methodology of the simulation).  The
figure shows that halos are significantly clustered toward the center
of \ion{H}{2} bubbles (seen as bright white spots) because these halos
are the sites of star formation and therefore the sources of ionizing
radiation.  These \ion{H}{2} bubbles have a size of 20--30\,Mpc
(comoving), corresponding to approximately 10\arcmin\ on the sky.

Such an SKA \hi\ map can be used in two ways.  First, we can look for
hyper-luminous galaxies and massive Pop~II forming galaxies inside
large \hii\ bubbles, which mark rare high-$\sigma$ peaks of the
background cosmic density field and harbor many star-forming galaxies.
Second, we can search for massive Pop~III forming galaxies {\it
  around} \hii\ bubbles.  The formation of massive Pop III galaxies
would require not only pristine metal-free environments (i.e., outside
\hii\ bubbles) but also strong UV radiation to suppress
\hh/\hi\ cooling and therefore star formation until minihalos are
assembled into galaxy-size halos (i.e., {\it around} \hii\ bubbles;
see the discussion in Section~\ref{overview}).

\subsection{Next Generation Optical/Near-Infrared Wide-Field Surveys}

\label{sec:next}

If accepted as the ESA M5 mission, \spica\ will be launched around
2030.  From the \spica\ high-redshift science point of view, this lead
time will allow us to take advantage of the next-generation
optical/near-infrared wide-field surveys for target selection, such as
LSST, \euclid, and \wfirst.  For example, the High Latitude Survey of
\wfirst\ is expected to detect around 100 bright ($m_{\rm 160,
  AB}$\,$<$\,26 mag) galaxies at $z$\,$>$\,10 over a survey area of
2,200\,deg$^2$ using the continuum drop-out technique
\citep{Spergel2015}.  Galaxies at such high redshift represent rare
peaks in the cosmic density field, marking the underlying regions of
large-scale density enhancement, where we can expect the existence of
many more objects at the same redshift \citep[e.g.,][]{Barkana2016}.
$z$\,$>$\,10 objects discovered by \wfirst\ (as well as \euclid) will
be strong rest-frame UV-continuum sources, which are likely unobscured
star-forming galaxies, but the regions marked by these sources would
likely harbor other types of high-redshift objects, such as
infrared-luminous (dusty/dust-obscured) galaxies and massive forming
galaxies.  Grism slitless spectroscopic surveys with \euclid\ and
\wfirst\ will also discover high-redshift \lya\ emitters
\citep[e.g.,][]{Bagley2017, Marchetti2017}, some of which may turn out
to be genuine Pop~III galaxies.  These wide-field surveys will also be
useful for finding exceptionally bright gravitationally-lensed
galaxies, which are expected to dominate the \wfirst\ sample of bright
$z$\,$>$\,10 galaxies \citep{Spergel2015}.

Another exciting possibility is that these next-generation deep
wide-field surveys may detect directly the explosions of massive
Pop~III SNe, so-called pair-instability SNe (PISNe; see
Appendix~\ref{sec:firstdust}), at $z$\,$\lesssim$\,7--8 with LSST, out
to $z$\,$\sim$\,15--20 with \wfirst, and out to $z$\,$\gtrsim$\,30
with JWST\footnote{Note, however, that at such high redshifts, the
  time-dilation effect is substantial.  For example, a 100-day
  transient event would become a 3-year event when observed at
  $z$\,$=$\,10, requiring a long-term monitoring effort.}
(\citealt{Whalen2013a}; see also \citealt{Wang2018}).  These
observations will catch PISNe in the initial UV-bright unobscured
phase, and if they are detected and studied, they will help us extend
our understanding to the subsequent dust-enshrouded phase, which is
directly relevant to SAFARI observations as further discussed in
Appendix~\ref{sec:firstdust}.  Furthermore, it is possible that some
supermassive stars ($\sim$\,$10^5$\,\msun) may explode as
thermonuclear supernovae with energies of $\sim$\,$10^{55}$\,erg, 100
times higher than those of Pop~III PISNe, making them the most
energetic explosions in the Universe \citep{Whalen2013b}.  Such
energetic events, if they exist, may also produce objects that can be
studied with SAFARI.

\subsection{SPICA/SMI Surveys}

\spica/SMI \citep{Kaneda2017,Roelfsema2018} has the ability to conduct
low-resolution (LR; $R$\,$=$\,50--120) multi-slit prism
spectroscopy covering 17--36\,\micron\ over a field of view of
12\arcmin$\times$10\arcmin.  It also has a slit-viewer camera (CAM),
which can perform 34\,\micron\ imaging over the same field of view.
These SMI modes will allow wide-field survey programs, and some
preliminary ideas have been presented in the companion papers, such as
the SMI/LR deep (1\,deg$^{2}$) and wide (10\,deg$^{2}$) spectroscopic
surveys \citep{Kaneda2017} and SMI/CAM ultra-deep (0.2\,deg$^{2}$),
deep (1\,deg$^{2}$), and shallow (600\,deg$^2$) imaging surveys
\citep{Gruppioni2017}.

For the purpose of finding bright $z$\,$>$\,5 targets for SAFARI,
these SMI surveys are particularly effective for detecting
AGN-dominated galaxies, which can be substantially brighter than
star-forming galaxies in the rest-frame mid-infrared.  At $z$\,$>$\,5,
SMI will observe $<$\,6\,\micron\ in the rest-frame, where the SEDs of
typical star-forming galaxies exhibit a broad trough between a stellar
continuum peaking in the near-infrared and a dust continuum peaking in
the far-infrared \citep[see,][]{Gruppioni2017}.  In the case of
AGN-dominated galaxies, however, a bright power-law AGN continuum
could fill this SED trough, and could even overwhelm the light from
the host galaxy all together.  As a result, the SMI-selected sample of
$z$\,$>$\,5 HyLIRGs is expected to be dominated by galaxies with
AGN\footnote{Note also that in the case of the SMI/LR surveys, all the
  major PAH features are redshifted out of the passband at
  $z$\,$>$\,5, except for the much fainter 3.3\,\micron\ feature,
  which makes SMI/LR detections of $z$\,$>$\,5 star-forming galaxies
  difficult.}  \citep{Gruppioni2017, Kaneda2017}.  Such a sample will
be particularly useful for studying the interplay between star
formation and black-hole accretion \citep{Spinoglio2017,
  Gruppioni2017}.

The SMI/CAM 600\,deg$^{2}$ shallow survey offers the potential to
discover exceptionally luminous $z$\,$>$\,5 galaxies for which SAFARI
follow-up spectroscopy will be possible.  Unlike
submillimeter/millimeter surveys, in which high-redshift galaxies
would often stand out in brightness, SMI mid-infrared surveys will see
a large number of foreground galaxies, among which a small number of
high-redshift objects will be hidden.  Selecting $z$\,$>$\,5 galaxies
will therefore require a detailed analysis of good-quality
multi-wavelength data covering the same fields like those mentioned in
Section~\ref{sec:next}.  Follow-up submillimeter/millimeter
observations will especially be useful for discriminating $z$\,$>$\,5
galaxies against lower-redshift galaxies.

It should also be noted that the SMI surveys have the potential to
identify $z$\,$>$\,5 HyLIRGs that may be missing in
submillimeter/millimeter surveys.  The all-sky mid-infrared survey
with the {\it Wide-Field Infrared Survey Explorer} (\wise) has
discovered hyper-luminous versions of dust-obscured galaxies (DOGs;
\citealt{Dey2008}), so-called Hot DOGs, up to a redshift of
$z$\,$=$\,4.59 \citep{Eisenhardt2012, Wu2012, Tsai2015}.  These
mid-infrared-selected Hot DOGs have bolometric luminosities of
$\gtrsim$\,$10^{14}$\,\lsun, as luminous as the brightest
submillimeter/millimeter-selected lensed infrared-luminous galaxies
listed in Table~\ref{tab:highz}.  However, most of these galaxies do
not appear to be lensed, suggesting that Hot DOGs are likely dominated
by AGN \citep{Tsai2015}.  Such AGN-dominated galaxies like Hot DOGs
may constitute an important population of infrared-luminous galaxies
at high redshift, which may be difficult to detect in the
submillimeter/millimeters surveys unless they are also forming stars
vigorously.  The SMI surveys will be sensitive to this type of
AGN-dominated galaxies, and can detect them to higher redshift.

\section{Summary}

\spica/SAFARI will be a uniquely powerful instrument, allowing us to
obtain good-quality rest-frame mid-infrared spectra of $z$\,$>$\,5
objects and to bridge the wavelength gap left unexplored by ALMA and
\jwst, which is so crucial for the studies of the high-redshift
Universe.  In this paper, we have examined the scientific potential of
SAFARI based on recent observational and theoretical studies.  The
following is a summary of the major points:
\begin{enumerate}
  \setlength{\itemsep}{5pt}

  \item SAFARI is capable of delivering good-quality rest-frame
    mid-infrared spectra for HyLIRGs (\lir\,$>$\,$10^{13}$\,\lsun) at
    $z$\,$=$\,5--10.  Currently, there are about a dozen such
    DSFGs known at $z$\,$=$\,5.2--6.9 (mostly lensed) and around 30 quasars
    at $z$\,$=$\,5.0--6.4.  The list of potential targets will grow over the
    coming years in terms of number and redshift range, allowing
    SAFARI to conduct the first rest-frame mid-infrared spectroscopic
    survey of $z$\,$>$\,5 galaxies in a significant number.

  \item To characterize the physical and chemical properties of
    $z$\,$>$\,5 galaxies, especially of those that are
    infrared-luminous (i.e., dusty/dusty-obscured), SAFARI's
    spectroscopic data will be essential, allowing us to examine a
    wealth of information, such as atomic fine-structure lines,
    molecular hydrogen lines (\hh), PAH features, and silicate
    emission/absorption features.  For example, low-metallicity
    galaxies will be easily recognized with distinct spectral
    characteristics such as weak PAH features, strong high-ionization
    lines, and a sharply rising mid-infrared continuum as observed in
    local BCDs.

  \item The rest-frame mid-infrared spectral range is particularly
    powerful for probing the first-generation objects because it
    contains, (1) key cooling lines of low-metallicity or metal-free
    gas (e.g., \sitwo, \fetwo, and \hh\ lines), and (2) emission
    features of solid compounds that are likely abundant in the
    remnants of Pop~III SNe (e.g., SiO$_{2}$; see
    Appendix~\ref{sec:firstdust}).  The feasibility of such SAFARI
    observations, however, is not yet clear, requiring more
    theoretical studies for further guidance, especially those focused
    on the most massive/luminous objects.  SAFARI detections of such
    spectral features, if successful, will open up a new frontier in
    the study of the early Universe, shedding light on the physical
    properties of the first galaxies and first stars.

   \item SAFARI's ability to explore the high-redshift Universe will
     be determined by the availability of sufficiently bright targets,
     whether intrinsically bright or gravitationally lensed.  In this
     regard, a series of wide-field surveys taking place over the
     coming decade (e.g., Advanced ACT, SPT3G, LSST, SKA, \euclid, and
     \wfirst) as well as the \spica/SMI surveys will greatly enhance
     our ability to discover extraordinary objects as
     exciting targets for SAFARI.

\end{enumerate}

If we take various theoretical predictions at their face value, there
will not be many first-generation objects that are bright enough for
SAFARI to detect and study.  However, it should also be remembered
that among the hundreds of millions of objects that the Sloan Digital
Sky Survey (SDSS) has cataloged, it only took the detections of a
handful of $z$\,$\gtrsim$\,6 quasars to completely change the
landscape in extragalactic astronomy and open up our view toward the
epoch of reionization \citep{Fan2001}.  With so many powerful
wide-field surveys on the horizon, which will explore a new parameter
space in terms of sensitivity and area coverage, we may be on the
verge of witnessing some totally unexpected discoveries, which have
always been the driver for astronomy.  With its expected launch around
2030, \spica\ is ideally positioned to take full advantage of these
upcoming wide-field surveys and their surprising discoveries.

\begin{figure*}[!htb]

\centerline{\includegraphics[width=1.0\textwidth]{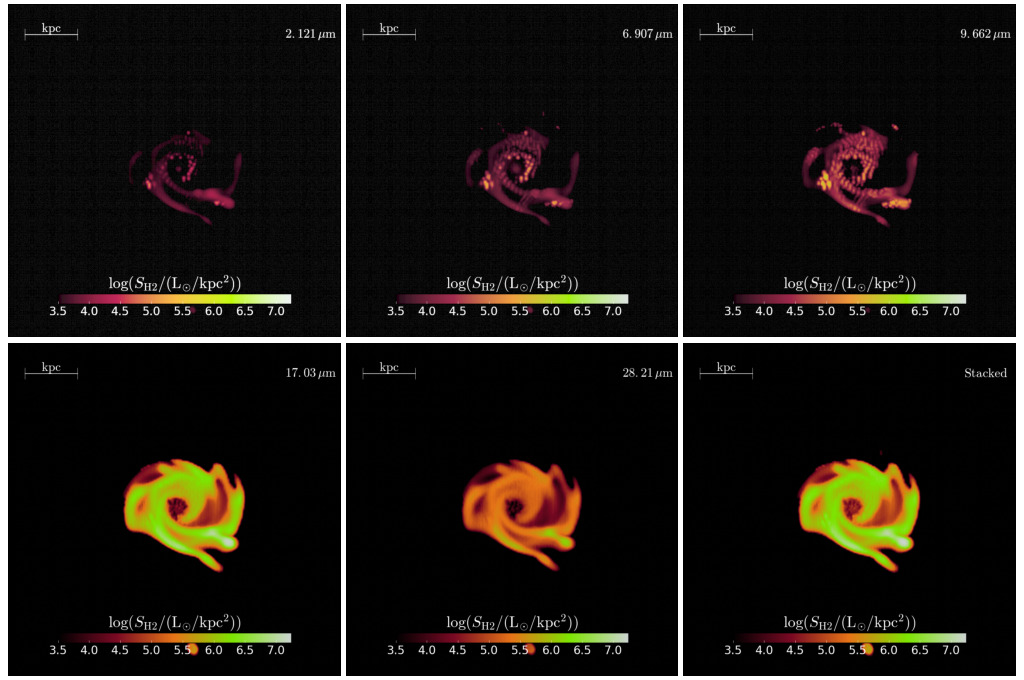}}

\caption{Synthetic H$_2$ line emission for the $z$\,$=$\,6 galaxy
  Alth{\ae}a \citep{Pallottini2017b}.  The lines shown are (from top
  left to bottom right) 1-0 S(1) 2.12\,\micron, 0-0 S(5)
  6.91\,\micron, 0-0 S(3) 9.66\,\micron, 0-0 S(1) 17.0\,\micron, and
  0-0 S(0) 28.2\,\micron.  The last panel at bottom right shows the
  stacked sum of all the lines, which is dominated by the 0-0 S(1)
  line.}
\label{H2_maps}

\end{figure*}

\begin{acknowledgements}

This paper is dedicated to the memory of Bruce Swinyard, who initiated
the \spica\ project in Europe, but unfortunately passed away on 22 May
2015 at the age of 52.  He was \iso-LWS calibration scientist,
\herschel-SPIRE instrument scientist, first European PI of \spica\ and
first design lead of SAFARI.

E.E.\ would like to thank Scuola Normale Superiore (Pisa, Italy) for
hosting his visit to discuss this work, and Justin Spilker for
providing information on some of the $z>5$ SPT galaxies.  He would
also like to thank Leon Koopmans, Jianwei Lyu, and Fengwu Sun for
helpful discussions.  L.V.\ acknowledges funding from the European
Union's Horizon 2020 research and innovation program under the Marie
Sk\l{}odowska-Curie Grant agreement No. 746119.  The research leading
to these results has received funding from the European Research
Council under the European Union Seventh Framework Programme
(FP/2007-2013) / ERC Grant Agreement no. 306476.

\end{acknowledgements}

\begin{appendix}
  \section{\hh\ Emission: Simulation for a \lowercase{$z$}\,$\simeq$\,6 LBG}

\label{h2}

Here, we use the model presented by \citet{Pallottini2017b} as an
example to illustrate the basic properties of the warm \hh\ gas
component in a typical $z$\,$\simeq$\,6 LBG.
The modeling methodology is fully described by \citet{Vallini2017},
\citet{Pallottini2017b}, and \citet{Vallini2018}.  The target halo has
a mass \mhalo\,$\simeq$\,$10^{11}$\,\msun\ and hosts a galaxy with a
stellar mass \mstar\,$\simeq$\,$10^{10}$\,\msun\ and SFR of about
100\,\msun\ yr$^{-1}$, consistent with typical $z$\,$\sim$\,6 LBGs
\citep[cf.,][]{Capak2015,Jiang2016}.  Following the convention of
\citet{Pallottini2017b}, we refer to this particular simulated galaxy
as ``Alth{\ae}a''.

The \hh\ emission maps of Alth{\ae}a are shown in
Figure~\ref{H2_maps}.  It has a total \hh\ mass of about
$10^{8}$\,M$_\odot$, which is seen to be mainly concentrated in a
disk-like structure (effective radius $\simeq$\,0.5\,kpc) that is
pressure-supported by radiation and composed of dense
($n$\,$\gtrsim$\,$10^{2.5}$\,cm$^{-3}$), metal-rich
($Z$\,$\simeq$\,0.5\,Z$_\odot$) gas.  The most luminous
pure-rotational \hh\ lines are the 0--0 S(1) line at 17\,$\mu$m and
0--0 S(0) line at 28\,$\mu$m.  About 95\,\% of the \hh\ gas is
characterized by a temperature of $T$\,$\simeq$\,100\,K, while 1\,\%
(4\,\%) has $T$\,$\simeq$\,1000\,K (20\,K).  Because of the dominance of
the $\simeq$\,100\,K component, 0--0 S(0) and 0--0 S(1) transitions are
the brightest, even though their oscillator strengths are small.
The spatially integrated luminosity of 0--0 S(1), the brightest
\hh\ line emitted by Alth{\ae}a, is 3.3$\times$$10^{6}$\,\lsun, a
factor of $\sim$\,500 below the SAFARI detection limit of
1.6$\times$$10^{9}$\,\lsun\ at $z$\,$=$\,6 (5\,$\sigma$, 10 hours).

  \section{Formation of the First Dust}

\label{sec:firstdust}

The first stars are predicted to be very massive, possibly with masses
over 100\,\msun\ \citep[e.g.,][]{Abel2002,Bromm2002,Omukai2003a}.
Given that the initial mass function (IMF) of Pop III stars was likely
top-heavy and extending to
$>$\,100\,\msun\ \citep[e.g.,][]{hirano2015,Stacy2016}, among the main
contributors of dust in the early Universe were likely the so-called
pair-instability supernovae\footnote{In these supernovae,
  electron-positron pair production serves as the energy sink,
  reducing the internal pressure and accelerating the gravitational
  collapse \citep{Barkat1967}.} (PISNe), whose progenitors have
main-sequence masses in the range of 140--260\,M$_{\odot}$
\citep{heger2002}.  Pop III stars in the mass ranges just above
($>$\,260\,\msun) or below (40--140\,\msun) are thought to collapse
directly into black holes, locking inside all the nucleosynthesis
products, while those in the 140--260\,\msun\ range are
thought to end their lives as PISNe with powerful explosions, leaving
no compact remnant behind and ejecting 15--30\,\% of the PISN progenitor
mass as dust (30--60\,\msun\ per star) into the surroundings
\citep{Nozawa2003,Schneider2004, Cherchneff2010}.  Considering such a
large dust production rate of PISNe, which is $>$\,10 times larger than
the value for typical core-collapse (i.e., Type II) SNe, it is
conceivable that clusters of Pop III stars would quickly enshroud
themselves with dust and become strong mid-infrared sources as PISNe
explode and the yet-to-explode less massive stars shine and heat the
ejected dust.  If such a scenario is correct, this implies that a
significant fraction of Pop III star clusters may be faint in the
optical/near-infrared (i.e., rest-frame UV/optical) but bright in the
far-infrared (i.e., rest-frame mid-infrared).  Far-infrared
spectroscopy of such dust-enshrouded Pop III star clusters could
detect spectral features produced by the first dust, revealing its
composition and production mechanisms.

Observations of young supernova remnants in the Milky Way and in the
Large Magellanic Cloud have provided evidence that dust can form in
the ejecta a few hundred days after the explosion
(e.g. \citealt{matsuura2009,dunne2009,gall2014,matsuura2015,
  delooze2017, temim2017}).  Theoretical models are able to reproduce
the observational data \citep{bocchio2016} and show that, depending on
the progenitor mass and metallicity, a variety of dust species can
form \citep{todini2001, Schneider2004, nozawa2007, bianchi2007,
  Cherchneff2010, marassi2014, marassi2015, Marassi2017, Sarangi2015}.

Since dust grains condense out of metals freshly synthesized by the
stellar progenitor, the process is expected to work even at very high
redshifts, right after the formation of the first generation of
Pop~III stars \citep{todini2001, Schneider2004, nozawa2007,
  marassi2014, marassi2015}.  While the validity of these predictions
relies on extrapolations of models that have been shown to work at
$Z$\,$\geq$\,0.4\,Z$_\odot$ \citep{bocchio2016}, indirect support for
an early phase of rapid dust enrichment comes from the analysis of the
large dust masses detected in the host galaxies of the first quasars
\citep{valiante2009, valiante2011, valiante2014} and in the color
evolution of normal star-forming galaxies at $z$\,$>$\,6
\citep{mancini2015, mancini2016}, as well as from the measurements of
dust extinction in $z>4$ quasars \citep[e.g.,][]{Maiolino2004,
  Gallerani2010}.  In addition, dust enrichment by Pop~III SNe
increases gas cooling at very low metallicity, opening a formation
pathway to the first low-mass stars \citep{schneider2006b,
  schneider2012a, schneider2012b} and matching the constraints from
stellar archaeology studies \citep{debennassuti2014,
  debennassuti2017}.

Dust enrichment by Pop~III SNe occurs on very short timescales,
comparable to the stellar evolutionary timescale of the most massive
SN progenitor present in the Pop~III star cluster.  This, in turn,
depends on the Pop~III IMF, which is still very poorly constrained by
ab-initio simulations \citep{hirano2015, Stacy2016}.  Dust production
starts to affect the colors of Pop~III galaxies a few Myrs after star
formation, when PISNe from 140--260\,\msun\ stars or core-collapse SNe
with $M$\,$<$\,40\,\msun\ release their chemical yields into the
surrounding medium.

By combining a Pop~III stellar population synthesis model
\citep[e.g.,][]{zackrisson2011} with dust production models of Pop~III
SNe, such as that of \citet{Schneider2004} for PISNe and
\citet{marassi2015} for Type-II SNe, it is possible to calculate the
infrared spectral evolution of a Pop III star cluster and examine its
spectral characteristics (R.~Schneider et al., in preparation).  In
the rest-frame mid-infrared range, particularly prominent will be
quartz (SiO$_2$) emission features at 9, 12.4, and 21.5\,\micron,
which can be quite strong because quartz is expected to be one of the
most abundant solid compounds formed in the ejecta of PISNe
\citep[e.g.,][]{Nozawa2003,Schneider2004}.  Note that the detection of
such spectral features, if successful, will allow us to investigate
directly the properties of the first dust freshly produced by the
first generation of SNe.

For SAFARI to be able to make such detection, the target Pop~III star
cluster needs to be extremely luminous.  In practice, this would mean
that the target star cluster must be maximally massive (the limit is
thought to be $\sim$\,$10^{5}$--$10^{6}$\,\msun), very young (ages up
to a few/several Myr), and gravitationally lensed by a large factor
($\gg$\,100$\times$).  The detectability of such a luminous lensed
Pop~III star cluster was discussed by \citet{Zackrisson2015}.
Considering the challenge of detecting such Pop~III objects at high
redshift, an alternative approach may be to search for lower-redshift
analogs (see Section~\ref{lowz_pop3}).

\end{appendix}


\bibliographystyle{pasa-mnras}
\bibliography{spica_highz}

\begin{thebibliography}{}
\makeatletter
\relax
\def\mn@urlcharsother{\let\do\@makeother \do\$\do\&\do\#\do\^\do\_\do\%\do\~}
\definecolor{darkblue}{rgb}{0,0,0.597656}
\def\mndoi{\begingroup\mn@urlcharsother \@ifnextchar [ {\mndoi@} {\mndoi@[]}}
\def\mndoi@[#1]#2{\def\@tempa{#1}\ifx\@tempa\@empty \href
  {http://dx.doi.org/#2} {\textcolor{darkblue}{doi:#2}}\else \href
  {http://dx.doi.org/#2} {\textcolor{darkblue}{#1}}\fi \endgroup}
\def\mn@eprint#1#2{\mn@eprint@#1:#2::\@nil}
\def\mn@eprint@arXiv#1{\href {http://arxiv.org/abs/#1} {{\tt arXiv:#1}}}
\def\mn@eprint@dblp#1{\href {http://dblp.uni-trier.de/rec/bibtex/#1.xml}
  {dblp:#1}}
\def\mn@eprint@#1:#2:#3:#4\@nil{\def\@tempa {#1}\def\@tempb {#2}\def\@tempc
  {#3}\ifx \@tempc \@empty \let \@tempc \@tempb \let \@tempb \@tempa \fi \ifx
  \@tempb \@empty \def\@tempb {arXiv}\fi \@ifundefined
  {mn@eprint@\@tempb}{\@tempb:\@tempc}{\expandafter \expandafter \csname
  mn@eprint@\@tempb\endcsname \expandafter{\@tempc}}}

\bibitem[\protect\citeauthoryear{{Abel}, {Bryan}  \& {Norman}}{{Abel}
  et~al.}{2002}]{Abel2002}
{Abel} T.,  {Bryan} G.~L.,   {Norman} M.~L.,  2002, \mndoi [Science]
  {10.1126/science.295.5552.93}, \href
  {http://adsabs.harvard.edu/abs/2002Sci...295...93A} {295, 93}

\bibitem[\protect\citeauthoryear{{Agarwal}, {Johnson}, {Zackrisson}, {Labbe},
  {van den Bosch}, {Natarajan}  \& {Khochfar}}{{Agarwal}
  et~al.}{2016}]{Agarwal2016}
{Agarwal} B.,  {Johnson} J.~L.,  {Zackrisson} E.,  {Labbe} I.,  {van den Bosch}
  F.~C.,  {Natarajan} P.,   {Khochfar} S.,  2016, \mndoi [\mnras]
  {10.1093/mnras/stw1173}, \href
  {http://adsabs.harvard.edu/abs/2016MNRAS.460.4003A} {460, 4003}

\bibitem[\protect\citeauthoryear{{Agarwal}, {Johnson}, {Khochfar},
  {Pellegrini}, {Rydberg}, {Klessen}  \& {Oesch}}{{Agarwal}
  et~al.}{2017}]{Agarwal2017}
{Agarwal} B.,  {Johnson} J.~L.,  {Khochfar} S.,  {Pellegrini} E.,  {Rydberg}
  C.-E.,  {Klessen} R.~S.,   {Oesch} P.,  2017, \mndoi [\mnras]
  {10.1093/mnras/stx794}, \href
  {http://adsabs.harvard.edu/abs/2017MNRAS.469..231A} {469, 231}

\bibitem[\protect\citeauthoryear{{{\'A}lvarez-M{\'a}rquez}
  et~al.,}{{{\'A}lvarez-M{\'a}rquez} et~al.}{2016}]{Alvarez2016}
{{\'A}lvarez-M{\'a}rquez} J.,  et~al., 2016, \mndoi [\aap]
  {10.1051/0004-6361/201527190}, \href
  {http://cdsads.u-strasbg.fr/abs/2016A%26A...587A.122A} {587, A122}

\bibitem[\protect\citeauthoryear{{Appleton} et~al.,}{{Appleton}
  et~al.}{2006}]{Appleton2006}
{Appleton} P.~N.,  et~al., 2006, \mndoi [\apjl] {10.1086/502646}, \href
  {http://adsabs.harvard.edu/abs/2006ApJ...639L..51A} {639, L51}

\bibitem[\protect\citeauthoryear{{Appleton} et~al.,}{{Appleton}
  et~al.}{2013}]{Appleton2013}
{Appleton} P.~N.,  et~al., 2013, \mndoi [\apj] {10.1088/0004-637X/777/1/66},
  \href {http://adsabs.harvard.edu/abs/2013ApJ...777...66A} {777, 66}

\bibitem[\protect\citeauthoryear{{Appleton} et~al.,}{{Appleton}
  et~al.}{2017}]{Appleton2017}
{Appleton} P.~N.,  et~al., 2017, \mndoi [\apj] {10.3847/1538-4357/836/1/76},
  \href {http://adsabs.harvard.edu/abs/2017ApJ...836...76A} {836, 76}

\bibitem[\protect\citeauthoryear{{Aravena} et~al.,}{{Aravena}
  et~al.}{2016}]{Aravena2016}
{Aravena} M.,  et~al., 2016, \mndoi [\mnras] {10.1093/mnras/stw275}, \href
  {http://adsabs.harvard.edu/abs/2016MNRAS.457.4406A} {457, 4406}

\bibitem[\protect\citeauthoryear{{Armus} et~al.,}{{Armus}
  et~al.}{2006}]{Armus2006}
{Armus} L.,  et~al., 2006, \mndoi [\apj] {10.1086/500040}, \href
  {http://cdsads.u-strasbg.fr/abs/2006ApJ...640..204A} {640, 204}

\bibitem[\protect\citeauthoryear{{Armus} et~al.,}{{Armus}
  et~al.}{2007}]{Armus2007}
{Armus} L.,  et~al., 2007, \mndoi [\apj] {10.1086/510107}, \href
  {http://adsabs.harvard.edu/abs/2007ApJ...656..148A} {656, 148}

\bibitem[\protect\citeauthoryear{{Asboth} et~al.,}{{Asboth}
  et~al.}{2016}]{Asboth2016}
{Asboth} V.,  et~al., 2016, \mndoi [\mnras] {10.1093/mnras/stw1769}, \href
  {http://adsabs.harvard.edu/abs/2016MNRAS.462.1989A} {462, 1989}

\bibitem[\protect\citeauthoryear{{Bagley} et~al.,}{{Bagley}
  et~al.}{2017}]{Bagley2017}
{Bagley} M.~B.,  et~al., 2017, \mndoi [\apj] {10.3847/1538-4357/837/1/11},
  \href {http://cdsads.u-strasbg.fr/abs/2017ApJ...837...11B} {837, 11}

\bibitem[\protect\citeauthoryear{{Barkana}}{{Barkana}}{2016}]{Barkana2016}
{Barkana} R.,  2016, \mndoi [\physrep] {10.1016/j.physrep.2016.06.006}, \href
  {http://adsabs.harvard.edu/abs/2016PhR...645....1B} {645, 1}

\bibitem[\protect\citeauthoryear{{Barkat}, {Rakavy}  \& {Sack}}{{Barkat}
  et~al.}{1967}]{Barkat1967}
{Barkat} Z.,  {Rakavy} G.,   {Sack} N.,  1967, \mndoi [Physical Review Letters]
  {10.1103/PhysRevLett.18.379}, \href
  {http://adsabs.harvard.edu/abs/1967PhRvL..18..379B} {18, 379}

\bibitem[\protect\citeauthoryear{{Behroozi} \& {Silk}}{{Behroozi} \&
  {Silk}}{2015}]{Behroozi2015}
{Behroozi} P.~S.,  {Silk} J.,  2015, \mndoi [\apj]
  {10.1088/0004-637X/799/1/32}, \href
  {http://adsabs.harvard.edu/abs/2015ApJ...799...32B} {799, 32}

\bibitem[\protect\citeauthoryear{{B{\'e}thermin} et~al.,}{{B{\'e}thermin}
  et~al.}{2017}]{Bethermin2017}
{B{\'e}thermin} M.,  et~al., 2017, \mndoi [\aap] {10.1051/0004-6361/201730866},
  \href {http://adsabs.harvard.edu/abs/2017A%26A...607A..89B} {607, A89}

\bibitem[\protect\citeauthoryear{{Bianchi} \& {Schneider}}{{Bianchi} \&
  {Schneider}}{2007}]{bianchi2007}
{Bianchi} S.,  {Schneider} R.,  2007, \mndoi [\mnras]
  {10.1111/j.1365-2966.2007.11829.x}, \href
  {http://adsabs.harvard.edu/abs/2007MNRAS.378..973B} {378, 973}

\bibitem[\protect\citeauthoryear{{Blain}, {Smail}, {Ivison}, {Kneib}  \&
  {Frayer}}{{Blain} et~al.}{2002}]{Blain2002}
{Blain} A.~W.,  {Smail} I.,  {Ivison} R.~J.,  {Kneib} J.-P.,   {Frayer} D.~T.,
  2002, \mndoi [\physrep] {10.1016/S0370-1573(02)00134-5}, \href
  {http://adsabs.harvard.edu/abs/2002PhR...369..111B} {369, 111}

\bibitem[\protect\citeauthoryear{{Bocchio}, {Marassi}, {Schneider}, {Bianchi},
  {Limongi}  \& {Chieffi}}{{Bocchio} et~al.}{2016}]{bocchio2016}
{Bocchio} M.,  {Marassi} S.,  {Schneider} R.,  {Bianchi} S.,  {Limongi} M.,
  {Chieffi} A.,  2016, \mndoi [A\&A] {10.1051/0004-6361/201527432}, \href
  {http://adsabs.harvard.edu/abs/2016A%26A...587A.157B} {587, A157}

\bibitem[\protect\citeauthoryear{{Bouwens} et~al.,}{{Bouwens}
  et~al.}{2012}]{Bouwens2012}
{Bouwens} R.~J.,  et~al., 2012, \mndoi [\apj] {10.1088/0004-637X/754/2/83},
  \href {http://adsabs.harvard.edu/abs/2012ApJ...754...83B} {754, 83}

\bibitem[\protect\citeauthoryear{{Bowler}, {McLure}, {Dunlop}, {McLeod},
  {Stanway}, {Eldridge}  \& {Jarvis}}{{Bowler} et~al.}{2017}]{Bowler2017}
{Bowler} R.~A.~A.,  {McLure} R.~J.,  {Dunlop} J.~S.,  {McLeod} D.~J.,
  {Stanway} E.~R.,  {Eldridge} J.~J.,   {Jarvis} M.~J.,  2017, \mndoi [\mnras]
  {10.1093/mnras/stx839}, \href
  {http://adsabs.harvard.edu/abs/2017MNRAS.469..448B} {469, 448}

\bibitem[\protect\citeauthoryear{{Bowler}, {Bourne}, {Dunlop}, {McLure}  \&
  {McLeod}}{{Bowler} et~al.}{2018}]{Bowler2018}
{Bowler} R.~A.~A.,  {Bourne} N.,  {Dunlop} J.~S.,  {McLure} R.~J.,   {McLeod}
  D.~J.,  2018, \mndoi [\mnras] {10.1093/mnras/sty2368}, \href
  {http://adsabs.harvard.edu/abs/2018MNRAS.481.1631B} {481, 1631}

\bibitem[\protect\citeauthoryear{{Bromm}}{{Bromm}}{2013}]{Bromm2013}
{Bromm} V.,  2013, \mndoi [Reports on Progress in Physics]
  {10.1088/0034-4885/76/11/112901}, \href
  {http://adsabs.harvard.edu/abs/2013RPPh...76k2901B} {76, 112901}

\bibitem[\protect\citeauthoryear{{Bromm} \& {Loeb}}{{Bromm} \&
  {Loeb}}{2003}]{Bromm2003b}
{Bromm} V.,  {Loeb} A.,  2003, \mndoi [\nat] {10.1038/nature02071}, \href
  {http://adsabs.harvard.edu/abs/2003Natur.425..812B} {425, 812}

\bibitem[\protect\citeauthoryear{{Bromm} \& {Yoshida}}{{Bromm} \&
  {Yoshida}}{2011}]{Bromm2011}
{Bromm} V.,  {Yoshida} N.,  2011, \mndoi [\araa]
  {10.1146/annurev-astro-081710-102608}, \href
  {http://adsabs.harvard.edu/abs/2011ARA%26A..49..373B} {49, 373}

\bibitem[\protect\citeauthoryear{{Bromm}, {Coppi}  \& {Larson}}{{Bromm}
  et~al.}{2002}]{Bromm2002}
{Bromm} V.,  {Coppi} P.~S.,   {Larson} R.~B.,  2002, \mndoi [\apj]
  {10.1086/323947}, \href {http://adsabs.harvard.edu/abs/2002ApJ...564...23B}
  {564, 23}

\bibitem[\protect\citeauthoryear{{Bromm}, {Yoshida}  \& {Hernquist}}{{Bromm}
  et~al.}{2003}]{Bromm2003a}
{Bromm} V.,  {Yoshida} N.,   {Hernquist} L.,  2003, \mndoi [\apjl]
  {10.1086/379359}, \href {http://adsabs.harvard.edu/abs/2003ApJ...596L.135B}
  {596, L135}

\bibitem[\protect\citeauthoryear{{Ca{\~n}ameras} et~al.,}{{Ca{\~n}ameras}
  et~al.}{2015}]{Canameras2015}
{Ca{\~n}ameras} R.,  et~al., 2015, \mndoi [\aap] {10.1051/0004-6361/201425128},
  \href {http://adsabs.harvard.edu/abs/2015A%26A...581A.105C} {581, A105}

\bibitem[\protect\citeauthoryear{{Capak} et~al.,}{{Capak}
  et~al.}{2011}]{Capak2011}
{Capak} P.~L.,  et~al., 2011, \mndoi [\nat] {10.1038/nature09681}, \href
  {http://adsabs.harvard.edu/abs/2011Natur.470..233C} {470, 233}

\bibitem[\protect\citeauthoryear{{Capak} et~al.,}{{Capak}
  et~al.}{2015}]{Capak2015}
{Capak} P.~L.,  et~al., 2015, \mndoi [\nat] {10.1038/nature14500}, \href
  {http://adsabs.harvard.edu/abs/2015Natur.522..455C} {522, 455}

\bibitem[\protect\citeauthoryear{{Casey}, {Narayanan}  \& {Cooray}}{{Casey}
  et~al.}{2014}]{Casey2014}
{Casey} C.~M.,  {Narayanan} D.,   {Cooray} A.,  2014, \mndoi [\physrep]
  {10.1016/j.physrep.2014.02.009}, \href
  {http://adsabs.harvard.edu/abs/2014PhR...541...45C} {541, 45}

\bibitem[\protect\citeauthoryear{{Cen} \& {Haiman}}{{Cen} \&
  {Haiman}}{2000}]{Cen2000}
{Cen} R.,  {Haiman} Z.,  2000, \mndoi [\apjl] {10.1086/312937}, \href
  {http://adsabs.harvard.edu/abs/2000ApJ...542L..75C} {542, L75}

\bibitem[\protect\citeauthoryear{{Cherchneff} \& {Dwek}}{{Cherchneff} \&
  {Dwek}}{2010}]{Cherchneff2010}
{Cherchneff} I.,  {Dwek} E.,  2010, \mndoi [\apj] {10.1088/0004-637X/713/1/1},
  \href {http://adsabs.harvard.edu/abs/2010ApJ...713....1C} {713, 1}

\bibitem[\protect\citeauthoryear{{Ciardi} \& {Ferrara}}{{Ciardi} \&
  {Ferrara}}{2001}]{Ciardi2001}
{Ciardi} B.,  {Ferrara} A.,  2001, \mndoi [\mnras]
  {10.1046/j.1365-8711.2001.04355.x}, \href
  {http://adsabs.harvard.edu/abs/2001MNRAS.324..648C} {324, 648}

\bibitem[\protect\citeauthoryear{{Ciardi} \& {Ferrara}}{{Ciardi} \&
  {Ferrara}}{2005}]{Ciardi2005}
{Ciardi} B.,  {Ferrara} A.,  2005, \mndoi [\ssr] {10.1007/s11214-005-3592-0},
  \href {http://adsabs.harvard.edu/abs/2005SSRv..116..625C} {116, 625}

\bibitem[\protect\citeauthoryear{{Cluver} et~al.,}{{Cluver}
  et~al.}{2010}]{Cluver2010}
{Cluver} M.~E.,  et~al., 2010, \mndoi [\apj] {10.1088/0004-637X/710/1/248},
  \href {http://adsabs.harvard.edu/abs/2010ApJ...710..248C} {710, 248}

\bibitem[\protect\citeauthoryear{{Combes} et~al.,}{{Combes}
  et~al.}{2012}]{Combes2012}
{Combes} F.,  et~al., 2012, \mndoi [\aap] {10.1051/0004-6361/201118750}, \href
  {http://adsabs.harvard.edu/abs/2012A%26A...538L...4C} {538, L4}

\bibitem[\protect\citeauthoryear{{Consiglio}, {Turner}, {Beck}  \&
  {Meier}}{{Consiglio} et~al.}{2016}]{Consiglio2016}
{Consiglio} S.~M.,  {Turner} J.~L.,  {Beck} S.,   {Meier} D.~S.,  2016, \mndoi
  [\apjl] {10.3847/2041-8205/833/1/L6}, \href
  {http://cdsads.u-strasbg.fr/abs/2016ApJ...833L...6C} {833, L6}

\bibitem[\protect\citeauthoryear{{Cooray} et~al.,}{{Cooray}
  et~al.}{2014}]{Cooray2014}
{Cooray} A.,  et~al., 2014, \mndoi [\apj] {10.1088/0004-637X/790/1/40}, \href
  {http://adsabs.harvard.edu/abs/2014ApJ...790...40C} {790, 40}

\bibitem[\protect\citeauthoryear{{Coppin} et~al.,}{{Coppin}
  et~al.}{2015}]{Coppin2015}
{Coppin} K.~E.~K.,  et~al., 2015, \mndoi [\mnras] {10.1093/mnras/stu2185},
  \href {http://adsabs.harvard.edu/abs/2015MNRAS.446.1293C} {446, 1293}

\bibitem[\protect\citeauthoryear{{Cormier} et~al.,}{{Cormier}
  et~al.}{2012}]{Cormier2012}
{Cormier} D.,  et~al., 2012, \mndoi [\aap] {10.1051/0004-6361/201219818}, \href
  {http://cdsads.u-strasbg.fr/abs/2012A%26A...548A..20C} {548, A20}

\bibitem[\protect\citeauthoryear{{Cormier} et~al.,}{{Cormier}
  et~al.}{2015}]{Cormier2015}
{Cormier} D.,  et~al., 2015, \mndoi [\aap] {10.1051/0004-6361/201425207}, \href
  {http://adsabs.harvard.edu/abs/2015A%26A...578A..53C} {578, A53}

\bibitem[\protect\citeauthoryear{{De Bennassuti}, {Schneider}, {Valiante}  \&
  {Salvadori}}{{De Bennassuti} et~al.}{2014}]{debennassuti2014}
{De Bennassuti} M.,  {Schneider} R.,  {Valiante} R.,   {Salvadori} S.,  2014,
  \mndoi [\mnras] {10.1093/mnras/stu1962}, \href
  {http://cdsads.u-strasbg.fr/abs/2014MNRAS.445.3039D} {445, 3039}

\bibitem[\protect\citeauthoryear{{De Bennassuti}, {Salvadori}, {Schneider},
  {Valiante}  \& {Omukai}}{{De Bennassuti} et~al.}{2017}]{debennassuti2017}
{De Bennassuti} M.,  {Salvadori} S.,  {Schneider} R.,  {Valiante} R.,
  {Omukai} K.,  2017, \mndoi [\mnras] {10.1093/mnras/stw2687}, \href
  {http://adsabs.harvard.edu/abs/2017MNRAS.465..926D} {465, 926}

\bibitem[\protect\citeauthoryear{{De Looze}, {Barlow}, {Swinyard}, {Rho},
  {Gomez}, {Matsuura}  \& {Wesson}}{{De Looze} et~al.}{2017}]{delooze2017}
{De Looze} I.,  {Barlow} M.~J.,  {Swinyard} B.~M.,  {Rho} J.,  {Gomez} H.~L.,
  {Matsuura} M.,   {Wesson} R.,  2017, \mndoi [\mnras] {10.1093/mnras/stw2837},
  \href {http://cdsads.u-strasbg.fr/abs/2017MNRAS.465.3309D} {465, 3309}

\bibitem[\protect\citeauthoryear{{Dey} et~al.,}{{Dey} et~al.}{2008}]{Dey2008}
{Dey} A.,  et~al., 2008, \mndoi [\apj] {10.1086/529516}, \href
  {http://adsabs.harvard.edu/abs/2008ApJ...677..943D} {677, 943}

\bibitem[\protect\citeauthoryear{{Dijkstra}, {Gronke}  \& {Sobral}}{{Dijkstra}
  et~al.}{2016}]{Dijkstra2016}
{Dijkstra} M.,  {Gronke} M.,   {Sobral} D.,  2016, \mndoi [\apj]
  {10.3847/0004-637X/823/2/74}, \href
  {http://adsabs.harvard.edu/abs/2016ApJ...823...74D} {823, 74}

\bibitem[\protect\citeauthoryear{{Dowell} et~al.,}{{Dowell}
  et~al.}{2014}]{Dowell2014}
{Dowell} C.~D.,  et~al., 2014, \mndoi [\apj] {10.1088/0004-637X/780/1/75},
  \href {http://adsabs.harvard.edu/abs/2014ApJ...780...75D} {780, 75}

\bibitem[\protect\citeauthoryear{{Dunlop}, {McLure}, {Robertson}, {Ellis},
  {Stark}, {Cirasuolo}  \& {de Ravel}}{{Dunlop} et~al.}{2012}]{Dunlop2012}
{Dunlop} J.~S.,  {McLure} R.~J.,  {Robertson} B.~E.,  {Ellis} R.~S.,  {Stark}
  D.~P.,  {Cirasuolo} M.,   {de Ravel} L.,  2012, \mndoi [\mnras]
  {10.1111/j.1365-2966.2011.20102.x}, \href
  {http://adsabs.harvard.edu/abs/2012MNRAS.420..901D} {420, 901}

\bibitem[\protect\citeauthoryear{{Dunlop} et~al.,}{{Dunlop}
  et~al.}{2017}]{Dunlop2017}
{Dunlop} J.~S.,  et~al., 2017, \mndoi [\mnras] {10.1093/mnras/stw3088}, \href
  {http://adsabs.harvard.edu/abs/2017MNRAS.466..861D} {466, 861}

\bibitem[\protect\citeauthoryear{{Dunne} et~al.,}{{Dunne}
  et~al.}{2009}]{dunne2009}
{Dunne} L.,  et~al., 2009, \mndoi [MNRAS] {10.1111/j.1365-2966.2009.14453.x},
  \href {http://adsabs.harvard.edu/abs/2009MNRAS.394.1307D} {394, 1307}

\bibitem[\protect\citeauthoryear{{Egami}, {Neugebauer}, {Soifer}, {Matthews},
  {Becklin}  \& {Ressler}}{{Egami} et~al.}{2006a}]{Egami2006}
{Egami} E.,  {Neugebauer} G.,  {Soifer} B.~T.,  {Matthews} K.,  {Becklin}
  E.~E.,   {Ressler} M.~E.,  2006a, \mndoi [\aj] {10.1086/499524}, \href
  {http://cdsads.u-strasbg.fr/abs/2006AJ....131.1253E} {131, 1253}

\bibitem[\protect\citeauthoryear{{Egami}, {Rieke}, {Fadda}  \& {Hines}}{{Egami}
  et~al.}{2006b}]{Egami2006b}
{Egami} E.,  {Rieke} G.~H.,  {Fadda} D.,   {Hines} D.~C.,  2006b, \mndoi
  [\apjl] {10.1086/509886}, \href
  {http://adsabs.harvard.edu/abs/2006ApJ...652L..21E} {652, L21}

\bibitem[\protect\citeauthoryear{{Egami} et~al.,}{{Egami}
  et~al.}{2010}]{Egami2010}
{Egami} E.,  et~al., 2010, \mndoi [\aap] {10.1051/0004-6361/201014696}, \href
  {http://adsabs.harvard.edu/abs/2010A%26A...518L..12E} {518, L12}

\bibitem[\protect\citeauthoryear{{Eisenhardt} et~al.,}{{Eisenhardt}
  et~al.}{2012}]{Eisenhardt2012}
{Eisenhardt} P.~R.~M.,  et~al., 2012, \mndoi [\apj]
  {10.1088/0004-637X/755/2/173}, \href
  {http://adsabs.harvard.edu/abs/2012ApJ...755..173E} {755, 173}

\bibitem[\protect\citeauthoryear{{Erb}, {Pettini}, {Shapley}, {Steidel}, {Law}
  \& {Reddy}}{{Erb} et~al.}{2010}]{Erb2010}
{Erb} D.~K.,  {Pettini} M.,  {Shapley} A.~E.,  {Steidel} C.~C.,  {Law} D.~R.,
  {Reddy} N.~A.,  2010, \mndoi [\apj] {10.1088/0004-637X/719/2/1168}, \href
  {http://cdsads.u-strasbg.fr/abs/2010ApJ...719.1168E} {719, 1168}

\bibitem[\protect\citeauthoryear{{Fan} et~al.,}{{Fan} et~al.}{2001}]{Fan2001}
{Fan} X.,  et~al., 2001, \mndoi [\aj] {10.1086/324111}, \href
  {http://adsabs.harvard.edu/abs/2001AJ....122.2833F} {122, 2833}

\bibitem[\protect\citeauthoryear{{Ferland} et~al.,}{{Ferland}
  et~al.}{2013}]{Ferland2013}
{Ferland} G.~J.,  et~al., 2013, RMXAA, \href
  {http://adsabs.harvard.edu/abs/2013RMxAA..49..137F} {49, 137}

\bibitem[\protect\citeauthoryear{{Fern{\'a}ndez-Ontiveros}, {Spinoglio},
  {Pereira-Santaella}, {Malkan}, {Andreani}  \&
  {Dasyra}}{{Fern{\'a}ndez-Ontiveros} et~al.}{2016}]{Fernandez2016}
{Fern{\'a}ndez-Ontiveros} J.~A.,  {Spinoglio} L.,  {Pereira-Santaella} M.,
  {Malkan} M.~A.,  {Andreani} P.,   {Dasyra} K.~M.,  2016, \mndoi [\apjs]
  {10.3847/0067-0049/226/2/19}, \href
  {http://adsabs.harvard.edu/abs/2016ApJS..226...19F} {226, 19}

\bibitem[\protect\citeauthoryear{{Fern{\'a}ndez-Ontiveros}
  et~al.,}{{Fern{\'a}ndez-Ontiveros} et~al.}{2017}]{Fernandez2017}
{Fern{\'a}ndez-Ontiveros} J.~A.,  et~al., 2017, \mndoi [\pasa]
  {10.1017/pasa.2017.43}, \href
  {http://adsabs.harvard.edu/abs/2017PASA...34...53F} {34, e053}

\bibitem[\protect\citeauthoryear{{Ferrara}}{{Ferrara}}{1998}]{Ferrara1998}
{Ferrara} A.,  1998, \mndoi [\apjl] {10.1086/311344}, \href
  {http://adsabs.harvard.edu/abs/1998ApJ...499L..17F} {499, L17}

\bibitem[\protect\citeauthoryear{{Finkelstein} et~al.,}{{Finkelstein}
  et~al.}{2012}]{Finkelstein2012}
{Finkelstein} S.~L.,  et~al., 2012, \mndoi [\apj]
  {10.1088/0004-637X/756/2/164}, \href
  {http://adsabs.harvard.edu/abs/2012ApJ...756..164F} {756, 164}

\bibitem[\protect\citeauthoryear{{Gall} et~al.,}{{Gall}
  et~al.}{2014}]{gall2014}
{Gall} C.,  et~al., 2014, \mndoi [\nat] {10.1038/nature13558}, \href
  {http://cdsads.u-strasbg.fr/abs/2014Natur.511..326G} {511, 326}

\bibitem[\protect\citeauthoryear{{Gallerani} et~al.,}{{Gallerani}
  et~al.}{2010}]{Gallerani2010}
{Gallerani} S.,  et~al., 2010, \mndoi [\aap] {10.1051/0004-6361/201014721},
  \href {http://adsabs.harvard.edu/abs/2010A%26A...523A..85G} {523, A85}

\bibitem[\protect\citeauthoryear{{Genzel} \& {Cesarsky}}{{Genzel} \&
  {Cesarsky}}{2000}]{Genzel2000}
{Genzel} R.,  {Cesarsky} C.~J.,  2000, \mndoi [\araa]
  {10.1146/annurev.astro.38.1.761}, \href
  {http://adsabs.harvard.edu/abs/2000ARA%26A..38..761G} {38, 761}

\bibitem[\protect\citeauthoryear{{Genzel} et~al.,}{{Genzel}
  et~al.}{1998}]{Genzel1998}
{Genzel} R.,  et~al., 1998, \mndoi [\apj] {10.1086/305576}, \href
  {http://adsabs.harvard.edu/abs/1998ApJ...498..579G} {498, 579}

\bibitem[\protect\citeauthoryear{{Gong}, {Cooray}  \& {Santos}}{{Gong}
  et~al.}{2013}]{Gong2013}
{Gong} Y.,  {Cooray} A.,   {Santos} M.~G.,  2013, \mndoi [\apj]
  {10.1088/0004-637X/768/2/130}, \href
  {http://adsabs.harvard.edu/abs/2013ApJ...768..130G} {768, 130}

\bibitem[\protect\citeauthoryear{{Gonz{\'a}lez-Alfonso}
  et~al.,}{{Gonz{\'a}lez-Alfonso} et~al.}{2017}]{Gonzalez2017}
{Gonz{\'a}lez-Alfonso} E.,  et~al., 2017, \mndoi [\pasa]
  {10.1017/pasa.2017.46}, \href
  {http://adsabs.harvard.edu/abs/2017PASA...34...54G} {34, e054}

\bibitem[\protect\citeauthoryear{{Greif}}{{Greif}}{2015}]{Greif2015}
{Greif} T.~H.,  2015, \mndoi [Computational Astrophysics and Cosmology]
  {10.1186/s40668-014-0006-2}, \href
  {http://adsabs.harvard.edu/abs/2015ComAC...2....3G} {2, 3}

\bibitem[\protect\citeauthoryear{{Gruppioni} et~al.,}{{Gruppioni}
  et~al.}{2017}]{Gruppioni2017}
{Gruppioni} C.,  et~al., 2017, \mndoi [\pasa] {10.1017/pasa.2017.49}, \href
  {http://adsabs.harvard.edu/abs/2017PASA...34...55G} {34, e055}

\bibitem[\protect\citeauthoryear{{Guillard}, {Boulanger}, {Pineau Des
  For{\^e}ts}  \& {Appleton}}{{Guillard} et~al.}{2009}]{Guillard2009}
{Guillard} P.,  {Boulanger} F.,  {Pineau Des For{\^e}ts} G.,   {Appleton}
  P.~N.,  2009, \mndoi [\aap] {10.1051/0004-6361/200811263}, \href
  {http://adsabs.harvard.edu/abs/2009A%26A...502..515G} {502, 515}

\bibitem[\protect\citeauthoryear{{Gullberg} et~al.,}{{Gullberg}
  et~al.}{2015}]{Gullberg2015}
{Gullberg} B.,  et~al., 2015, \mndoi [\mnras] {10.1093/mnras/stv372}, \href
  {http://adsabs.harvard.edu/abs/2015MNRAS.449.2883G} {449, 2883}

\bibitem[\protect\citeauthoryear{{Haiman}}{{Haiman}}{2002}]{Haiman2002}
{Haiman} Z.,  2002, \mndoi [\apjl] {10.1086/343101}, \href
  {http://adsabs.harvard.edu/abs/2002ApJ...576L...1H} {576, L1}

\bibitem[\protect\citeauthoryear{{Haiman}, {Thoul}  \& {Loeb}}{{Haiman}
  et~al.}{1996}]{Haiman1996}
{Haiman} Z.,  {Thoul} A.~A.,   {Loeb} A.,  1996, \mndoi [\apj]
  {10.1086/177343}, \href {http://adsabs.harvard.edu/abs/1996ApJ...464..523H}
  {464, 523}

\bibitem[\protect\citeauthoryear{{Hao} et~al.,}{{Hao} et~al.}{2005}]{Hao2005}
{Hao} L.,  et~al., 2005, \mndoi [\apjl] {10.1086/431227}, \href
  {http://adsabs.harvard.edu/abs/2005ApJ...625L..75H} {625, L75}

\bibitem[\protect\citeauthoryear{{Harrington} et~al.,}{{Harrington}
  et~al.}{2016}]{Harrington2016}
{Harrington} K.~C.,  et~al., 2016, \mndoi [\mnras] {10.1093/mnras/stw614},
  \href {http://adsabs.harvard.edu/abs/2016MNRAS.458.4383H} {458, 4383}

\bibitem[\protect\citeauthoryear{{Hashimoto} et~al.,}{{Hashimoto}
  et~al.}{2018a}]{Hashimoto2018b}
{Hashimoto} T.,  et~al., 2018a, preprint, \href
  {http://adsabs.harvard.edu/abs/2018arXiv180600486H} {} (\mn@eprint {arXiv}
  {1806.00486})

\bibitem[\protect\citeauthoryear{Hashimoto et~al.,}{Hashimoto
  et~al.}{2018b}]{Hashimoto2018}
Hashimoto T.,  et~al., 2018b, \mndoi [Nature] {10.1038/s41586-018-0117-z}, 557,
  392

\bibitem[\protect\citeauthoryear{{Heger} \& {Woosley}}{{Heger} \&
  {Woosley}}{2002}]{heger2002}
{Heger} A.,  {Woosley} S.~E.,  2002, \mndoi [\apj] {10.1086/338487}, \href
  {http://adsabs.harvard.edu/abs/2002ApJ...567..532H} {567, 532}

\bibitem[\protect\citeauthoryear{{Higdon}, {Armus}, {Higdon}, {Soifer}  \&
  {Spoon}}{{Higdon} et~al.}{2006}]{Higdon2006}
{Higdon} S.~J.~U.,  {Armus} L.,  {Higdon} J.~L.,  {Soifer} B.~T.,   {Spoon}
  H.~W.~W.,  2006, \mndoi [\apj] {10.1086/505701}, \href
  {http://adsabs.harvard.edu/abs/2006ApJ...648..323H} {648, 323}

\bibitem[\protect\citeauthoryear{{Hirano}, {Hosokawa}, {Yoshida}, {Omukai}  \&
  {Yorke}}{{Hirano} et~al.}{2015}]{hirano2015}
{Hirano} S.,  {Hosokawa} T.,  {Yoshida} N.,  {Omukai} K.,   {Yorke} H.~W.,
  2015, \mndoi [MNRAS] {10.1093/mnras/stv044}, \href
  {http://adsabs.harvard.edu/abs/2015MNRAS.448..568H} {448, 568}

\bibitem[\protect\citeauthoryear{{Hirasawa}}{{Hirasawa}}{1969}]{Hirasawa1969}
{Hirasawa} T.,  1969, \mndoi [Progress of Theoretical Physics]
  {10.1143/PTP.42.523}, \href
  {http://adsabs.harvard.edu/abs/1969PThPh..42..523H} {42, 523}

\bibitem[\protect\citeauthoryear{{Ho} \& {Keto}}{{Ho} \& {Keto}}{2007}]{Ho2007}
{Ho} L.~C.,  {Keto} E.,  2007, \mndoi [\apj] {10.1086/511260}, \href
  {http://adsabs.harvard.edu/abs/2007ApJ...658..314H} {658, 314}

\bibitem[\protect\citeauthoryear{{Hughes} et~al.,}{{Hughes}
  et~al.}{1998}]{Hughes1998}
{Hughes} D.~H.,  et~al., 1998, \mndoi [\nat] {10.1038/28328}, \href
  {http://adsabs.harvard.edu/abs/1998Natur.394..241H} {394, 241}

\bibitem[\protect\citeauthoryear{{Hunt}, {Bianchi}  \& {Maiolino}}{{Hunt}
  et~al.}{2005}]{Hunt2005}
{Hunt} L.,  {Bianchi} S.,   {Maiolino} R.,  2005, \mndoi [\aap]
  {10.1051/0004-6361:20042157}, \href
  {http://adsabs.harvard.edu/abs/2005A%26A...434..849H} {434, 849}

\bibitem[\protect\citeauthoryear{{Jiang} et~al.,}{{Jiang}
  et~al.}{2010}]{Jiang2010}
{Jiang} L.,  et~al., 2010, \mndoi [\nat] {10.1038/nature08877}, \href
  {http://adsabs.harvard.edu/abs/2010Natur.464..380J} {464, 380}

\bibitem[\protect\citeauthoryear{{Jiang} et~al.,}{{Jiang}
  et~al.}{2013}]{Jiang2013}
{Jiang} L.,  et~al., 2013, \mndoi [\apj] {10.1088/0004-637X/772/2/99}, \href
  {http://adsabs.harvard.edu/abs/2013ApJ...772...99J} {772, 99}

\bibitem[\protect\citeauthoryear{{Jiang} et~al.,}{{Jiang}
  et~al.}{2016}]{Jiang2016}
{Jiang} L.,  et~al., 2016, \mndoi [\apj] {10.3847/0004-637X/816/1/16}, \href
  {http://adsabs.harvard.edu/abs/2016ApJ...816...16J} {816, 16}

\bibitem[\protect\citeauthoryear{{Jimenez} \& {Haiman}}{{Jimenez} \&
  {Haiman}}{2006}]{jimenez2006}
{Jimenez} R.,  {Haiman} Z.,  2006, \mndoi [\nat] {10.1038/nature04580}, \href
  {http://adsabs.harvard.edu/abs/2006Natur.440..501J} {440, 501}

\bibitem[\protect\citeauthoryear{{Johnson}}{{Johnson}}{2010}]{Johnson2010}
{Johnson} J.~L.,  2010, \mndoi [\mnras] {10.1111/j.1365-2966.2010.16351.x},
  \href {http://adsabs.harvard.edu/abs/2010MNRAS.404.1425J} {404, 1425}

\bibitem[\protect\citeauthoryear{{Johnson}, {Greif}  \& {Bromm}}{{Johnson}
  et~al.}{2008}]{Johnson2008}
{Johnson} J.~L.,  {Greif} T.~H.,   {Bromm} V.,  2008, \mndoi [\mnras]
  {10.1111/j.1365-2966.2008.13381.x}, \href
  {http://adsabs.harvard.edu/abs/2008MNRAS.388...26J} {388, 26}

\bibitem[\protect\citeauthoryear{{Johnson}, {Dalla Vecchia}  \&
  {Khochfar}}{{Johnson} et~al.}{2013}]{Johnson2013}
{Johnson} J.~L.,  {Dalla Vecchia} C.,   {Khochfar} S.,  2013, \mndoi [\mnras]
  {10.1093/mnras/sts011}, \href
  {http://adsabs.harvard.edu/abs/2013MNRAS.428.1857J} {428, 1857}

\bibitem[\protect\citeauthoryear{{Kaneda} et~al.,}{{Kaneda}
  et~al.}{2017}]{Kaneda2017}
{Kaneda} H.,  et~al., 2017, \mndoi [\pasa] {10.1017/pasa.2017.56}, \href
  {http://adsabs.harvard.edu/abs/2017PASA...34...59K} {34, e059}

\bibitem[\protect\citeauthoryear{{Kehrig}, {V{\'{\i}}lchez},
  {P{\'e}rez-Montero}, {Iglesias-P{\'a}ramo}, {Brinchmann}, {Kunth}, {Durret}
  \& {Bayo}}{{Kehrig} et~al.}{2015}]{Kehrig2015}
{Kehrig} C.,  {V{\'{\i}}lchez} J.~M.,  {P{\'e}rez-Montero} E.,
  {Iglesias-P{\'a}ramo} J.,  {Brinchmann} J.,  {Kunth} D.,  {Durret} F.,
  {Bayo} F.~M.,  2015, \mndoi [\apjl] {10.1088/2041-8205/801/2/L28}, \href
  {http://adsabs.harvard.edu/abs/2015ApJ...801L..28K} {801, L28}

\bibitem[\protect\citeauthoryear{{Kepley}, {Leroy}, {Johnson}, {Sandstrom}  \&
  {Chen}}{{Kepley} et~al.}{2016}]{Kepley2016}
{Kepley} A.~A.,  {Leroy} A.~K.,  {Johnson} K.~E.,  {Sandstrom} K.,   {Chen}
  C.-H.~R.,  2016, \mndoi [\apj] {10.3847/0004-637X/828/1/50}, \href
  {http://cdsads.u-strasbg.fr/abs/2016ApJ...828...50K} {828, 50}

\bibitem[\protect\citeauthoryear{{Knudsen}, {Watson}, {Frayer}, {Christensen},
  {Gallazzi}, {Micha{\l}owski}, {Richard}  \& {Zavala}}{{Knudsen}
  et~al.}{2017}]{Knudsen2017}
{Knudsen} K.~K.,  {Watson} D.,  {Frayer} D.,  {Christensen} L.,  {Gallazzi} A.,
   {Micha{\l}owski} M.~J.,  {Richard} J.,   {Zavala} J.,  2017, \mndoi [\mnras]
  {10.1093/mnras/stw3066}, \href
  {http://adsabs.harvard.edu/abs/2017MNRAS.466..138K} {466, 138}

\bibitem[\protect\citeauthoryear{{Koopmans} et~al.,}{{Koopmans}
  et~al.}{2015}]{Koopmans2015}
{Koopmans} L.,  et~al., 2015, Advancing Astrophysics with the Square Kilometre
  Array (AASKA14), \href {http://adsabs.harvard.edu/abs/2015aska.confE...1K}
  {p.~1}

\bibitem[\protect\citeauthoryear{{Koprowski} et~al.,}{{Koprowski}
  et~al.}{2016}]{Koprowski2016}
{Koprowski} M.~P.,  et~al., 2016, \mndoi [\apjl] {10.3847/2041-8205/828/2/L21},
  \href {http://adsabs.harvard.edu/abs/2016ApJ...828L..21K} {828, L21}

\bibitem[\protect\citeauthoryear{{Laporte} et~al.,}{{Laporte}
  et~al.}{2017}]{Laporte2017}
{Laporte} N.,  et~al., 2017, \mndoi [\apjl] {10.3847/2041-8213/aa62aa}, \href
  {http://adsabs.harvard.edu/abs/2017ApJ...837L..21L} {837, L21}

\bibitem[\protect\citeauthoryear{{Lebouteiller}, {Barry}, {Spoon},
  {Bernard-Salas}, {Sloan}, {Houck}  \& {Weedman}}{{Lebouteiller}
  et~al.}{2011}]{Lebouteiller2011}
{Lebouteiller} V.,  {Barry} D.~J.,  {Spoon} H.~W.~W.,  {Bernard-Salas} J.,
  {Sloan} G.~C.,  {Houck} J.~R.,   {Weedman} D.~W.,  2011, \mndoi [\apjs]
  {10.1088/0067-0049/196/1/8}, \href
  {http://adsabs.harvard.edu/abs/2011ApJS..196....8L} {196, 8}

\bibitem[\protect\citeauthoryear{{Leipski} et~al.,}{{Leipski}
  et~al.}{2014}]{Leipski2014}
{Leipski} C.,  et~al., 2014, \mndoi [\apj] {10.1088/0004-637X/785/2/154}, \href
  {http://adsabs.harvard.edu/abs/2014ApJ...785..154L} {785, 154}

\bibitem[\protect\citeauthoryear{{Lutz}, {Sturm}, {Genzel}, {Spoon},
  {Moorwood}, {Netzer}  \& {Sternberg}}{{Lutz} et~al.}{2003}]{Lutz2003}
{Lutz} D.,  {Sturm} E.,  {Genzel} R.,  {Spoon} H.~W.~W.,  {Moorwood} A.~F.~M.,
  {Netzer} H.,   {Sternberg} A.,  2003, \mndoi [\aap]
  {10.1051/0004-6361:20031165}, \href
  {http://cdsads.u-strasbg.fr/abs/2003A%26A...409..867L} {409, 867}

\bibitem[\protect\citeauthoryear{{Lyu}, {Rieke}  \& {Alberts}}{{Lyu}
  et~al.}{2016}]{Lyu2016}
{Lyu} J.,  {Rieke} G.~H.,   {Alberts} S.,  2016, \mndoi [\apj]
  {10.3847/0004-637X/816/2/85}, \href
  {http://adsabs.harvard.edu/abs/2016ApJ...816...85L} {816, 85}

\bibitem[\protect\citeauthoryear{{Ma} et~al.,}{{Ma} et~al.}{2015}]{Ma2015}
{Ma} J.,  et~al., 2015, \mndoi [\apj] {10.1088/0004-637X/812/1/88}, \href
  {http://adsabs.harvard.edu/abs/2015ApJ...812...88M} {812, 88}

\bibitem[\protect\citeauthoryear{{Ma} et~al.,}{{Ma} et~al.}{2016}]{Ma2016}
{Ma} J.,  et~al., 2016, \mndoi [\apj] {10.3847/0004-637X/832/2/114}, \href
  {http://adsabs.harvard.edu/abs/2016ApJ...832..114M} {832, 114}

\bibitem[\protect\citeauthoryear{{Madau} \& {Dickinson}}{{Madau} \&
  {Dickinson}}{2014}]{Madau2014}
{Madau} P.,  {Dickinson} M.,  2014, \mndoi [\araa]
  {10.1146/annurev-astro-081811-125615}, \href
  {http://adsabs.harvard.edu/abs/2014ARA%26A..52..415M} {52, 415}

\bibitem[\protect\citeauthoryear{{Madden}, {Galliano}, {Jones}  \&
  {Sauvage}}{{Madden} et~al.}{2006}]{Madden2006}
{Madden} S.~C.,  {Galliano} F.,  {Jones} A.~P.,   {Sauvage} M.,  2006, \mndoi
  [\aap] {10.1051/0004-6361:20053890}, \href
  {http://adsabs.harvard.edu/abs/2006A%26A...446..877M} {446, 877}

\bibitem[\protect\citeauthoryear{{Magdis} et~al.,}{{Magdis}
  et~al.}{2017}]{Magdis2017}
{Magdis} G.~E.,  et~al., 2017, \mndoi [\aap] {10.1051/0004-6361/201731037},
  \href {http://adsabs.harvard.edu/abs/2017A%26A...603A..93M} {603, A93}

\bibitem[\protect\citeauthoryear{{Mainali}, {Kollmeier}, {Stark}, {Simcoe},
  {Walth}, {Newman}  \& {Miller}}{{Mainali} et~al.}{2017}]{Mainali2007}
{Mainali} R.,  {Kollmeier} J.~A.,  {Stark} D.~P.,  {Simcoe} R.~A.,  {Walth} G.,
   {Newman} A.~B.,   {Miller} D.~R.,  2017, \mndoi [\apjl]
  {10.3847/2041-8213/836/1/L14}, \href
  {http://cdsads.u-strasbg.fr/abs/2017ApJ...836L..14M} {836, L14}

\bibitem[\protect\citeauthoryear{{Maiolino}, {Schneider}, {Oliva}, {Bianchi},
  {Ferrara}, {Mannucci}, {Pedani}  \& {Roca Sogorb}}{{Maiolino}
  et~al.}{2004}]{Maiolino2004}
{Maiolino} R.,  {Schneider} R.,  {Oliva} E.,  {Bianchi} S.,  {Ferrara} A.,
  {Mannucci} F.,  {Pedani} M.,   {Roca Sogorb} M.,  2004, \mndoi [\nat]
  {10.1038/nature02930}, \href
  {http://adsabs.harvard.edu/abs/2004Natur.431..533M} {431, 533}

\bibitem[\protect\citeauthoryear{{Mancini}, {Schneider}, {Graziani},
  {Valiante}, {Dayal}, {Maio}, {Ciardi}  \& {Hunt}}{{Mancini}
  et~al.}{2015}]{mancini2015}
{Mancini} M.,  {Schneider} R.,  {Graziani} L.,  {Valiante} R.,  {Dayal} P.,
  {Maio} U.,  {Ciardi} B.,   {Hunt} L.~K.,  2015, \mndoi [\mnras]
  {10.1093/mnrasl/slv070}, \href
  {http://adsabs.harvard.edu/abs/2015MNRAS.451L..70M} {451, L70}

\bibitem[\protect\citeauthoryear{{Mancini}, {Schneider}, {Graziani},
  {Valiante}, {Dayal}, {Maio}  \& {Ciardi}}{{Mancini}
  et~al.}{2016}]{mancini2016}
{Mancini} M.,  {Schneider} R.,  {Graziani} L.,  {Valiante} R.,  {Dayal} P.,
  {Maio} U.,   {Ciardi} B.,  2016, \mndoi [\mnras] {10.1093/mnras/stw1783},
  \href {http://adsabs.harvard.edu/abs/2016MNRAS.462.3130M} {462, 3130}

\bibitem[\protect\citeauthoryear{{Marassi}, {Chiaki}, {Schneider}, {Limongi},
  {Omukai}, {Nozawa}, {Chieffi}  \& {Yoshida}}{{Marassi}
  et~al.}{2014}]{marassi2014}
{Marassi} S.,  {Chiaki} G.,  {Schneider} R.,  {Limongi} M.,  {Omukai} K.,
  {Nozawa} T.,  {Chieffi} A.,   {Yoshida} N.,  2014, \mndoi [\apj]
  {10.1088/0004-637X/794/2/100}, \href
  {http://adsabs.harvard.edu/abs/2014ApJ...794..100M} {794, 100}

\bibitem[\protect\citeauthoryear{{Marassi}, {Schneider}, {Limongi}, {Chieffi},
  {Bocchio}  \& {Bianchi}}{{Marassi} et~al.}{2015}]{marassi2015}
{Marassi} S.,  {Schneider} R.,  {Limongi} M.,  {Chieffi} A.,  {Bocchio} M.,
  {Bianchi} S.,  2015, \mndoi [MNRAS] {10.1093/mnras/stv2267}, \href
  {http://adsabs.harvard.edu/abs/2015MNRAS.454.4250M} {454, 4250}

\bibitem[\protect\citeauthoryear{{Marassi}, {Schneider}, {Limongi}, {Chieffi},
  {Graziani}, {Bocchio}  \& {Bianchi}}{{Marassi} et~al.}{2017}]{Marassi2017}
{Marassi} S.,  {Schneider} R.,  {Limongi} M.,  {Chieffi} A.,  {Graziani} L.,
  {Bocchio} M.,   {Bianchi} S.,  2017, MNRAS, in prep

\bibitem[\protect\citeauthoryear{{Marchetti}, {Serjeant}  \&
  {Vaccari}}{{Marchetti} et~al.}{2017}]{Marchetti2017}
{Marchetti} L.,  {Serjeant} S.,   {Vaccari} M.,  2017, \mndoi [\mnras]
  {10.1093/mnras/stx1553}, \href
  {http://cdsads.u-strasbg.fr/abs/2017MNRAS.470.5007M} {470, 5007}

\bibitem[\protect\citeauthoryear{{Marrone} et~al.,}{{Marrone}
  et~al.}{2018}]{Marrone2018}
{Marrone} D.~P.,  et~al., 2018, \mndoi [\nat] {10.1038/nature24629}, \href
  {http://adsabs.harvard.edu/abs/2018Natur.553...51M} {553, 51}

\bibitem[\protect\citeauthoryear{{Marsden} et~al.,}{{Marsden}
  et~al.}{2014}]{Marsden2014}
{Marsden} D.,  et~al., 2014, \mndoi [\mnras] {10.1093/mnras/stu001}, \href
  {http://adsabs.harvard.edu/abs/2014MNRAS.439.1556M} {439, 1556}

\bibitem[\protect\citeauthoryear{{Matsuda}, {Sat{\=o}}  \& {Takeda}}{{Matsuda}
  et~al.}{1969}]{Matsuda1969}
{Matsuda} T.,  {Sat{\=o}} H.,   {Takeda} H.,  1969, \mndoi [Progress of
  Theoretical Physics] {10.1143/PTP.42.219}, \href
  {http://adsabs.harvard.edu/abs/1969PThPh..42..219M} {42, 219}

\bibitem[\protect\citeauthoryear{{Matsuura} et~al.,}{{Matsuura}
  et~al.}{2009}]{matsuura2009}
{Matsuura} M.,  et~al., 2009, \mndoi [MNRAS]
  {10.1111/j.1365-2966.2009.14743.x}, \href
  {http://adsabs.harvard.edu/abs/2009MNRAS.396..918M} {396, 918}

\bibitem[\protect\citeauthoryear{{Matsuura} et~al.,}{{Matsuura}
  et~al.}{2015}]{matsuura2015}
{Matsuura} M.,  et~al., 2015, \mndoi [\apj] {10.1088/0004-637X/800/1/50}, \href
  {http://adsabs.harvard.edu/abs/2015ApJ...800...50M} {800, 50}

\bibitem[\protect\citeauthoryear{{Matthee} et~al.,}{{Matthee}
  et~al.}{2017}]{Matthee2017}
{Matthee} J.,  et~al., 2017, \mndoi [\apj] {10.3847/1538-4357/aa9931}, \href
  {http://adsabs.harvard.edu/abs/2017ApJ...851..145M} {851, 145}

\bibitem[\protect\citeauthoryear{{Mizusawa}, {Omukai}  \& {Nishi}}{{Mizusawa}
  et~al.}{2005}]{Mizusawa2005}
{Mizusawa} H.,  {Omukai} K.,   {Nishi} R.,  2005, \mndoi [\pasj]
  {10.1093/pasj/57.6.951}, \href
  {http://adsabs.harvard.edu/abs/2005PASJ...57..951M} {57, 951}

\bibitem[\protect\citeauthoryear{{Mocanu} et~al.,}{{Mocanu}
  et~al.}{2013}]{Mocanu2013}
{Mocanu} L.~M.,  et~al., 2013, \mndoi [\apj] {10.1088/0004-637X/779/1/61},
  \href {http://adsabs.harvard.edu/abs/2013ApJ...779...61M} {779, 61}

\bibitem[\protect\citeauthoryear{{Nayyeri} et~al.,}{{Nayyeri}
  et~al.}{2016}]{Nayyeri2016}
{Nayyeri} H.,  et~al., 2016, \mndoi [\apj] {10.3847/0004-637X/823/1/17}, \href
  {http://adsabs.harvard.edu/abs/2016ApJ...823...17N} {823, 17}

\bibitem[\protect\citeauthoryear{{Negrello} et~al.,}{{Negrello}
  et~al.}{2010}]{Negrello2010}
{Negrello} M.,  et~al., 2010, \mndoi [Science] {10.1126/science.1193420}, \href
  {http://adsabs.harvard.edu/abs/2010Sci...330..800N} {330, 800}

\bibitem[\protect\citeauthoryear{{Negrello} et~al.,}{{Negrello}
  et~al.}{2017}]{Negrello2017}
{Negrello} M.,  et~al., 2017, \mndoi [\mnras] {10.1093/mnras/stw2911}, \href
  {http://adsabs.harvard.edu/abs/2017MNRAS.465.3558N} {465, 3558}

\bibitem[\protect\citeauthoryear{{Nguyen} et~al.,}{{Nguyen}
  et~al.}{2010}]{Nguyen2010}
{Nguyen} H.~T.,  et~al., 2010, \mndoi [\aap] {10.1051/0004-6361/201014680},
  \href {http://adsabs.harvard.edu/abs/2010A%26A...518L...5N} {518, L5}

\bibitem[\protect\citeauthoryear{{Nozawa}, {Kozasa}, {Umeda}, {Maeda}  \&
  {Nomoto}}{{Nozawa} et~al.}{2003}]{Nozawa2003}
{Nozawa} T.,  {Kozasa} T.,  {Umeda} H.,  {Maeda} K.,   {Nomoto} K.,  2003,
  \mndoi [\apj] {10.1086/379011}, \href
  {http://adsabs.harvard.edu/abs/2003ApJ...598..785N} {598, 785}

\bibitem[\protect\citeauthoryear{{Nozawa}, {Kozasa}, {Habe}, {Dwek}, {Umeda},
  {Tominaga}, {Maeda}  \& {Nomoto}}{{Nozawa} et~al.}{2007}]{nozawa2007}
{Nozawa} T.,  {Kozasa} T.,  {Habe} A.,  {Dwek} E.,  {Umeda} H.,  {Tominaga} N.,
   {Maeda} K.,   {Nomoto} K.,  2007, \mndoi [ApJ] {10.1086/520621}, \href
  {http://adsabs.harvard.edu/abs/2007ApJ...666..955N} {666, 955}

\bibitem[\protect\citeauthoryear{{Ogle}, {Boulanger}, {Guillard}, {Evans},
  {Antonucci}, {Appleton}, {Nesvadba}  \& {Leipski}}{{Ogle}
  et~al.}{2010}]{Ogle2010}
{Ogle} P.,  {Boulanger} F.,  {Guillard} P.,  {Evans} D.~A.,  {Antonucci} R.,
  {Appleton} P.~N.,  {Nesvadba} N.,   {Leipski} C.,  2010, \mndoi [\apj]
  {10.1088/0004-637X/724/2/1193}, \href
  {http://adsabs.harvard.edu/abs/2010ApJ...724.1193O} {724, 1193}

\bibitem[\protect\citeauthoryear{{Ogle}, {Davies}, {Appleton}, {Bertincourt},
  {Seymour}  \& {Helou}}{{Ogle} et~al.}{2012}]{Ogle2012}
{Ogle} P.,  {Davies} J.~E.,  {Appleton} P.~N.,  {Bertincourt} B.,  {Seymour}
  N.,   {Helou} G.,  2012, \mndoi [\apj] {10.1088/0004-637X/751/1/13}, \href
  {http://adsabs.harvard.edu/abs/2012ApJ...751...13O} {751, 13}

\bibitem[\protect\citeauthoryear{{Oh} \& {Haiman}}{{Oh} \&
  {Haiman}}{2002}]{Oh2002}
{Oh} S.~P.,  {Haiman} Z.,  2002, \mndoi [\apj] {10.1086/339393}, \href
  {http://adsabs.harvard.edu/abs/2002ApJ...569..558O} {569, 558}

\bibitem[\protect\citeauthoryear{{Omukai} \& {Kitayama}}{{Omukai} \&
  {Kitayama}}{2003}]{Omukai2003}
{Omukai} K.,  {Kitayama} T.,  2003, \mndoi [\apj] {10.1086/379282}, \href
  {http://adsabs.harvard.edu/abs/2003ApJ...599..738O} {599, 738}

\bibitem[\protect\citeauthoryear{{Omukai} \& {Nishi}}{{Omukai} \&
  {Nishi}}{1999}]{Omukai1999}
{Omukai} K.,  {Nishi} R.,  1999, \mndoi [\apj] {10.1086/307285}, \href
  {http://adsabs.harvard.edu/abs/1999ApJ...518...64O} {518, 64}

\bibitem[\protect\citeauthoryear{{Omukai} \& {Palla}}{{Omukai} \&
  {Palla}}{2003}]{Omukai2003a}
{Omukai} K.,  {Palla} F.,  2003, \mndoi [\apj] {10.1086/374810}, \href
  {http://adsabs.harvard.edu/abs/2003ApJ...589..677O} {589, 677}

\bibitem[\protect\citeauthoryear{{Oteo} et~al.,}{{Oteo}
  et~al.}{2013}]{Oteo2013}
{Oteo} I.,  et~al., 2013, \mndoi [\aap] {10.1051/0004-6361/201321478}, \href
  {http://cdsads.u-strasbg.fr/abs/2013A%26A...554L...3O} {554, L3}

\bibitem[\protect\citeauthoryear{{Ouchi} et~al.,}{{Ouchi}
  et~al.}{2018}]{Ouchi2018}
{Ouchi} M.,  et~al., 2018, \mndoi [\pasj] {10.1093/pasj/psx074}, \href
  {http://adsabs.harvard.edu/abs/2018PASJ...70S..13O} {70, S13}

\bibitem[\protect\citeauthoryear{{Pacucci}, {Pallottini}, {Ferrara}  \&
  {Gallerani}}{{Pacucci} et~al.}{2017}]{Pacucci2017}
{Pacucci} F.,  {Pallottini} A.,  {Ferrara} A.,   {Gallerani} S.,  2017, \mndoi
  [\mnras] {10.1093/mnrasl/slx029}, \href
  {http://adsabs.harvard.edu/abs/2017MNRAS.468L..77P} {468, L77}

\bibitem[\protect\citeauthoryear{{Pallottini}, {Ferrara}, {Gallerani},
  {Salvadori}  \& {D'Odorico}}{{Pallottini} et~al.}{2014}]{Pallottini2014}
{Pallottini} A.,  {Ferrara} A.,  {Gallerani} S.,  {Salvadori} S.,   {D'Odorico}
  V.,  2014, \mndoi [\mnras] {10.1093/mnras/stu451}, \href
  {http://adsabs.harvard.edu/abs/2014MNRAS.440.2498P} {440, 2498}

\bibitem[\protect\citeauthoryear{{Pallottini}, {Gallerani}, {Ferrara}, {Yue},
  {Vallini}, {Maiolino}  \& {Feruglio}}{{Pallottini}
  et~al.}{2015a}]{Pallottini2015}
{Pallottini} A.,  {Gallerani} S.,  {Ferrara} A.,  {Yue} B.,  {Vallini} L.,
  {Maiolino} R.,   {Feruglio} C.,  2015a, \mndoi [\mnras]
  {10.1093/mnras/stv1788}, \href
  {http://adsabs.harvard.edu/abs/2015MNRAS.453.1898P} {453, 1898}

\bibitem[\protect\citeauthoryear{{Pallottini} et~al.,}{{Pallottini}
  et~al.}{2015b}]{Pallottini2015b}
{Pallottini} A.,  et~al., 2015b, \mndoi [\mnras] {10.1093/mnras/stv1795}, \href
  {http://adsabs.harvard.edu/abs/2015MNRAS.453.2465P} {453, 2465}

\bibitem[\protect\citeauthoryear{{Pallottini}, {Ferrara}, {Gallerani},
  {Vallini}, {Maiolino}  \& {Salvadori}}{{Pallottini}
  et~al.}{2017a}]{Pallottini2017}
{Pallottini} A.,  {Ferrara} A.,  {Gallerani} S.,  {Vallini} L.,  {Maiolino} R.,
    {Salvadori} S.,  2017a, \mndoi [\mnras] {10.1093/mnras/stw2847}, \href
  {http://adsabs.harvard.edu/abs/2017MNRAS.465.2540P} {465, 2540}

\bibitem[\protect\citeauthoryear{{Pallottini}, {Ferrara}, {Bovino}, {Vallini},
  {Gallerani}, {Maiolino}  \& {Salvadori}}{{Pallottini}
  et~al.}{2017b}]{Pallottini2017b}
{Pallottini} A.,  {Ferrara} A.,  {Bovino} S.,  {Vallini} L.,  {Gallerani} S.,
  {Maiolino} R.,   {Salvadori} S.,  2017b, \mndoi [\mnras]
  {10.1093/mnras/stx1792}, \href
  {http://cdsads.u-strasbg.fr/abs/2017MNRAS.471.4128P} {471, 4128}

\bibitem[\protect\citeauthoryear{{Pavesi} et~al.,}{{Pavesi}
  et~al.}{2018}]{Pavesi2018}
{Pavesi} R.,  et~al., 2018, \mndoi [\apj] {10.3847/1538-4357/aac6b6}, \href
  {http://adsabs.harvard.edu/abs/2018ApJ...861...43P} {861, 43}

\bibitem[\protect\citeauthoryear{{Peebles} \& {Dicke}}{{Peebles} \&
  {Dicke}}{1968}]{Peebles1968}
{Peebles} P.~J.~E.,  {Dicke} R.~H.,  1968, \mndoi [\apj] {10.1086/149811},
  \href {http://adsabs.harvard.edu/abs/1968ApJ...154..891P} {154, 891}

\bibitem[\protect\citeauthoryear{{Penzias} \& {Wilson}}{{Penzias} \&
  {Wilson}}{1965}]{Penzias1965}
{Penzias} A.~A.,  {Wilson} R.~W.,  1965, \mndoi [\apj] {10.1086/148307}, \href
  {http://adsabs.harvard.edu/abs/1965ApJ...142..419P} {142, 419}

\bibitem[\protect\citeauthoryear{{Planck Collaboration} et~al.,}{{Planck
  Collaboration} et~al.}{2016}]{Planck2016}
{Planck Collaboration} et~al., 2016, \mndoi [\aap]
  {10.1051/0004-6361/201525830}, \href
  {http://cdsads.u-strasbg.fr/abs/2016A%26A...594A..13P} {594, A13}

\bibitem[\protect\citeauthoryear{{Pope} et~al.,}{{Pope}
  et~al.}{2008}]{Pope2008}
{Pope} A.,  et~al., 2008, \mndoi [\apj] {10.1086/527030}, \href
  {http://adsabs.harvard.edu/abs/2008ApJ...675.1171P} {675, 1171}

\bibitem[\protect\citeauthoryear{{Press} \& {Schechter}}{{Press} \&
  {Schechter}}{1974}]{Press1974}
{Press} W.~H.,  {Schechter} P.,  1974, \mndoi [\apj] {10.1086/152650}, \href
  {http://adsabs.harvard.edu/abs/1974ApJ...187..425P} {187, 425}

\bibitem[\protect\citeauthoryear{{Rawle} et~al.,}{{Rawle}
  et~al.}{2014}]{Rawle2014}
{Rawle} T.~D.,  et~al., 2014, \mndoi [\apj] {10.1088/0004-637X/783/1/59}, \href
  {http://adsabs.harvard.edu/abs/2014ApJ...783...59R} {783, 59}

\bibitem[\protect\citeauthoryear{{Rawle} et~al.,}{{Rawle}
  et~al.}{2016}]{Rawle2016}
{Rawle} T.~D.,  et~al., 2016, \mndoi [\mnras] {10.1093/mnras/stw712}, \href
  {http://adsabs.harvard.edu/abs/2016MNRAS.459.1626R} {459, 1626}

\bibitem[\protect\citeauthoryear{{R{\'e}my-Ruyer} et~al.,}{{R{\'e}my-Ruyer}
  et~al.}{2015}]{Remy-Ruyer2015}
{R{\'e}my-Ruyer} A.,  et~al., 2015, \mndoi [\aap]
  {10.1051/0004-6361/201526067}, \href
  {http://adsabs.harvard.edu/abs/2015A%26A...582A.121R} {582, A121}

\bibitem[\protect\citeauthoryear{{Ricotti}, {Gnedin}  \& {Shull}}{{Ricotti}
  et~al.}{2008}]{ricotti2008}
{Ricotti} M.,  {Gnedin} N.~Y.,   {Shull} J.~M.,  2008, \mndoi [\apj]
  {10.1086/590901}, \href {http://adsabs.harvard.edu/abs/2008ApJ...685...21R}
  {685, 21}

\bibitem[\protect\citeauthoryear{{Riechers} et~al.,}{{Riechers}
  et~al.}{2010}]{Riechers2010}
{Riechers} D.~A.,  et~al., 2010, \mndoi [\apjl] {10.1088/2041-8205/720/2/L131},
  \href {http://adsabs.harvard.edu/abs/2010ApJ...720L.131R} {720, L131}

\bibitem[\protect\citeauthoryear{{Riechers} et~al.,}{{Riechers}
  et~al.}{2013}]{Riechers2013}
{Riechers} D.~A.,  et~al., 2013, \mndoi [\nat] {10.1038/nature12050}, \href
  {http://adsabs.harvard.edu/abs/2013Natur.496..329R} {496, 329}

\bibitem[\protect\citeauthoryear{{Riechers} et~al.,}{{Riechers}
  et~al.}{2014}]{Riechers2014a}
{Riechers} D.~A.,  et~al., 2014, \mndoi [\apj] {10.1088/0004-637X/786/1/31},
  \href {http://adsabs.harvard.edu/abs/2014ApJ...786...31R} {786, 31}

\bibitem[\protect\citeauthoryear{{Riechers} et~al.,}{{Riechers}
  et~al.}{2017}]{Riechers2017}
{Riechers} D.~A.,  et~al., 2017, \mndoi [\apj] {10.3847/1538-4357/aa8ccf},
  \href {http://adsabs.harvard.edu/abs/2017ApJ...850....1R} {850, 1}

\bibitem[\protect\citeauthoryear{{Rieke}, {Alonso-Herrero}, {Weiner},
  {P{\'e}rez-Gonz{\'a}lez}, {Blaylock}, {Donley}  \& {Marcillac}}{{Rieke}
  et~al.}{2009}]{Rieke2009}
{Rieke} G.~H.,  {Alonso-Herrero} A.,  {Weiner} B.~J.,  {P{\'e}rez-Gonz{\'a}lez}
  P.~G.,  {Blaylock} M.,  {Donley} J.~L.,   {Marcillac} D.,  2009, \mndoi
  [\apj] {10.1088/0004-637X/692/1/556}, \href
  {http://adsabs.harvard.edu/abs/2009ApJ...692..556R} {692, 556}

\bibitem[\protect\citeauthoryear{{Rigby} et~al.,}{{Rigby}
  et~al.}{2008}]{Rigby2008}
{Rigby} J.~R.,  et~al., 2008, \mndoi [\apj] {10.1086/525273}, \href
  {http://adsabs.harvard.edu/abs/2008ApJ...675..262R} {675, 262}

\bibitem[\protect\citeauthoryear{{Rigopoulou}, {Kunze}, {Lutz}, {Genzel}  \&
  {Moorwood}}{{Rigopoulou} et~al.}{2002}]{Rigopoulou2002}
{Rigopoulou} D.,  {Kunze} D.,  {Lutz} D.,  {Genzel} R.,   {Moorwood} A.~F.~M.,
  2002, \mndoi [\aap] {10.1051/0004-6361:20020607}, \href
  {http://adsabs.harvard.edu/abs/2002A%26A...389..374R} {389, 374}

\bibitem[\protect\citeauthoryear{{Roelfsema} et~al.,}{{Roelfsema}
  et~al.}{2018}]{Roelfsema2018}
{Roelfsema} P.~R.,  et~al., 2018, \mndoi [\pasa] {10.1017/pasa.2018.15}, \href
  {http://adsabs.harvard.edu/abs/2018PASA...35...30R} {35, e030}

\bibitem[\protect\citeauthoryear{{Rujopakarn}, {Rieke}, {Weiner},
  {P{\'e}rez-Gonz{\'a}lez}, {Rex}, {Walth}  \& {Kartaltepe}}{{Rujopakarn}
  et~al.}{2013}]{Rujopakarn2013}
{Rujopakarn} W.,  {Rieke} G.~H.,  {Weiner} B.~J.,  {P{\'e}rez-Gonz{\'a}lez} P.,
   {Rex} M.,  {Walth} G.~L.,   {Kartaltepe} J.~S.,  2013, \mndoi [\apj]
  {10.1088/0004-637X/767/1/73}, \href
  {http://adsabs.harvard.edu/abs/2013ApJ...767...73R} {767, 73}

\bibitem[\protect\citeauthoryear{{Sanders} \& {Mirabel}}{{Sanders} \&
  {Mirabel}}{1996}]{Sanders1996}
{Sanders} D.~B.,  {Mirabel} I.~F.,  1996, \mndoi [\araa]
  {10.1146/annurev.astro.34.1.749}, \href
  {http://adsabs.harvard.edu/abs/1996ARA%26A..34..749S} {34, 749}

\bibitem[\protect\citeauthoryear{{Santoro} \& {Shull}}{{Santoro} \&
  {Shull}}{2006}]{Santoro2006}
{Santoro} F.,  {Shull} J.~M.,  2006, \mndoi [\apj] {10.1086/501518}, \href
  {http://adsabs.harvard.edu/abs/2006ApJ...643...26S} {643, 26}

\bibitem[\protect\citeauthoryear{{Sarangi} \& {Cherchneff}}{{Sarangi} \&
  {Cherchneff}}{2015}]{Sarangi2015}
{Sarangi} A.,  {Cherchneff} I.,  2015, A\&A, 575, A95

\bibitem[\protect\citeauthoryear{{Sargent} \& {Searle}}{{Sargent} \&
  {Searle}}{1970}]{Sargent1970}
{Sargent} W.~L.~W.,  {Searle} L.,  1970, \mndoi [\apjl] {10.1086/180644}, \href
  {http://adsabs.harvard.edu/abs/1970ApJ...162L.155S} {162, L155}

\bibitem[\protect\citeauthoryear{{Saslaw} \& {Zipoy}}{{Saslaw} \&
  {Zipoy}}{1967}]{Saslaw1967}
{Saslaw} W.~C.,  {Zipoy} D.,  1967, \mndoi [\nat] {10.1038/216976a0}, \href
  {http://adsabs.harvard.edu/abs/1967Natur.216..976S} {216, 976}

\bibitem[\protect\citeauthoryear{{Scannapieco}, {Schneider}  \&
  {Ferrara}}{{Scannapieco} et~al.}{2003}]{scannapieco2003}
{Scannapieco} E.,  {Schneider} R.,   {Ferrara} A.,  2003, \mndoi [\apj]
  {10.1086/374412}, \href {http://adsabs.harvard.edu/abs/2003ApJ...589...35S}
  {589, 35}

\bibitem[\protect\citeauthoryear{{Schneider}, {Ferrara}  \&
  {Salvaterra}}{{Schneider} et~al.}{2004}]{Schneider2004}
{Schneider} R.,  {Ferrara} A.,   {Salvaterra} R.,  2004, \mndoi [\mnras]
  {10.1111/j.1365-2966.2004.07876.x}, \href
  {http://adsabs.harvard.edu/abs/2004MNRAS.351.1379S} {351, 1379}

\bibitem[\protect\citeauthoryear{{Schneider}, {Salvaterra}, {Ferrara}  \&
  {Ciardi}}{{Schneider} et~al.}{2006a}]{schneider2006a}
{Schneider} R.,  {Salvaterra} R.,  {Ferrara} A.,   {Ciardi} B.,  2006a, \mndoi
  [\mnras] {10.1111/j.1365-2966.2006.10331.x}, \href
  {http://adsabs.harvard.edu/abs/2006MNRAS.369..825S} {369, 825}

\bibitem[\protect\citeauthoryear{{Schneider}, {Omukai}, {Inoue}  \&
  {Ferrara}}{{Schneider} et~al.}{2006b}]{schneider2006b}
{Schneider} R.,  {Omukai} K.,  {Inoue} A.~K.,   {Ferrara} A.,  2006b, \mndoi
  [\mnras] {10.1111/j.1365-2966.2006.10391.x}, \href
  {http://adsabs.harvard.edu/abs/2006MNRAS.369.1437S} {369, 1437}

\bibitem[\protect\citeauthoryear{{Schneider}, {Omukai}, {Bianchi}  \&
  {Valiante}}{{Schneider} et~al.}{2012a}]{schneider2012a}
{Schneider} R.,  {Omukai} K.,  {Bianchi} S.,   {Valiante} R.,  2012a, \mndoi
  [\mnras] {10.1111/j.1365-2966.2011.19818.x}, \href
  {http://adsabs.harvard.edu/abs/2012MNRAS.419.1566S} {419, 1566}

\bibitem[\protect\citeauthoryear{{Schneider}, {Omukai}, {Limongi}, {Ferrara},
  {Salvaterra}, {Chieffi}  \& {Bianchi}}{{Schneider}
  et~al.}{2012b}]{schneider2012b}
{Schneider} R.,  {Omukai} K.,  {Limongi} M.,  {Ferrara} A.,  {Salvaterra} R.,
  {Chieffi} A.,   {Bianchi} S.,  2012b, \mndoi [\mnras]
  {10.1111/j.1745-3933.2012.01257.x}, \href
  {http://adsabs.harvard.edu/abs/2012MNRAS.423L..60S} {423, L60}

\bibitem[\protect\citeauthoryear{{Schneider}, {Bianchi}, {Valiante}, {Risaliti}
   \& {Salvadori}}{{Schneider} et~al.}{2015}]{Schneider2015b}
{Schneider} R.,  {Bianchi} S.,  {Valiante} R.,  {Risaliti} G.,   {Salvadori}
  S.,  2015, \mndoi [\aap] {10.1051/0004-6361/201526105}, \href
  {http://adsabs.harvard.edu/abs/2015A%26A...579A..60S} {579, A60}

\bibitem[\protect\citeauthoryear{{Shi} et~al.,}{{Shi} et~al.}{2006}]{Shi2006}
{Shi} Y.,  et~al., 2006, \mndoi [\apj] {10.1086/508737}, \href
  {http://adsabs.harvard.edu/abs/2006ApJ...653..127S} {653, 127}

\bibitem[\protect\citeauthoryear{{Shi} et~al.,}{{Shi} et~al.}{2007}]{Shi2007}
{Shi} Y.,  et~al., 2007, \mndoi [\apj] {10.1086/521594}, \href
  {http://adsabs.harvard.edu/abs/2007ApJ...669..841S} {669, 841}

\bibitem[\protect\citeauthoryear{{Shi}, {Rieke}, {Ogle}, {Jiang}  \&
  {Diamond-Stanic}}{{Shi} et~al.}{2009}]{Shi2009}
{Shi} Y.,  {Rieke} G.~H.,  {Ogle} P.,  {Jiang} L.,   {Diamond-Stanic} A.~M.,
  2009, \mndoi [\apj] {10.1088/0004-637X/703/1/1107}, \href
  {http://adsabs.harvard.edu/abs/2009ApJ...703.1107S} {703, 1107}

\bibitem[\protect\citeauthoryear{{Shi}, {Rieke}, {Ogle}, {Su}  \&
  {Balog}}{{Shi} et~al.}{2014}]{Shi2014}
{Shi} Y.,  {Rieke} G.~H.,  {Ogle} P.~M.,  {Su} K.~Y.~L.,   {Balog} Z.,  2014,
  \mndoi [\apjs] {10.1088/0067-0049/214/2/23}, \href
  {http://adsabs.harvard.edu/abs/2014ApJS..214...23S} {214, 23}

\bibitem[\protect\citeauthoryear{{Shibuya} et~al.,}{{Shibuya}
  et~al.}{2018}]{Shibuya2018}
{Shibuya} T.,  et~al., 2018, \mndoi [\pasj] {10.1093/pasj/psx107}, \href
  {http://adsabs.harvard.edu/abs/2018PASJ...70S..15S} {70, S15}

\bibitem[\protect\citeauthoryear{{Siana}, {Teplitz}, {Chary}, {Colbert}  \&
  {Frayer}}{{Siana} et~al.}{2008}]{Siana2008}
{Siana} B.,  {Teplitz} H.~I.,  {Chary} R.-R.,  {Colbert} J.,   {Frayer} D.~T.,
  2008, \mndoi [\apj] {10.1086/592682}, \href
  {http://cdsads.u-strasbg.fr/abs/2008ApJ...689...59S} {689, 59}

\bibitem[\protect\citeauthoryear{{Siana} et~al.,}{{Siana}
  et~al.}{2009}]{Siana2009}
{Siana} B.,  et~al., 2009, \mndoi [\apj] {10.1088/0004-637X/698/2/1273}, \href
  {http://cdsads.u-strasbg.fr/abs/2009ApJ...698.1273S} {698, 1273}

\bibitem[\protect\citeauthoryear{{Siebenmorgen}, {Haas}, {Kr{\"u}gel}  \&
  {Schulz}}{{Siebenmorgen} et~al.}{2005}]{Siebenmorgen2005}
{Siebenmorgen} R.,  {Haas} M.,  {Kr{\"u}gel} E.,   {Schulz} B.,  2005, \mndoi
  [\aap] {10.1051/0004-6361:200500109}, \href
  {http://adsabs.harvard.edu/abs/2005A%26A...436L...5S} {436, L5}

\bibitem[\protect\citeauthoryear{{Smith} et~al.,}{{Smith}
  et~al.}{2007}]{Smith2007}
{Smith} J.~D.~T.,  et~al., 2007, \mndoi [\apj] {10.1086/510549}, \href
  {http://adsabs.harvard.edu/abs/2007ApJ...656..770S} {656, 770}

\bibitem[\protect\citeauthoryear{{Smith}, {Bromm}  \& {Loeb}}{{Smith}
  et~al.}{2016}]{Smith2016}
{Smith} A.,  {Bromm} V.,   {Loeb} A.,  2016, \mndoi [\mnras]
  {10.1093/mnras/stw1129}, \href
  {http://adsabs.harvard.edu/abs/2016MNRAS.460.3143S} {460, 3143}

\bibitem[\protect\citeauthoryear{{Smol{\v c}i{\'c}} et~al.,}{{Smol{\v c}i{\'c}}
  et~al.}{2015}]{Smolcic2015}
{Smol{\v c}i{\'c}} V.,  et~al., 2015, \mndoi [\aap]
  {10.1051/0004-6361/201424996}, \href
  {http://adsabs.harvard.edu/abs/2015A%26A...576A.127S} {576, A127}

\bibitem[\protect\citeauthoryear{{Sobacchi} \& {Mesinger}}{{Sobacchi} \&
  {Mesinger}}{2014}]{Sobacchi2014}
{Sobacchi} E.,  {Mesinger} A.,  2014, \mndoi [\mnras] {10.1093/mnras/stu377},
  \href {http://adsabs.harvard.edu/abs/2014MNRAS.440.1662S} {440, 1662}

\bibitem[\protect\citeauthoryear{{Sobral}, {Matthee}, {Darvish}, {Schaerer},
  {Mobasher}, {R{\"o}ttgering}, {Santos}  \& {Hemmati}}{{Sobral}
  et~al.}{2015}]{Sobral2015}
{Sobral} D.,  {Matthee} J.,  {Darvish} B.,  {Schaerer} D.,  {Mobasher} B.,
  {R{\"o}ttgering} H.~J.~A.,  {Santos} S.,   {Hemmati} S.,  2015, \mndoi [\apj]
  {10.1088/0004-637X/808/2/139}, \href
  {http://adsabs.harvard.edu/abs/2015ApJ...808..139S} {808, 139}

\bibitem[\protect\citeauthoryear{{Sobral} et~al.,}{{Sobral}
  et~al.}{2017}]{Sobral2018}
{Sobral} D.,  et~al., 2017, preprint, \href
  {http://adsabs.harvard.edu/abs/2017arXiv171008422S} {} (\mn@eprint {arXiv}
  {1710.08422})

\bibitem[\protect\citeauthoryear{{Soifer}, {Helou}  \& {Werner}}{{Soifer}
  et~al.}{2008}]{Soifer2008}
{Soifer} B.~T.,  {Helou} G.,   {Werner} M.,  2008, \mndoi [\araa]
  {10.1146/annurev.astro.46.060407.145144}, \href
  {http://adsabs.harvard.edu/abs/2008ARA%26A..46..201S} {46, 201}

\bibitem[\protect\citeauthoryear{{Spergel} et~al.,}{{Spergel}
  et~al.}{2015}]{Spergel2015}
{Spergel} D.,  et~al., 2015, preprint, \href
  {http://adsabs.harvard.edu/abs/2015arXiv150303757S} {} (\mn@eprint {arXiv}
  {1503.03757})

\bibitem[\protect\citeauthoryear{{Spilker} et~al.,}{{Spilker}
  et~al.}{2016}]{Spilker2016}
{Spilker} J.~S.,  et~al., 2016, \mndoi [\apj] {10.3847/0004-637X/826/2/112},
  \href {http://adsabs.harvard.edu/abs/2016ApJ...826..112S} {826, 112}

\bibitem[\protect\citeauthoryear{{Spinoglio} et~al.,}{{Spinoglio}
  et~al.}{2017}]{Spinoglio2017}
{Spinoglio} L.,  et~al., 2017, \mndoi [\pasa] {10.1017/pasa.2017.48}, \href
  {http://adsabs.harvard.edu/abs/2017PASA...34...57S} {34, e057}

\bibitem[\protect\citeauthoryear{{Stacy}, {Bromm}  \& {Lee}}{{Stacy}
  et~al.}{2016}]{Stacy2016}
{Stacy} A.,  {Bromm} V.,   {Lee} A.~T.,  2016, \mndoi [\mnras]
  {10.1093/mnras/stw1728}, \href
  {http://adsabs.harvard.edu/abs/2016MNRAS.462.1307S} {462, 1307}

\bibitem[\protect\citeauthoryear{{Stark} et~al.,}{{Stark}
  et~al.}{2014}]{Stark2014}
{Stark} D.~P.,  et~al., 2014, \mndoi [\mnras] {10.1093/mnras/stu1618}, \href
  {http://cdsads.u-strasbg.fr/abs/2014MNRAS.445.3200S} {445, 3200}

\bibitem[\protect\citeauthoryear{{Stark} et~al.,}{{Stark}
  et~al.}{2015a}]{Stark2015a}
{Stark} D.~P.,  et~al., 2015a, \mndoi [\mnras] {10.1093/mnras/stv688}, \href
  {http://cdsads.u-strasbg.fr/abs/2015MNRAS.450.1846S} {450, 1846}

\bibitem[\protect\citeauthoryear{{Stark} et~al.,}{{Stark}
  et~al.}{2015b}]{Stark2015b}
{Stark} D.~P.,  et~al., 2015b, \mndoi [\mnras] {10.1093/mnras/stv1907}, \href
  {http://cdsads.u-strasbg.fr/abs/2015MNRAS.454.1393S} {454, 1393}

\bibitem[\protect\citeauthoryear{{Stark} et~al.,}{{Stark}
  et~al.}{2017}]{Stark2007}
{Stark} D.~P.,  et~al., 2017, \mndoi [\mnras] {10.1093/mnras/stw2233}, \href
  {http://cdsads.u-strasbg.fr/abs/2017MNRAS.464..469S} {464, 469}

\bibitem[\protect\citeauthoryear{{Stiavelli} \& {Trenti}}{{Stiavelli} \&
  {Trenti}}{2010}]{Stiavelli2010}
{Stiavelli} M.,  {Trenti} M.,  2010, \mndoi [\apjl]
  {10.1088/2041-8205/716/2/L190}, \href
  {http://adsabs.harvard.edu/abs/2010ApJ...716L.190S} {716, L190}

\bibitem[\protect\citeauthoryear{{Strandet} et~al.,}{{Strandet}
  et~al.}{2016}]{Strandet2016}
{Strandet} M.~L.,  et~al., 2016, \mndoi [\apj] {10.3847/0004-637X/822/2/80},
  \href {http://adsabs.harvard.edu/abs/2016ApJ...822...80S} {822, 80}

\bibitem[\protect\citeauthoryear{{Strandet} et~al.,}{{Strandet}
  et~al.}{2017}]{Strandet2017}
{Strandet} M.~L.,  et~al., 2017, \mndoi [\apjl] {10.3847/2041-8213/aa74b0},
  \href {http://adsabs.harvard.edu/abs/2017ApJ...842L..15S} {842, L15}

\bibitem[\protect\citeauthoryear{{Sulentic}, {Rosado}, {Dultzin-Hacyan},
  {Verdes-Montenegro}, {Trinchieri}, {Xu}  \& {Pietsch}}{{Sulentic}
  et~al.}{2001}]{Sulentic2001}
{Sulentic} J.~W.,  {Rosado} M.,  {Dultzin-Hacyan} D.,  {Verdes-Montenegro} L.,
  {Trinchieri} G.,  {Xu} C.,   {Pietsch} W.,  2001, \mndoi [\aj]
  {10.1086/324455}, \href {http://adsabs.harvard.edu/abs/2001AJ....122.2993S}
  {122, 2993}

\bibitem[\protect\citeauthoryear{{Swinbank} et~al.,}{{Swinbank}
  et~al.}{2010}]{Swinbank2010}
{Swinbank} A.~M.,  et~al., 2010, \mndoi [\nat] {10.1038/nature08880}, \href
  {http://adsabs.harvard.edu/abs/2010Natur.464..733S} {464, 733}

\bibitem[\protect\citeauthoryear{{Takeda}, {Sat{\=o}}  \& {Matsuda}}{{Takeda}
  et~al.}{1969}]{Takeda1969}
{Takeda} H.,  {Sat{\=o}} H.,   {Matsuda} T.,  1969, \mndoi [Progress of
  Theoretical Physics] {10.1143/PTP.41.840}, \href
  {http://adsabs.harvard.edu/abs/1969PThPh..41..840T} {41, 840}

\bibitem[\protect\citeauthoryear{{Tamura} et~al.,}{{Tamura}
  et~al.}{2018}]{Tamura2018}
{Tamura} Y.,  et~al., 2018, preprint, \href
  {http://adsabs.harvard.edu/abs/2018arXiv180604132T} {} (\mn@eprint {arXiv}
  {1806.04132})

\bibitem[\protect\citeauthoryear{{Tegmark}, {Silk}, {Rees}, {Blanchard}, {Abel}
   \& {Palla}}{{Tegmark} et~al.}{1997}]{Tegmark1997}
{Tegmark} M.,  {Silk} J.,  {Rees} M.~J.,  {Blanchard} A.,  {Abel} T.,   {Palla}
  F.,  1997, \mndoi [\apj] {10.1086/303434}, \href
  {http://adsabs.harvard.edu/abs/1997ApJ...474....1T} {474, 1}

\bibitem[\protect\citeauthoryear{{Temim}, {Dwek}, {Arendt}, {Borkowski},
  {Reynolds}, {Slane}, {Gelfand}  \& {Raymond}}{{Temim}
  et~al.}{2017}]{temim2017}
{Temim} T.,  {Dwek} E.,  {Arendt} R.~G.,  {Borkowski} K.~J.,  {Reynolds} S.~P.,
   {Slane} P.,  {Gelfand} J.~D.,   {Raymond} J.~C.,  2017, \mndoi [\apj]
  {10.3847/1538-4357/836/1/129}, \href
  {http://adsabs.harvard.edu/abs/2017ApJ...836..129T} {836, 129}

\bibitem[\protect\citeauthoryear{{Thornley}, {Schreiber}, {Lutz}, {Genzel},
  {Spoon}, {Kunze}  \& {Sternberg}}{{Thornley} et~al.}{2000}]{Thornley2000}
{Thornley} M.~D.,  {Schreiber} N.~M.~F.,  {Lutz} D.,  {Genzel} R.,  {Spoon}
  H.~W.~W.,  {Kunze} D.,   {Sternberg} A.,  2000, \mndoi [\apj]
  {10.1086/309261}, \href {http://adsabs.harvard.edu/abs/2000ApJ...539..641T}
  {539, 641}

\bibitem[\protect\citeauthoryear{{Thoul} \& {Weinberg}}{{Thoul} \&
  {Weinberg}}{1995}]{Thoul1995}
{Thoul} A.~A.,  {Weinberg} D.~H.,  1995, \mndoi [\apj] {10.1086/175455}, \href
  {http://adsabs.harvard.edu/abs/1995ApJ...442..480T} {442, 480}

\bibitem[\protect\citeauthoryear{{Todini} \& {Ferrara}}{{Todini} \&
  {Ferrara}}{2001}]{todini2001}
{Todini} P.,  {Ferrara} A.,  2001, \mndoi [\mnras]
  {10.1046/j.1365-8711.2001.04486.x}, \href
  {http://adsabs.harvard.edu/abs/2001MNRAS.325..726T} {325, 726}

\bibitem[\protect\citeauthoryear{{Togi} \& {Smith}}{{Togi} \&
  {Smith}}{2016}]{Togi2016}
{Togi} A.,  {Smith} J.~D.~T.,  2016, \mndoi [\apj]
  {10.3847/0004-637X/830/1/18}, \href
  {http://adsabs.harvard.edu/abs/2016ApJ...830...18T} {830, 18}

\bibitem[\protect\citeauthoryear{{Tornatore}, {Ferrara}  \&
  {Schneider}}{{Tornatore} et~al.}{2007}]{tornatore2007}
{Tornatore} L.,  {Ferrara} A.,   {Schneider} R.,  2007, \mndoi [\mnras]
  {10.1111/j.1365-2966.2007.12215.x}, \href
  {http://adsabs.harvard.edu/abs/2007MNRAS.382..945T} {382, 945}

\bibitem[\protect\citeauthoryear{{Trenti} \& {Stiavelli}}{{Trenti} \&
  {Stiavelli}}{2009}]{Trenti2009}
{Trenti} M.,  {Stiavelli} M.,  2009, \mndoi [\apj]
  {10.1088/0004-637X/694/2/879}, \href
  {http://adsabs.harvard.edu/abs/2009ApJ...694..879T} {694, 879}

\bibitem[\protect\citeauthoryear{{Tsai} et~al.,}{{Tsai}
  et~al.}{2015}]{Tsai2015}
{Tsai} C.-W.,  et~al., 2015, \mndoi [\apj] {10.1088/0004-637X/805/2/90}, \href
  {http://adsabs.harvard.edu/abs/2015ApJ...805...90T} {805, 90}

\bibitem[\protect\citeauthoryear{{Ueda} et~al.,}{{Ueda}
  et~al.}{2018}]{Ueda2018}
{Ueda} Y.,  et~al., 2018, \mndoi [\apj] {10.3847/1538-4357/aa9f10}, \href
  {http://adsabs.harvard.edu/abs/2018ApJ...853...24U} {853, 24}

\bibitem[\protect\citeauthoryear{{Valiante}, {Schneider}, {Bianchi}  \&
  {Andersen}}{{Valiante} et~al.}{2009}]{valiante2009}
{Valiante} R.,  {Schneider} R.,  {Bianchi} S.,   {Andersen} A.~C.,  2009,
  \mndoi [MNRAS] {10.1111/j.1365-2966.2009.15076.x}, \href
  {http://adsabs.harvard.edu/abs/2009MNRAS.397.1661V} {397, 1661}

\bibitem[\protect\citeauthoryear{{Valiante}, {Schneider}, {Salvadori}  \&
  {Bianchi}}{{Valiante} et~al.}{2011}]{valiante2011}
{Valiante} R.,  {Schneider} R.,  {Salvadori} S.,   {Bianchi} S.,  2011, \mndoi
  [\mnras] {10.1111/j.1365-2966.2011.19168.x}, \href
  {http://adsabs.harvard.edu/abs/2011MNRAS.416.1916V} {416, 1916}

\bibitem[\protect\citeauthoryear{{Valiante}, {Schneider}, {Salvadori}  \&
  {Gallerani}}{{Valiante} et~al.}{2014}]{valiante2014}
{Valiante} R.,  {Schneider} R.,  {Salvadori} S.,   {Gallerani} S.,  2014,
  \mndoi [\mnras] {10.1093/mnras/stu1613}, \href
  {http://adsabs.harvard.edu/abs/2014MNRAS.444.2442V} {444, 2442}

\bibitem[\protect\citeauthoryear{{Vallini}, {Dayal}  \& {Ferrara}}{{Vallini}
  et~al.}{2012}]{Vallini2012}
{Vallini} L.,  {Dayal} P.,   {Ferrara} A.,  2012, \mndoi [\mnras]
  {10.1111/j.1365-2966.2012.20551.x}, \href
  {http://cdsads.u-strasbg.fr/abs/2012MNRAS.421.3266V} {421, 3266}

\bibitem[\protect\citeauthoryear{{Vallini}, {Gallerani}, {Ferrara}  \&
  {Baek}}{{Vallini} et~al.}{2013}]{Vallini2013}
{Vallini} L.,  {Gallerani} S.,  {Ferrara} A.,   {Baek} S.,  2013, \mndoi
  [\mnras] {10.1093/mnras/stt828}, \href
  {http://cdsads.u-strasbg.fr/abs/2013MNRAS.433.1567V} {433, 1567}

\bibitem[\protect\citeauthoryear{{Vallini}, {Gallerani}, {Ferrara},
  {Pallottini}  \& {Yue}}{{Vallini} et~al.}{2015}]{Vallini2015}
{Vallini} L.,  {Gallerani} S.,  {Ferrara} A.,  {Pallottini} A.,   {Yue} B.,
  2015, \mndoi [\apj] {10.1088/0004-637X/813/1/36}, \href
  {http://adsabs.harvard.edu/abs/2015ApJ...813...36V} {813, 36}

\bibitem[\protect\citeauthoryear{{Vallini}, {Ferrara}, {Pallottini}  \&
  {Gallerani}}{{Vallini} et~al.}{2017}]{Vallini2017}
{Vallini} L.,  {Ferrara} A.,  {Pallottini} A.,   {Gallerani} S.,  2017, \mndoi
  [\mnras] {10.1093/mnras/stx180}, \href
  {http://adsabs.harvard.edu/abs/2017MNRAS.467.1300V} {467, 1300}

\bibitem[\protect\citeauthoryear{{Vallini}, {Pallottini}, {Ferrara},
  {Gallerani}, {Sobacchi}  \& {Behrens}}{{Vallini} et~al.}{2018}]{Vallini2018}
{Vallini} L.,  {Pallottini} A.,  {Ferrara} A.,  {Gallerani} S.,  {Sobacchi} E.,
    {Behrens} C.,  2018, \mndoi [\mnras] {10.1093/mnras/stx2376}, \href
  {http://adsabs.harvard.edu/abs/2018MNRAS.473..271V} {473, 271}

\bibitem[\protect\citeauthoryear{{Van der Tak} et~al.,}{{Van der Tak}
  et~al.}{2018}]{Tak2018}
{Van der Tak} F.~F.~S.,  et~al., 2018, \mndoi [\pasa] {10.1017/pasa.2017.67},
  \href {http://adsabs.harvard.edu/abs/2018PASA...35....2V} {35, e002}

\bibitem[\protect\citeauthoryear{{Vieira} et~al.,}{{Vieira}
  et~al.}{2013}]{Vieira2013}
{Vieira} J.~D.,  et~al., 2013, \mndoi [\nat] {10.1038/nature12001}, \href
  {http://adsabs.harvard.edu/abs/2013Natur.495..344V} {495, 344}

\bibitem[\protect\citeauthoryear{{Visbal}, {Bryan}  \& {Haiman}}{{Visbal}
  et~al.}{2017}]{Visbal2017}
{Visbal} E.,  {Bryan} G.~L.,   {Haiman} Z.,  2017, \mndoi [\mnras]
  {10.1093/mnras/stx909}, \href
  {http://adsabs.harvard.edu/abs/2017MNRAS.469.1456V} {469, 1456}

\bibitem[\protect\citeauthoryear{{Walter} et~al.,}{{Walter}
  et~al.}{2012}]{Walter2012}
{Walter} F.,  et~al., 2012, \mndoi [\nat] {10.1038/nature11073}, \href
  {http://adsabs.harvard.edu/abs/2012Natur.486..233W} {486, 233}

\bibitem[\protect\citeauthoryear{{Wang} et~al.,}{{Wang}
  et~al.}{2016}]{Wang2016}
{Wang} T.,  et~al., 2016, \mndoi [\apj] {10.3847/0004-637X/816/2/84}, \href
  {http://adsabs.harvard.edu/abs/2016ApJ...816...84W} {816, 84}

\bibitem[\protect\citeauthoryear{{Wang} et~al.,}{{Wang}
  et~al.}{2017}]{Wang2018}
{Wang} L.,  et~al., 2017, preprint, \href
  {http://adsabs.harvard.edu/abs/2017arXiv171007005W} {} (\mn@eprint {arXiv}
  {1710.07005})

\bibitem[\protect\citeauthoryear{{Watson} et~al.,}{{Watson}
  et~al.}{2011}]{Watson2011}
{Watson} D.,  et~al., 2011, \mndoi [\apj] {10.1088/0004-637X/741/1/58}, \href
  {http://cdsads.u-strasbg.fr/abs/2011ApJ...741...58W} {741, 58}

\bibitem[\protect\citeauthoryear{{Wei{\ss}} et~al.,}{{Wei{\ss}}
  et~al.}{2013}]{Weiss2013}
{Wei{\ss}} A.,  et~al., 2013, \mndoi [\apj] {10.1088/0004-637X/767/1/88}, \href
  {http://adsabs.harvard.edu/abs/2013ApJ...767...88W} {767, 88}

\bibitem[\protect\citeauthoryear{{Whalen} et~al.,}{{Whalen}
  et~al.}{2013a}]{Whalen2013a}
{Whalen} D.~J.,  et~al., 2013a, \mndoi [\apj] {10.1088/0004-637X/777/2/110},
  \href {http://adsabs.harvard.edu/abs/2013ApJ...777..110W} {777, 110}

\bibitem[\protect\citeauthoryear{{Whalen} et~al.,}{{Whalen}
  et~al.}{2013b}]{Whalen2013b}
{Whalen} D.~J.,  et~al., 2013b, \mndoi [\apj] {10.1088/0004-637X/778/1/17},
  \href {http://adsabs.harvard.edu/abs/2013ApJ...778...17W} {778, 17}

\bibitem[\protect\citeauthoryear{{Wolfire}, {Hollenbach}  \& {McKee}}{{Wolfire}
  et~al.}{2010}]{Wolfire2010}
{Wolfire} M.~G.,  {Hollenbach} D.,   {McKee} C.~F.,  2010, \mndoi [\apj]
  {10.1088/0004-637X/716/2/1191}, \href
  {http://adsabs.harvard.edu/abs/2010ApJ...716.1191W} {716, 1191}

\bibitem[\protect\citeauthoryear{{Wu}, {Charmandaris}, {Hao}, {Brandl},
  {Bernard-Salas}, {Spoon}  \& {Houck}}{{Wu} et~al.}{2006}]{Wu2006}
{Wu} Y.,  {Charmandaris} V.,  {Hao} L.,  {Brandl} B.~R.,  {Bernard-Salas} J.,
  {Spoon} H.~W.~W.,   {Houck} J.~R.,  2006, \mndoi [\apj] {10.1086/499226},
  \href {http://adsabs.harvard.edu/abs/2006ApJ...639..157W} {639, 157}

\bibitem[\protect\citeauthoryear{{Wu} et~al.,}{{Wu} et~al.}{2012}]{Wu2012}
{Wu} J.,  et~al., 2012, \mndoi [\apj] {10.1088/0004-637X/756/1/96}, \href
  {http://adsabs.harvard.edu/abs/2012ApJ...756...96W} {756, 96}

\bibitem[\protect\citeauthoryear{{Xie}, {Li}  \& {Hao}}{{Xie}
  et~al.}{2017}]{Xie2017}
{Xie} Y.,  {Li} A.,   {Hao} L.,  2017, \mndoi [\apjs]
  {10.3847/1538-4365/228/1/6}, \href
  {http://adsabs.harvard.edu/abs/2017ApJS..228....6X} {228, 6}

\bibitem[\protect\citeauthoryear{{Xu}, {Norman}, {O'Shea}  \& {Wise}}{{Xu}
  et~al.}{2016}]{xu2016}
{Xu} H.,  {Norman} M.~L.,  {O'Shea} B.~W.,   {Wise} J.~H.,  2016, \mndoi [\apj]
  {10.3847/0004-637X/823/2/140}, \href
  {http://adsabs.harvard.edu/abs/2016ApJ...823..140X} {823, 140}

\bibitem[\protect\citeauthoryear{{Yajima} \& {Khochfar}}{{Yajima} \&
  {Khochfar}}{2017}]{Yajima2017}
{Yajima} H.,  {Khochfar} S.,  2017, \mndoi [\mnras] {10.1093/mnrasl/slw249},
  \href {http://adsabs.harvard.edu/abs/2017MNRAS.467L..51Y} {467, L51}

\bibitem[\protect\citeauthoryear{{Yajima}, {Shlosman}, {Romano-D{\'{\i}}az}  \&
  {Nagamine}}{{Yajima} et~al.}{2015}]{Yajima2015}
{Yajima} H.,  {Shlosman} I.,  {Romano-D{\'{\i}}az} E.,   {Nagamine} K.,  2015,
  \mndoi [\mnras] {10.1093/mnras/stv974}, \href
  {http://cdsads.u-strasbg.fr/abs/2015MNRAS.451..418Y} {451, 418}

\bibitem[\protect\citeauthoryear{{Yoshida}, {Abel}, {Hernquist}  \&
  {Sugiyama}}{{Yoshida} et~al.}{2003}]{Yoshida2003}
{Yoshida} N.,  {Abel} T.,  {Hernquist} L.,   {Sugiyama} N.,  2003, \mndoi
  [\apj] {10.1086/375810}, \href
  {http://adsabs.harvard.edu/abs/2003ApJ...592..645Y} {592, 645}

\bibitem[\protect\citeauthoryear{{Yoshida}, {Hosokawa}  \& {Omukai}}{{Yoshida}
  et~al.}{2012}]{Yoshida2012}
{Yoshida} N.,  {Hosokawa} T.,   {Omukai} K.,  2012, \mndoi [Progress of
  Theoretical and Experimental Physics] {10.1093/ptep/pts022}, \href
  {http://adsabs.harvard.edu/abs/2012PTEP.2012aA305Y} {2012, 01A305}

\bibitem[\protect\citeauthoryear{{Younger} et~al.,}{{Younger}
  et~al.}{2007}]{Younger2007}
{Younger} J.~D.,  et~al., 2007, \mndoi [\apj] {10.1086/522776}, \href
  {http://adsabs.harvard.edu/abs/2007ApJ...671.1531Y} {671, 1531}

\bibitem[\protect\citeauthoryear{{Zackrisson}, {Rydberg}, {Schaerer},
  {{\"O}stlin}  \& {Tuli}}{{Zackrisson} et~al.}{2011}]{zackrisson2011}
{Zackrisson} E.,  {Rydberg} C.-E.,  {Schaerer} D.,  {{\"O}stlin} G.,   {Tuli}
  M.,  2011, \mndoi [\apj] {10.1088/0004-637X/740/1/13}, \href
  {http://adsabs.harvard.edu/abs/2011ApJ...740...13Z} {740, 13}

\bibitem[\protect\citeauthoryear{{Zackrisson}, {Gonz{\'a}lez}, {Eriksson},
  {Asadi}, {Safranek-Shrader}, {Trenti}  \& {Inoue}}{{Zackrisson}
  et~al.}{2015}]{Zackrisson2015}
{Zackrisson} E.,  {Gonz{\'a}lez} J.,  {Eriksson} S.,  {Asadi} S.,
  {Safranek-Shrader} C.,  {Trenti} M.,   {Inoue} A.~K.,  2015, \mndoi [\mnras]
  {10.1093/mnras/stv492}, \href
  {http://adsabs.harvard.edu/abs/2015MNRAS.449.3057Z} {449, 3057}

\bibitem[\protect\citeauthoryear{{Zavala} et~al.,}{{Zavala}
  et~al.}{2018}]{Zavala2018}
{Zavala} J.~A.,  et~al., 2018, \mndoi [Nature Astronomy]
  {10.1038/s41550-017-0297-8}, \href
  {http://adsabs.harvard.edu/abs/2018NatAs...2...56Z} {2, 56}

\makeatother
\end{thebibliography}

\end{document}